\documentclass[12pt,a4paper]{article}

\usepackage[a4paper,text={170mm,257mm},centering]{geometry}

\usepackage[english]{babel}

\usepackage[latin1]{inputenc}
\usepackage[toc,page]{appendix}
\usepackage{amsfonts,amsbsy,bm,euscript,mathrsfs}
\usepackage{amssymb,stmaryrd,faktor,slashed}
\usepackage{color}
\usepackage[tbtags]{amsmath}
\usepackage[bookmarks=true,colorlinks=true,linkcolor=black,citecolor=black,urlcolor=black,bookmarksnumbered]{hyperref}

\usepackage[nosort]{cite}

\usepackage{authblk, physics, multirow, comment}
\usepackage{graphicx}
\usepackage[verbose]{wrapfig}
\usepackage{caption}

\DeclareMathAlphabet{\mathpzc}{OT1}{pzc}{m}{it}

\numberwithin{equation}{section}

\newcommand{\bF}{\mathbf{F}}
\newcommand{\bR}{\mathbf{R}}
\newcommand{\be}{\mathbf{e}}
\newcommand{\bs}{\mathbf{s}}
\newcommand{\bg}{\mathbf{g}}
\newcommand{\bH}{\mathbf{H}}
\newcommand{\bC}{\mathbf{C}}
\newcommand{\bA}{\mathbf{A}}
\newcommand{\bB}{\mathbf{B}}
\newcommand{\bQ}{\mathbf{Q}}
\newcommand{\bP}{\mathbf{P}}
\newcommand{\bV}{\mathbf{V}}
\newcommand{\bu}{\mathbf{u}}
\newcommand{\bv}{\mathbf{v}}
\newcommand{\bG}{\mathbf{G}}
\newcommand{\bM}{\mathbf{M}}
\newcommand{\bTh}{\mathbf{\Theta}}
\newcommand{\bPsi}{\mathbf{\Psi}}

\newcommand{\bbPsi}{\bar{\mathbf{\Psi}}}
\newcommand{\bGa}{\mathbf{\Gamma}}
\newcommand{\bPhi}{\mathbf{\Phi}}
\newcommand{\bom}{\bm{\omega}}
\newcommand{\btau}{\bm{\tau}}
\newcommand{\bla}{\bm{\lambda}}
\newcommand{\bbla}{\bar{\bm{\lambda}}}
\newcommand{\bep}{\bm{\epsilon}}
\newcommand{\bbep}{\bar{\bm{\epsilon}}}

\newcommand{\cF}{\mathcal{F}}
\newcommand{\cA}{\mathcal{A}}
\newcommand{\cg}{\mathpzc{g}}
\newcommand{\cR}{\mathcal{R}}
\newcommand{\ce}{\mathpzc{e}} 
\newcommand{\cPsi}{\Psi^{\text{c}}{}}
\newcommand{\cbPsi}{\overline{\Psi^{\text{c}}}{}}
\newcommand{\com}{\mathpzc{w}}
\newcommand{\cep}{\zeta}

\newcommand{\ie}{{\it i.e.\ }}

\newcommand{\eg}{{\it e.g.\ }}

\DeclareMathOperator{\sgn}{sgn}

\begin{document}

\begin{flushright}\small{Imperial/TP/2024/KS/{01}}\\ \small{DESY-24-179}\end{flushright}
\vspace{0.1cm}

\begin{center}
			{\Large\bf Compactification on Calabi-Yau threefolds: \\Consistent truncation to pure supergravity 
		 
			\vspace{0.1cm}
		}
		\vspace{1.0cm}		
		{Jieming Lin$^{a,}$\footnote{jieming.lin22@imperial.ac.uk}, Torben Skrzypek$^{b,}$\footnote{torben.skrzypek@desy.de} and 
			K.S. Stelle$^{a,}$\footnote{k.stelle@imperial.ac.uk}
		}	
		
		\vspace{0.3cm}		
		
		{\em
			$^{a}$Abdus Salam Centre for Theoretical Physics, Imperial College London,\\ Prince Consort Road, London, SW7 2AZ, UK\\
			$^{b}$ Deutsches Elektronen-Synchrotron DESY, Notkestra{\ss}e 85, 22607 Hamburg, Germany
		}		
\end{center}	
\vspace{0.5cm}
	
\begin{abstract}
We study compactifications of eleven- and ten-dimensional maximal supergravity on Calabi-Yau threefolds. We explicitly construct truncations to pure supergravity with eight supercharges in five and four dimensions and show that they are consistent, \ie that every solution of the lower-dimensional equations of motion fully solves the higher-dimensional ones. We furthermore match the supersymmetry transformations and demonstrate the consistency to full non-linear order in fermions. Our construction is independent of the choice of Calabi-Yau threefold and only involves the universal structures such as the K\"ahler form and the holomorphic three-form, in agreement with implicit constructions in the generalised geometry literature. As an immediate application, we embed four-dimensional extremal black holes in the higher-dimensional supergravities. We furthermore propose Ans\"atze for consistent truncations on all universal structures, leading to supergravities with additional matter multiples. An extensive list of equations of motion and supersymmetry transformations for various supergravity theories is provided in the appendix.
\end{abstract}

\newpage
\tableofcontents
\newpage

\setcounter{footnote}{0}

\section{Introduction}\label{intro}

Compactifications on Calabi-Yau threefolds (CY3s) are frequently studied in string phenomenology, allowing one to construct realistic effective field theories (EFTs) in four dimensions that are related to a stringy UV-completion (see \eg \cite{Grana:2005jc,Becker:2006dvp,Tomasiello:2022dwe} for reviews). For example, one may take type IIB supergravity, which is the low-energy EFT of type IIB string theory, and consider it on a spacetime of the vacuum form
\begin{equation}
\mathbb{R}^{1,3}\cross \text{CY}3\,.
\end{equation}
At energies far below the characteristic (inverse length) scale of the CY3, one may integrate out the resulting massive Kaluza-Klein modes encoding the dependence of fields on the CY3-space and treat the lowest-mass (\ie massless) modes as a 4d EFT. The character of this lowest-modes EFT then comes into question. Generically, the process of integrating out massive Kaluza-Klein modes generates higher-derivative terms.\footnote{This effect was studied in \cite{Duff:1989cr}, making use of 
the structure imposed by the amount of supersymmetry surviving in the Calabi-Yau compactification. In compactifications of maximal (32 supercharge) $D=10$ or $D=11$ supergravities, higher-derivative corrections will be generated while in compactifications of the $D=10$ half-maximal (16 supercharge) heterotic supergravity, it was argued that there could already be EFT corrections at the two-derivative level. }
These can be disregarded at sufficiently low energies, \eg when constructing phenomenologically relevant solutions. Since CY3s allow for a covariantly constant Killing spinor, this procedure retains $\mathcal{N}=2$ supersymmetry and results in a 4d $\mathcal{N}=2$ supergravity theory with various matter multiplets. However, solutions of these theories will generally not solve the higher-derivative-corrected equations of motion and therefore will not lift to valid solutions of the higher-dimensional theories.

The question now arises whether such lifting can be achieved at least for some subset of fields in the 4d supergravity, forming a {\em consistent truncation} of the low-energy EFT. In a truncation, the retained modes are taken not to depend on the CY3 space at all. A consistent truncation is one in which arbitrary solutions of the resulting lower-dimensional field equations are, nonetheless, valid solutions of the original higher-dimensional theory when expressed back in the original higher dimension theory using the structure of the Kaluza-Klein Ansatz. The program of finding consistent truncations has a long history in a variety of compactification scenarios \cite{Duff:1984hn, Nastase:1999kf, Gauntlett:2007ma, Gauntlett:2007sm, Gauntlett:2009zw, Gauntlett:2010vu, Donos:2010ax,Baguet:2015sma,Varela:2015ywx, MatthewCheung:2019ehr, Josse:2021put, Cheung:2022ilc, Couzens:2022lvg, Faedo:2022rqx}.

In \cite{Cassani:2019vcl}, it was argued that a restriction of the 10d field decomposition to invariant tensors under the structure group $SU(3)$ of the CY3 yields a consistent truncation including the 4d $\mathcal{N}=2$ supergravity multiplet and two matter multiplets. The consistency of this construction is guaranteed because no combination of the retained $SU(3)$-singlets can source any of the non-singlet representations that were truncated away. It should be noted that such arguments based on reduction-manifold structure groups are a significant extension of the well-known prescription \cite{DeWitt1965dynamical} for consistent truncations based on reduction to the invariant sector of a Killing symmetry of the reduction manifold. In cases of compact Calabi-Yau manifold reductions, there are no such continuous Killing symmetries, however, so a detailed analysis of the formal consistency argument based instead on the CY {\em structure group} is in order.\footnote{At the bosonic level, Refs \cite{Terrisse_2019, tsimpis2020} constructed an embedding of the $SU(3)$-invariant sector into type IIA supergravity. This was done together with an analysis of other backgrounds and the types of higher-derivative corrections resulting from more general (inconsistent) reductions.}

In this paper we perform an explicit construction of a CY3 truncation in ordinary supergravity language and give a subsequent proof of consistency. We restrict our attention to the pure supergravity subsector which only involves the supergravity multiplet and which itself constitutes a consistent truncation.\footnote{In \cite{Leung:2022nhy} it was conjectured that a consistent truncation to pure supergravity is always possible in supersymmetric braneworld scenarios. The present paper confirms this conjecture in the analogous context of Calabi-Yau compactifications.} We comment on adding the additional hypermultiplets but leave a detailed analysis thereof to future work.

Most situations of interest only require the spectrum of excitations around a classical bosonic background and therefore do not depend on terms which are non-linear in fermions. At this level, one can formulate a na{\"\i}ve truncation Ansatz which solves the bosonic equations of motion consistently. However, in order to discuss interactions and to prove consistency of supersymmetry, we have to study the full non-linear equations of motions and supersymmetry transformations. To achieve consistency also in these terms, we have to correct our na{\"\i}ve Ansatz by fermion bilinears. The correct bilinears can be determined by studying supercovariant quantities. Take for example the 3-form field strength $\bF_{MNP}$ in the IIB supergravity, which we na{\"\i}vely want to truncate out. This field strength is not covariant with respect to the gauged supersymmetry transformations, so we have to supplement it by fermion bilinears to arrive at the supercovariant quantity
\begin{equation}
\hat{\bF}_{ABC}= \bF_{ABC}-3\overline{\bPsi^c}_{[A}\Gamma_{BC]}\bla-3\bar{\bPsi}_{[A}\Gamma_{B}\bPsi_{C]}\,.
\end{equation}
Given that the gravitino is retained in the truncation, one can identify the correct fermion bilinear to which $\bF_{ABC}$ must be equated in the truncation:
\begin{equation}
    \hat{\bF}_{ABC}=0 \qquad \rightarrow \qquad \bF_{ABC}=3\bar{\bPsi}_{[A}\Gamma_{B}\bPsi_{C]}\,.
\end{equation}
Note that this solves the Bianchi ``identity'' only on-shell, \ie up to the gravitino equation of motion; this is not incompatible with the consistency of the truncation, however.\\

Following the above general strategy, we gather the necessary Ans\"atze for Calabi-Yau reductions to pure supergravity in Section \ref{Ansatz}. We first consider 11d supergravity and reduce it to pure 5d supergravity with 8 supercharges. We then consider an additional $S^1$-reduction to pure 4d $\mathcal{N}=2$ supergravity. The corresponding Ansatz is given in Section \ref{Ansatz11d}. Alternatively, one may first dimensionally reduce 11d supergravity down to 10d type IIA supergravity and then perform the reduction on a CY3. We present the relevant IIA Ansatz in Section \ref{AnsatzIIA}. Finally, we consider the CY3-reduction of type IIB supergravity down to pure 4d $\mathcal{N}=2$ supergravity in Section \ref{AnsatzIIB}.

In Section \ref{proof} we show that these Ans\"atze lead to consistent truncations. We first demonstrate the consistency of the equations of motion at bosonic level. We then move on to check that the supersymmetry transformations in higher dimensions induce the lower-dimensional ones on-shell. Although these consistency analyses are convincing in themselves, we present the final check of consistency for the full equations of motion including fermions in Appendix \ref{fermionEoM}, concluding the proof. 

In Section \ref{BH} we present an application of these results by embedding extremal BPS black holes in the 4d $\mathcal{N}=2$ supergravity and uplifting this solution to 11d and 10d. We comment on their interpretation as extended objects wrapping cycles in the CY3. In Section \ref{Matter} we propose, in accordance with Ref.\ \cite{Cassani:2019vcl}, an embedding Ansatz including the vector multiplet and hypermultiplets in the truncation. This Ansatz is guided by the same principles as for the pure supergravity Ansatz but this has not yet been checked for consistency. We conclude in Section \ref{conclusion} and point towards future directions. 

Our conventions are summarised in Appendix \ref{conventions}, where we also list the equations of motion and supersymmetry transformations, including all fermion terms, of all supergravity theories discussed in this paper. We hope this appendix will prove useful for comparing our results with work in different conventions.

\section{Ans\"atze for consistent truncations}\label{Ansatz}
In this section\footnote{Details on field content, equations of motion and supersymmetry transformations of the various theories discussed here can be found in Appendix \ref{conventions}, which also provides an overview of our naming and fermion conventions. Let us just mention here that bold letters refer to 11/10d fields, curly letters to 5d and regular ones to 4d if not explicitly part of the CY3 data. Greek indices denote 5d/4d coordinates, capital Latin indices denote 11/10d coordinates and lowercase Latin indices refer to the CY3.}  we collate the Ans\"atze for embedding the effective 5 and 4d fields of our truncation into the 11 and 10d fields on the full spacetime $\mathcal{M}_{4/5}\cross \text{CY}3 $. We will show their consistency in the following section. Before listing the embedding Ans\"atze, however, let us first review some facts about Calabi-Yau 3-folds which will be useful in the following.

A Calabi-Yau n-fold is a compact K\"ahler manifold with vanishing first Chern class. As such, its metric can (at least locally) be generated by a single real function $\mathcal{K}$, the K\"ahler potential. In complex coordinates, this K\"ahler potential specifies the metric $g_{i\bar{\jmath}}=\partial_i\partial_{\bar{\jmath}}\mathcal{K}$, with all other entries vanishing. The associated K\"ahler form is given by $\omega_{i\bar{\jmath}}=ig_{i\bar{\jmath}}dy^i\wedge d\bar{y}^{\bar{\jmath}}$. 

An important characteristic of Calabi-Yau n-folds is their $SU(n)$ structure group. In three complex dimensions, a manifold with an $SU(3)$ structure group has invariant tensors $\{\omega, \Omega\}$, where $\omega$ is the real K\"ahler $(1,1)$-form and $\Omega$ a holomorphic 3-form. For a CY3, the invariant tensors should satisfy the integrability conditions
\begin{equation}
    \dd\omega = \dd\Omega = \omega\wedge\Omega = 0\,.
\end{equation}
Furthermore, the $SU(3)$ structure group allows for a covariantly constant Killing spinor $\eta_+$ and its charge-conjugate $\eta_-$. They are Weyl spinors of opposite chirality. The invariant tensors can be rewritten in terms of spinor bilinears as 
\begin{equation}
    \omega_{mn}=\pm i\eta_\mp^\dagger\Sigma_{mn}\eta_\mp\,,\qquad \Omega_{mnp}=-\eta_-^\dagger\Sigma_{mnp}\eta_+\,,\qquad \bar{\Omega}_{mnp}=\eta_+^\dagger\Sigma_{mnp}\eta_-\,,
\end{equation}
where $\Sigma$ denotes the 6d Clifford algebra as discussed in Appendix \ref{conventions}.
We choose the normalisation
\begin{equation}
    \eta^\dagger_\pm \eta_\pm = 1\,,\qquad \text{vol}_{\text{CY3}}=-\frac{i}{8}\Omega\wedge\bar{\Omega}=-\frac16\omega\wedge\omega\wedge\omega\,,
    \label{eq:CY-geo}
\end{equation}
and note some other useful relations
\begin{align}\label{relation}
    &* \omega = -\frac12 \omega\wedge\omega\,,\qquad \omega_{pq}\omega^{pq}=6\,,\qquad \omega_{m}{}^p \omega_{np} = g_{mn}\,,\\
    &* \Omega = i\Omega\,,\qquad * \bar{\Omega} = -i\bar{\Omega}\,.\label{relation2}
\end{align}
Another important fact is 
\begin{equation}
    \eta_a^\dagger \Sigma_m \eta_b = 0\,,\quad a,b = \pm\,,
    \label{relation3}
\end{equation}
which does not admit a Killing vector on a generic irreducible CY3. More explicitly, the left-hand side vanishes due to a chirality mismatch when $a=b$, and antisymmetry of $\Sigma_m$ when $a\neq b$.

If we decompose the spinor, 2-form, and 3-form representations of $SO(6)$ on CY3 into $SU(3)$ representations, we find that the invariant tensors $\{\eta,\omega,\Omega\}$ are precisely the relevant $SU(3)$-singlets. More specifically, the spinor $\mathbf{4}$, 2-from $\mathbf{15}$, and 3-form $\mathbf{20}$ representations of $SO(6)$ in the internal 6d space decompose under  $SU(3)$ as \begin{equation}
\mathbf{4}\rightarrow\mathbf{3}+\mathbf{1}\,,\qquad\mathbf{15}\rightarrow\mathbf{8}+\mathbf{3}+\bar{\mathbf{3}}+\mathbf{1}\,,\qquad\mathbf{20}\rightarrow\mathbf{6}+\bar{\mathbf{6}}+\mathbf{3}+\bar{\mathbf{3}}+\mathbf{1}+\mathbf{1}\,.
\end{equation}
A similar decomposition of the vector fields instead yields no singlet at all, providing another perspective on the relation \eqref{relation3}. 
As stated above and in \cite{Cassani:2019vcl}, the Ans\"atze for consistent truncations should only involve the structures $\{\eta,\omega,\Omega\}$. This furthermore makes them independent of the specific details of the chosen CY3, such as, for example, the Hodge numbers $h^{1,1}$ and $h^{2,1}$. This allows us to essentially guess the correct embedding Ansatz at least to leading order in fermions.
\subsection{From 11d to pure 5d and 4d supergravity}\label{Ansatz11d}

Let us first compactify 11d supergravity on an arbitrary CY3. Our na{\"\i}ve bosonic embedding Ansatz is 
\begin{equation}
    \dd\bs_{11}^2= \cg_{\mu\nu}(x)\dd x^\mu \dd x^\nu + \partial_{i}\partial_{\bar{\jmath}}\mathcal{K}(y)\dd y^i \dd\bar{y}^{\bar{\jmath}}\,,\qquad \bA_{\mu mn}= - \sqrt{\frac13} \cA_\mu(x)\cdot\omega_{mn}\,,
    \label{eq:11d-5d-bosonic-ansatz}
\end{equation}
where $\cg_{\mu\nu}$ and $\cA_\mu$ denote the 5d metric and gauge field, respectively, and all other components are set to $0$. The gravitino takes a factorised form 
\begin{equation}
    \bPsi_{\mu} = \Psi_\mu \otimes \eta + \cPsi_\mu\otimes \tilde{\eta}\,,
    \label{eq:11d-5d-fermionic-ansatz}
\end{equation}
where $\Psi_\mu$ is the spin-3/2 gravitino in 5d with 4 complex components and 
\begin{equation}\label{eta}
    \eta=\frac{1}{\sqrt{2}}(\eta_+ + \eta_-),\quad \tilde{\eta}=\frac{i}{\sqrt{2}}\left(\eta_{-}-\eta_{+}\right)
\end{equation}
are related to the Killing spinor on the CY3. The $\bPsi_m$ components vanish. Similarly, the supersymmetry transformation parameter is factorised as $\bep = \cep \otimes \eta + \cep^c \otimes \tilde{\eta}$. The 11d gamma matrices and intertwiners (defined in Appendix \ref{conventionsFermions}) are factorised according to \eqref{eq:11-5intertwiner}. 

Further explanation is needed to see that \eqref{eq:11d-5d-fermionic-ansatz} constitutes a Majorana gravitino. 
Notice that both charge conjugation operators $\tilde{D}^{(5)}$ and $D^{(6)}$ generate symplectic Majorana representations, resulting in an overall Majorana representation for the 11d gravitino, as is required. In this context $\tilde{\eta} = (D^{(6)}\eta)^*$ is the symplectic Majorana conjugate to $\eta$, justifying our parametrisation \eqref{eta}.

The Ansatz \eqref{eq:11d-5d-bosonic-ansatz} and \eqref{eq:11d-5d-fermionic-ansatz} is enough to uplift any solution of 5d supergravity. However, if we want to show consistency at the full non-linear level in fermions, we have to introduce further spinor bilinear corrections to the Ansatz, such that the full equations of motion remain satisfied and 11d supersymmetry transformations reduce to the 5d ones. We can infer the appropriate spinor bilinears by requiring a matching of supercovariant quantities. Indeed, we can see that under the Ansatz \eqref{eq:11d-5d-bosonic-ansatz} and \eqref{eq:11d-5d-fermionic-ansatz}, the 11d supercovariant spin connection and field strength (\ref{eq:11d-supercov}) reduce to the 5d ones (\ref{eq:5d-supercov}) 
\begin{equation}
    \hat{\bom}_{\mu \underline{\nu\rho}}=\hat{\com}_{\mu \underline{\nu\rho}}\,,\quad \hat{\bom}_{m\underline{np}} = \omega_{m\underline{np}}\,,\quad \hat{\bF}_{\mu\nu mn}= - \frac{1}{\sqrt{3}}\hat{\cF}_{\mu\nu}\omega_{mn}\,,
\end{equation}
except for the $\mu\nu\rho\sigma$-component of $\hat{\bF}_{(4)}$, which picks up a term
\begin{equation}
    \hat{\bF}_{\mu\nu\rho\sigma}=-3\bar{\bPsi}_{[\mu}\Gamma_{\nu\rho}\bPsi_{\sigma]}\,.
    \label{eq:11d-ansatz-nonvanishing}
\end{equation}
We see that our na{\"\i}ve assumption of $\bF_{\mu\nu\rho\sigma}=0$ was not a good supercovariant choice, so we should instead require $\hat\bF_{\mu\nu\rho\sigma}=0$. 
We therefore improve our Ansatz by 
\begin{equation}
    \bF_{\mu\nu\rho\sigma}=0\qquad \rightarrow \qquad \bF_{\mu\nu\rho\sigma}=3\bar{\bPsi}_{[\mu}\Gamma_{\nu\rho}\bPsi_{\sigma]}\,,
    \label{eq:11d-ansatz-corrected}
\end{equation}
so that $\hat{\bF}_{\mu\nu\rho\sigma}=0$ is the correct Ansatz truncation. In terms of 5d fields, this component takes the form
\begin{equation}
    \bF_{\mu\nu\rho\sigma}=3\left(\bar{\Psi}_{[\mu}\gamma_{\nu\rho}\Psi_{\sigma]} + \cPsi_{[\mu}\gamma_{\nu\rho}\cPsi_{\sigma]}\right)\,,
    \label{eq:11d-4form}
\end{equation}
and by making use of the 5d gravitino equation of motion, one can show that this 4-form is closed. This allows us to implicitly include the generating potential $\bA_{\mu\nu\rho}$ in our Ansatz. 
In Section \ref{Consistency11-4}, we will show that the full embedding Ansatz \eqref{eq:11d-5d-bosonic-ansatz}, \eqref{eq:11d-5d-fermionic-ansatz} and \eqref{eq:11d-ansatz-corrected} is indeed consistent.

\

It is now straightforward to compactify the 5d  supergravity on a circle, achieving a 4d $\mathcal{N}=2$ supergravity with a vector multiplet $\{B_\mu, \lambda^I, \varphi^I\}$ \cite{Chamseddine:1980mpx}. A further truncation is needed to get to pure 4d $\mathcal{N}=2$ supergravity (if that is the desired result). We recall the compactification procedure on a circle in a purely bosonic background. The embedding Ansatz is
\begin{equation}
    \dd s_5^2=e^{-\frac{1}{\sqrt{3}}\varphi}\dd s_4^2 + e^{\frac{2}{\sqrt{3}}\varphi}\left(\dd z+B^1_\mu \dd x^\mu \right)^2\,,\qquad \cA_{(1)} = B^2_{(1)} + \chi \dd z\,.
    \label{eq:5d-4d-1}
\end{equation}
$B^1_\mu$ is the Kaluza-Klein vector with field strength $G^1_{(2)}=\dd B_{(1)}^1$. We define the field strength of the reduced vector field as 
\begin{equation}
    G^2_{(2)} = \dd B^2_{(1)} - \dd \chi \wedge B^1_{(1)}\,.
    \label{eq:5d-4d-2}
\end{equation}
The fermionic embedding Ansatz is 
\begin{equation}
    \begin{aligned}
        &\Psi_\mu = e^{\frac{1}{4\sqrt{3}}\varphi}\psi_\mu + \frac{1}{2\sqrt{2}}\gamma_\mu\gamma^{(4)}e^{-\frac{\sqrt{3}}{4}\varphi}\lambda\,,\quad \Psi_4 = \frac{i}{\sqrt{2}}e^{-\frac{\sqrt{3}}{4}}\lambda\,,\quad \cep = e^{\frac{1}{4\sqrt{3}}\varphi}\epsilon\,, \\ 
    \end{aligned}
    \label{eq:5d-4d-3}
\end{equation}
The factorisation of gamma matrices and intertwiners are chosen according to \eqref{eq:5-4-gamma matrices} and \eqref{eq:5-4-intertwiners}.

If we want to truncate to pure 4d $\mathcal{N}=2$ supergravity, we can set the scalars and dilatini to zero $(\varphi=\chi=\lambda=0)$ and fix the two gauge fields to satisfy the following duality relation 
\begin{equation}
     G^2_{(2)} = -*\sqrt{3}G^1_{(2)} = -*\frac{\sqrt{3}}{2}F_{(2)}\,,
    \label{eq:4dN2-trunc-vector}
\end{equation} 
with $F_{(2)} = \dd A_{(1)}$.
This argument provides us with a na{\"\i}ve embedding Ansatz. As before, we should compare the  supercovariant spin connection and field strengths in 5d \eqref{eq:5d-supercov} with the ones in 4d \eqref{eq:4d-supercov} in order to find the appropriate fermionic completion. We find a consistent truncation with the following embedding 
\begin{equation}
    \begin{aligned}
        &\dd s_5^2 = \dd s_4^2 + \left(\dd z+\frac12 A_\mu  \dd x^\mu \right)^2\,,\qquad \Psi_\mu = \psi_\mu = \psi^1_\mu + \psi_{2\mu} \,,\qquad \cep=\epsilon = \epsilon^1 + \epsilon_{2}\,, \\
        &\cF_{\mu\nu} = -\frac{\sqrt{3}}{4}\epsilon_{\mu\nu\rho\sigma}\left(F^{\rho\sigma} - \frac{i}{2}\epsilon^{\rho\sigma\lambda\eta}\left(\bar{\psi}_\lambda^I \psi_\eta^J\varepsilon_{IJ} - \bar{\psi}_{I\lambda} \psi_{J\eta}\varepsilon^{IJ} \right) + \left(\bar{\psi}^{\rho I} \psi^{\sigma J}\varepsilon_{IJ} + \bar{\psi}_I^\rho\psi_J^\sigma\varepsilon^{IJ} \right)\right)\,.
    \end{aligned}
    \label{eq:5d-4d-ansatz-corrected}
\end{equation}
We separate the Dirac spinors in 4d into Weyl spinors and use the upper or lower placement of the $SU(2)$ indices $I,J$ to denote chirality of the Weyl spinors (see Appendix \ref{conventionsFermions}). Under this Ansatz, the supercovariant spin connection and field strength reduce as 
\begin{equation}
     \hat{\com}_{\mu\underline{\nu\rho}}=\hat{\omega}_{\mu\underline{\nu\rho}}\,,\quad \hat{\com}_{\underline{\nu}5\underline{\mu}} = \frac14 \hat{F}_{\underline{\mu\nu}}\,,\quad \hat{\com}_{5\underline{\nu\mu}}=\frac14 \hat{F}_{\underline{\mu\nu}}\,, \quad  \hat{\com}_{m\underline{np}} = \omega_{m\underline{np}}\,, \quad \hat{\cF}_{\mu\nu} = -\frac{\sqrt{3}}{4} \epsilon_{\mu\nu\rho\sigma}\hat{F}^{\rho\sigma}\,.
\end{equation}
In Section \ref{Consistency11-4} (and Appendix \ref{fermionEoM}) we will show the consistency of this embedding Ansatz.

\subsection{From IIA to pure 4d supergravity}\label{AnsatzIIA}

Following the same logic as in the previous case we arrive at an embedding Ansatz with non-vanishing components given by
\begin{equation}
    \begin{aligned}
        &\dd\bs^2_{10} = g_{\mu\nu}(x) \dd x^\mu \dd x^\nu + \partial_i\partial_{\overline{\jmath}}\mathcal{K}(y)\dd y^i\dd \overline{y}^{\overline{\jmath}}\,,\qquad \bC_\mu=\frac{1}{2}A_\mu\,, \\
        &\bF_{\mu\nu mn} = \frac14\epsilon_{\mu\nu\rho\sigma}\left(F^{\rho\sigma} - \frac{i}{2}\epsilon^{\rho\sigma\lambda\eta}\left(\bar{\psi}^I_\lambda \psi_\eta^J\varepsilon_{IJ} - \bar{\psi}_{I\lambda} \psi_{J\eta}\varepsilon^{IJ} \right) + \left(\bar{\psi}^{I\rho} \psi^{\sigma J}\varepsilon_{IJ} + \bar{\psi}_I^\rho\psi_J^\sigma\varepsilon^{IJ} \right) \right) \omega_{mn}\,,\\
        &\bH_{\mu\nu\rho}=3\left(\bar{\psi}^1_{[\mu}\gamma_\nu\psi_{\rho]1} -  \bar{\psi}^2_{[\mu}\gamma_\nu\psi_{\rho]2}\right)\,,\quad \bF_{\mu\nu\rho\sigma} = -6\left( \bar{\psi}^1_{[\mu}\gamma_{\nu\rho}\psi^2_{\sigma]} + \bar{\psi}_{1[\mu}\gamma_{\nu\rho}\psi_{\sigma]2}\right) - 4\bC_{[\mu}\bH_{\nu\rho\sigma]}\,, \\
        &\bPsi_\mu  = \frac{1}{\sqrt{2}}\left(\psi^1_\mu\otimes\eta_+ +\psi_{1\mu}\otimes\eta_- + \psi_{2\mu}\otimes\eta_+ +\psi^2_{\mu}\otimes\eta_-\right)\,,\\
        &\bep  = \frac{1}{\sqrt{2}}\left(\epsilon^1\otimes\eta_+ +\epsilon_1 \otimes\eta_-  + \epsilon_2\otimes\eta_+ +\epsilon^2 \otimes\eta_- \right)\,.\\
    \end{aligned}
    \label{eq:IIA-4d-ansztz-corrected}
\end{equation}
Here, $\{\psi^I_\mu\,,\psi_{I\mu}\}$ and $\{\epsilon^I\,,\epsilon_{I}\}$ are two pairs of charge conjugated spinors, and hence the Ansatz \eqref{eq:IIA-4d-ansztz-corrected} satisfies the appropriate Majorana conditions for 10d gravitino and dilatino.
The gamma matrices and intertwiners should factorise according to \eqref{eq:10-4intertwiner}.

Under this embedding Ansatz, the 10d supercovariant quantities reduce to 4d supercovariant quantities as
\begin{equation}
    \hat{\bF}_{\mu\nu} = \frac12 \hat{F}_{\mu\nu},\quad \hat{\bF}_{\mu\nu mn}=\frac14\epsilon_{\mu\nu\rho\sigma }\hat{F}^{\rho\sigma }\omega_{mn},\quad \hat{\bom}_{\mu\underline{\nu\rho}} = \hat{\omega}_{\mu\underline{\nu\rho}}\,,\quad \hat{\bom}_{m\underline{np}} = \omega_{m\underline{np}}.
\end{equation}
We will show consistency of this Ansatz in Section \ref{ConsistencyIIA-4}.

\subsection{From IIB to pure 4d supergravity}\label{AnsatzIIB}
Following the same logic as in the previous cases we arrive at an embedding Ansatz with non-vanishing components given by
\begin{equation}
    \begin{aligned}
        &\dd\bs_{10}^{2}=g_{\mu\nu}(x)\dd x^{\mu}\dd x^{\nu}+\partial_{i}\partial_{\bar{\jmath}}\mathcal{K}(y)\dd y^{i}\dd y^{\bar{\jmath}}\,,\\
        &\bF_{\mu\nu\rho}=-\frac{3}{2}\left(\bar{\psi}^{1}_{[\mu}\gamma_{\nu}\psi_{1\rho]}-\bar{\psi}^{2}_{[\mu}\gamma_{\nu}\psi_{2\rho]}\right)-\frac{3}{2}i\left(\bar{\psi}^{1}_{[\mu}\gamma_{\nu}\psi_{2\rho]}+\bar{\psi}_{1[\mu}\gamma_{\nu}\psi_{\rho]}^{2}\right)\,,\\
        &\bF_{\mu\nu\rho mn}=12i\left(\bar{\psi}^{1}_{[\mu}\gamma_{\nu}\psi_{2\rho]}-\bar{\psi}_{1[\mu}\gamma_{\nu}\psi_{\rho]}^{2}\right)\omega_{mn}\,,\\
        &\bF_{\mu\nu ijk}=\frac{1}{2}\left(F_{\mu\nu}^{+}+\frac{1}{2}\left(\bar{\psi}_{I\mu}\psi_{J\nu}\varepsilon^{IJ}-\bar{\psi}^{I}_{\mu}\psi_{\nu}^{J}\varepsilon_{IJ}\right)+\frac{i}{4}\epsilon_{\mu\nu\rho\sigma}\left(\bar{\psi}_{I}^{\rho}\psi_{J}^{\sigma}\varepsilon^{IJ}+\bar{\psi}^{I\rho}\psi^{J\sigma}\varepsilon_{IJ}\right)\right)\Omega_{ijk}\,,\\
        &\bF_{\mu\nu\bar{\imath}\bar{\jmath}\bar{k}}=\frac{1}{2}\left(F_{\mu\nu}^{-}+\frac{1}{2}\left(\bar{\psi}^{I}_{\mu}\psi_{\nu}^{J}\varepsilon_{IJ}-\bar{\psi}_{I\mu}\psi_{J\nu}\varepsilon^{IJ}\right)-\frac{i}{4}\epsilon_{\mu\nu\rho\sigma}\left(\bar{\psi}_{I}^{\rho}\psi_{J}^{\sigma}\varepsilon^{IJ}+\bar{\psi}^{I\rho}\psi^{J\sigma}\varepsilon_{IJ}\right)\right)\bar{\Omega}_{\bar{\imath}\bar{\jmath}\bar{k}}\,,\\
        &\bPsi_\mu = \frac{1}{2}\left(\psi_{\mu}^{1}\otimes\eta_{+}+\psi_{1\mu}\otimes\eta_{-}\right) + \frac{i}{2}\left(\psi_{\mu}^{2}\otimes\eta_{+}+\psi_{2\mu}\otimes\eta_{-}\right) \,,\\
        &\bep = \frac{1}{2}\left(\epsilon^{1}\otimes\eta_{+}+\epsilon_{1}\otimes\eta_{-}\right) + \frac{i}{2}\left(\epsilon^{2}\otimes\eta_{+}+\epsilon_{2}\otimes\eta_{-}\right)\,,\\
    \end{aligned}
    \label{eq:IIB-4-ansatz-corrected}
\end{equation}
where we define the (anti-)self-dual field strength in 4d as 
\begin{equation}
    F^\pm_{(2)} \equiv \frac12 \left(F_{(2)} \pm i*F_{(2)}\right)\,,
\end{equation}
satisfying $F^\pm_{(2)} = \pm i*F^\pm_{(2)}$.
The gamma matrices and intertwiners factorise according to \eqref{eq:10-4intertwiner}.
Under this Ansatz, only the supercovariant spin connection and supercovariant 5-form field strength survive as 
\begin{equation}
    \hat{\bom}_{\mu \underline{\nu\rho}} = \hat{\omega}_{\mu \underline{\nu\rho}}\,,\quad \hat{\bom}_{m\underline{np}} = \omega_{m\underline{np}}\,,\quad\hat{\tilde{\bF}}_{(5)} = \frac12 \hat{F}^+_{(2)}\wedge\Omega_{(3)} + \frac12 \hat{F}^-_{(2)}\wedge\bar{\Omega}_{(3)}\,.
\end{equation}
We will show consistency of this Ansatz in Section \ref{ConsistencyIIB-4}.

\section{Demonstration of consistency}\label{proof}

Having established our Ans\"atze for various scenarios, we now have to show that they indeed lead to consistent truncations. A first preliminary check is to consider the bosonic 11d/10d equations of motion and to show that with our embedding Ansatz they reduce to the 5d/4d equations of motion. If our aim is to treat consistent truncation just as a technique to generate supergravity solutions, this level of detail is already sufficient. 

However, if we want to show consistency of the full truncated theory, we need to take the fermions into account as well. The first non-trivial check in this context is to consider supersymmetry transformations of the various fields involved. In a consistent truncation, we expect the 11d/10d supersymmetry transformations to reduce to the 5d/4d ones. We will check this explicitly for all theories considered. 

The final ingredient to prove consistency would then be also to check the equations of motion to full non-linear order in fermions. This step, which we outline in Appendix \ref{fermionEoM}, is slightly redundant as the supersymmetry transformations transform equations of motion into each other and having shown consistency of the supersymmetry transformations, we only have to show consistency of a subset of the equations of motion. In Appendix \ref{fermionEoM} we therefore focus on the equations of motion up to spin $\tfrac{3}{2}$, leaving Einstein's equations implicit.

\subsection{From 11d to pure 5d and 4d supergravity}\label{Consistency11-4}

For a purely bosonic background (\ie $\bPsi=0$) the non-trivial part of the equations of motions (\ref{eq:11d-eom}) and the Bianchi identity \eqref{11dbianchi} take the form  
\begin{equation}
    \begin{aligned}
        &\dd\bF_{(4)}=0\,,\qquad \dd*\bF_{(4)}+\frac12 \bF_{(4)}\wedge \bF_{(4)}=0\,,\\
        &\bR_{MN}=\frac{1}{12}\left(\bF_{MPQR}\bF_{N}{}^{PQR}-\frac{1}{12}\bg_{MN}\bF_{PQRS}\bF^{PQRS}\right)\,.
    \end{aligned}
\end{equation}
Under the Ansatz (\ref{eq:11d-5d-bosonic-ansatz}) and (\ref{eq:11d-5d-fermionic-ansatz}) the 11-dimensional Bianchi identity simply reduces to the 5d Bianchi identity $d\cF_{(2)}=0$. In our Ansatz, the 11d spacetime is a direct product of 5d spacetime and a CY3, which guarantees that Hodge dualisation factorises into external and internal space. We may furthermore make use of the relation \eqref{relation}. The field strength equation then becomes 
\begin{equation}
    \left(\dd*\cF_{(2)}  + \frac{1}{\sqrt{3}} \cF_{(2)}\wedge\cF_{(2)}\right)\wedge \omega\wedge\omega=0\,,
\end{equation}
which is the purely bosonic equation of motion for the 5d field strength \eqref{eq:5d-eom}.  The Ricci tensor factorises as $\bR_{\mu\nu} = \cR_{\mu\nu}$, $\bR_{mn} = R^{\text{CY3}}_{mn}=0$, and $\bR_{\mu m} = 0$. The $\mu\nu$-component of Einstein's equation reduces to
\begin{equation}
    \cR_{\mu\nu} = \frac{1}{12}\left(\cF_{\mu\lambda}\cF_{\nu}{}^\lambda  - \frac16 \cg_{\mu\nu}\cF_{\rho\sigma}\cF^{\rho\sigma} \right)\omega_{mn}\omega^{mn}\,.
\end{equation}
Making use of the normalisation $\omega_{mn}\omega^{mn}=6$, we indeed find the 5d  equation \eqref{eq:5d-eom}. The $mn$-component is purely geometrical and is satisfied due to \eqref{relation}. The $\mu m$-component is trivially satisfied, ending the proof of consistency at the level of the bosonic equations of motion.

If we want to move on to 4d, we may simply compactify on a circle as discussed in the previous section, resulting in a 4d $\mathcal{N}=2$ supergravity with a vector multiplet. The full equations of motion take the form
\begin{equation}
    \begin{aligned}
        &\dd*\dd\varphi = -\frac{1}{\sqrt{3}}e^{-\frac{2}{\sqrt{3}}\varphi} \dd\chi\wedge *\dd\chi + \frac{\sqrt{3}}{2} e^{\sqrt{3}\varphi} G^1_{(2)}\wedge *G^1_{(2)} + \frac{1}{2\sqrt{3}}e^{\frac{1}{\sqrt{3}}\varphi}G^2_{(2)}\wedge *G^2_{(2)}\,, \\
        &\dd\left( e^{-\frac{2}{\sqrt{3}}\varphi} *\dd\chi\right) = \frac{1}{\sqrt{3}}G^2_{(2)}\wedge G^2_{(2)} + e^{\frac{1}{\sqrt{3}}\varphi} G^1_{(2)}\wedge *G^2_{(2)}\,,\\
        &\dd\left( e^{\sqrt{3}\varphi} *G^1_{(2)} \right) = -e^{\frac{1}{\sqrt{3}}\varphi} \dd\chi\wedge *G^1_{(2)}\,,\quad \dd\left( e^{\frac{1}{\sqrt{3}}} *G^2_{(2)} \right) = -\frac{2}{\sqrt{3}}\dd\chi\wedge G^2_{(2)}\,, \\
        &R_{\mu\nu} = \frac12 \left(\partial_\mu\varphi\partial_\nu\varphi + e^{\frac{1}{\sqrt{3}}\varphi}\partial_\mu\chi\partial_\nu\chi\right) \\
        &\qquad + \frac12 e^{\sqrt{3}\varphi}\left(G^1_{\mu\lambda}G^1_{\nu}{}^\lambda - \frac14 g_{\mu\nu}G^1_{\rho\sigma}G^{1\rho\sigma} \right)+\frac12 e^{\frac{1}{\sqrt{3}}\varphi}\left( G^2_{\mu\lambda}G^2_{\nu}{}^\lambda - \frac14 g_{\mu\nu}G^2_{\rho\sigma}G^{2\rho\sigma} \right)\,.
    \end{aligned}
\end{equation}
The truncation (\ref{eq:4dN2-trunc-vector}) cancels the contribution from field strengths in the scalar equations, allowing us to consistently set the scalar fields to $0$. This leaves us with the non-trivial equations
\begin{equation}
    \begin{aligned}
        &\dd*F_{(2)} = 0\,,\qquad \dd F_{(2)} = 0\,, \\
        &R_{\mu\nu} = \frac{1}{2}F_{\mu\rho}F_{\nu}{}^{\rho}-\frac{1}{8}g_{\mu\nu}F_{\rho\sigma}F^{\rho\sigma}\,,
    \end{aligned}
\end{equation}
which are indeed the equations of motion and Bianchi identity of pure 4d $\mathcal{N}=2$ supergravity.

\

Next, we will demonstrate the consistency of these truncations, now including fermions, by matching the supersymmetry transformations \eqref{eq:11d-susy}. The fact that our embedding Ansatz relates supercovariant quantities in different dimensions simplifies this task significantly. Let us first consider the transformation of the 11d gravitino (\ref{eq:11d-susy}). The $\mu$-component is 
\begin{equation}\label{gravitino-mu-component}
    \begin{aligned}
        \delta\bPsi_\mu =& \delta\Psi_\mu \otimes \eta + \delta\cPsi_\mu \otimes \tilde{\eta} \\
        =& D_\mu(\hat{\com})\cep \otimes \eta + D_\mu(\hat{\com})\cep^c \otimes \tilde{\eta} \\
        &+ \frac{1}{48}\frac{1}{\sqrt{3}}\hat{\cF}_{\nu\rho}\omega_{mn} \left[\left(\gamma^{\nu\rho}{}_\mu - 4\delta_\mu^\nu \gamma^{\rho}\right)\otimes \Sigma^{mn}\right] \left( \cep \otimes \eta + \cep^c \otimes \tilde{\eta} \right)\,,
    \end{aligned}
\end{equation}
where $D_\mu(\hat{\com})$ is the covariant derivative with respect to the supercovariant spin connection in 5d \eqref{eq:5d-supercov}.
Using the Fierz identity (\ref{eq:Fierz-identity}) in the internal space, we can deduce
\begin{equation}
    \eta_+ \otimes \eta_+^\dagger  - \eta_- \otimes \eta_-^\dagger = -\frac{i}{8}\left(\omega_{mn} \Sigma^{mn} + 2 \Sigma^{(6)} \right)
\end{equation}
and thus find the identity
\begin{equation}
    \omega_{mn}\Sigma^{mn}\eta_\pm = \pm 6i\eta_\pm\,.
    \label{eq:omega-gamma-projection}
\end{equation}
This allows us to project \eqref{gravitino-mu-component} onto $\eta_+$ and $\eta_-$ in the internal space and we find precisely the supersymmetry transformation (\ref{eq:5d-susy-trans}) of the 5d gravitino as their coefficients.

Given that $\eta_\pm$ is a covariantly conserved spinor in the internal space, the $m$-component of the 11d gravitino transformation reduces to
\begin{equation}
    \delta \bPsi_{m} = \frac{1}{48}\frac{1}{\sqrt{3}}\left(\tilde{\cF}_{\mu\nu}\gamma^{\mu\nu}\otimes\left(\omega_{pq}\Sigma^{pq}{}_m - 4\omega_{mn}\Sigma^n \right) \right) \left( \cep \otimes \eta + \cep^c \otimes \tilde{\eta} \right)\,.
\end{equation}
We may transform $\omega_{pq}\Sigma^{pq}{}_m - 4\omega_{mn}\Sigma^n = \frac32 \omega_{pq}\Sigma^{pq}\Sigma_m -\frac12 \omega_{pq}\Sigma_m\Sigma^{pq} $\,, which acting on $\eta_\pm$ gives zero. Thus, supersymmetry transformations do not generate a $m$-component of the gravitino, which is consistent with our Ansatz.

We now consider supersymmetry transformations of the bosonic fields. The only non-trivial transformation of the vielbein (\ref{eq:11d-susy}) is the $\mu \underline{\nu}$-component
\begin{equation}
        \delta \be_\mu{}^{\underline{\nu}} = \delta \ce_\mu{}^{\underline{\nu}} = \frac{i}{2}\left(-\bar{\cep}\gamma^{\underline{\nu}} \Psi_\mu^c + \overline{\cep^c}\gamma^{\underline{\nu}}\Psi_\mu\right)\,,
\end{equation}
which is the 5d transformation given in \eqref{eq:5d-susy-trans}.
Here, we make use of the orthonormality conditions $\bar{\eta} \eta  = \bar{\tilde{\eta}}\tilde{\eta} = 1$ and $\bar{\eta} \tilde{\eta}  = \bar{\tilde{\eta}}\eta = 0$. The $mn$-component is trivial as we have set $\bPsi_m = 0$. The $\mu m$-component vanishes due to relation \eqref{relation3}.

The $\mu mn$-component of the 3-form potential transforms as (\ref{eq:11d-susy})
\begin{equation}
    \begin{aligned}
        \delta \bA_{\mu mn} =& -\frac{1}{\sqrt{3}} \delta (\cA_\mu \omega_{mn}) \\
        =&\frac12 \left(\bar{\cep}\otimes\bar{\eta} + \overline{\cep^c}\otimes\bar{\tilde{\eta}}\right) \left(\mathbf{1} 
        \otimes  \Sigma_{mn} \right) \left(\Psi_{\mu}\otimes\eta + \cPsi_{\mu} \otimes\tilde{\eta}\right) \\
        =&-\frac12\left( \overline{\cep^c}\Psi_\mu - \bar{\cep}\Psi_\mu^c \right)\omega_{mn}\,.
    \end{aligned}
\end{equation}
This gives the 5d supersymmetry transformation of the potential in (\ref{eq:5d-susy-trans}) while reflecting invariance of the K\"ahler potential $\delta \omega_{mn} = 0$. The only other non-trivial transformation is that of the $\mu\nu\rho$-component. It evaluates to
\begin{equation}
    \delta \bA_{\mu\nu\rho} = \frac{3}{2} \left(\bar{\cep}\gamma_{[\mu\nu}\Psi_{\rho]}  + \overline{\cep^c}\gamma_{[\mu\nu}\Psi^c_{\rho]} \right)\,,
\end{equation}
which is the supersymmetry transformation of the potential defined implicitly below \eqref{eq:11d-4form}.
This non-vanishing transformation is required to cancel the transformation of the spinor bilinear that we introduced in \eqref{eq:11d-ansatz-corrected} so that $\delta \hat{\bF}_{\mu\nu\rho\sigma} = 0$. To see this, consider the $\mu\nu\rho\sigma$-component of (\ref{eq:11d-susy-supercov}), 
\begin{equation}
    \delta \hat{\bF}_{\mu\nu\rho\sigma} = 6 \bar{\cep}\gamma_{[\nu\rho}\hat{D}_\mu(\hat{\com})\Psi_{\sigma]} +6 \overline{\cep^c}\gamma_{[\nu\rho}\hat{D}_\mu(\hat{\com})\Psi^c_{\sigma]}\,,
\end{equation}
where the left-hand side vanishes identically according to our Ansatz (\ref{eq:11d-ansatz-corrected}). The right-hand side vanishes on-shell once the 5d gravitino equation of motion \eqref{eq:5d-eom} is imposed
\begin{equation}
    \gamma_{[\nu_1\nu_2}\hat{D}_{\nu_3}(\hat{\com})\Psi_{\nu_4]}\sim \epsilon_{\nu_1\cdots \nu_4\lambda}\gamma^{\mu_3\mu_4 \lambda} \hat{D}_{\mu_3}(\hat{\com})\Psi_{\mu_4} = 0.
\end{equation}

We again note that the $\mu\nu mn$-component of the supersymmetry transformation of the supercovariant field strength (\ref{eq:11d-susy-supercov}) gives the transformation of the supercovariant field strength in 5d (\ref{eq:5d-susy-supercov-field-strength}) induced by the supersymmetry transformation of $\bA_{\mu mn}$ and the gravitino. 

We have now shown that under our Ansatz all supersymmetry transformations of the 11d theory reduce to the supersymmetry transformations of the truncated 5d theory.

We now move on to the subsequent truncation to 4d. 
Let us start with the gravitino transformation in (\ref{eq:5d-susy-trans}). The $\mu$-component is 
\begin{equation}
    \begin{aligned}
        \delta\Psi_\mu =& D_\mu(\hat{\omega})\epsilon + \frac18 \hat{F}_{\mu\nu}\gamma^\nu(i\gamma^{(4)})\epsilon + \frac{i}{8\sqrt{3}}\left(\frac{\sqrt{3}}{4} \varepsilon_{\rho\sigma \lambda\eta}\hat{F}^{\lambda\eta} \right)\left(\gamma_\mu{}^{\rho\sigma} - 4\delta^\rho_\mu \gamma^\sigma\right)\epsilon \\
        =& \left(D_\mu(\hat{\omega})-\frac18 \hat{F}_{\rho\sigma}\gamma^{\rho\sigma}\gamma_\mu i\gamma^{(4)}\right)\epsilon
    \end{aligned}
\end{equation}
where we used $\gamma^{a_1\cdots a_r}\gamma^{(4)} = \frac{1}{(4-r)!} \varepsilon^{a_r\cdots a_1 b_1\cdots b_{4-r}} \gamma_{b_1\cdots b_{4-r}}$ in the final step. Using the position of the $SU(2)$ index to indicate chirality, we can rewrite the Dirac spinors in terms of Weyl spinors, which gives exactly the expected supersymmetry transformation (\ref{eq:4d-susy}) in 4d. The $4$-component of the transformation vanishes identically as 
\begin{equation}
     \delta\Psi_4 = -\frac{1}{16}\hat{F}_{\mu\nu} \gamma^{\mu\nu} \epsilon + \frac{i}{8\sqrt{3}}\left(-\frac{\sqrt{3}}{4} \varepsilon_{\mu\nu\rho\sigma}\hat{F}^{\rho\sigma} \right) \left(-i\gamma^{(4)} \gamma^{\mu\nu}\right) \epsilon = 0\,.
\end{equation}
The supersymmetry transformation of the graviton is 
\begin{equation}
    \begin{aligned}
        \delta \ce^{\underline{\nu}}{}_\mu &= \delta e^{\underline{\nu}}{}_\mu = \frac{1}{2}\left(\bar{\epsilon}^I\gamma^{\underline{\nu}}\psi_{I\mu} + \bar{\epsilon}_I\gamma^{\underline{\nu}}\psi^I_{\mu}\right)\,, \\
        \delta \ce^{\underline{4}}{}_\mu &= \delta\left(\frac12 A_\mu \right) = -\frac12 \left(\bar{\zeta}^I\psi_\mu^J\varepsilon_{IJ} + \bar{\zeta}_I\psi_{\mu J}\varepsilon^{IJ} \right)\,
    \end{aligned}
\end{equation}
and $\delta e^{\underline{\mu}}{}_4 = \delta e^{\underline{4}}{}_4 = 0$ trivially. We thus arrive at the transformation
\begin{equation}
    \delta e^{\underline{\nu}}{}_\mu = \frac12\left(\bar{\epsilon}^I\gamma^{\underline{\nu}}\psi_{I\mu} + \bar{\epsilon}_I\gamma^{\underline{\nu}}\psi^I_{\mu}\right),\quad \delta A_\mu = - \left(\bar{\zeta}^I\psi_\mu^J\varepsilon_{IJ} + \bar{\zeta}_I\psi_{\mu J}\varepsilon^{IJ} \right),
\end{equation}
which is exactly (\ref{eq:4d-susy}). The transformation of the 5d gauge potential is 
\begin{equation}
    \delta {\cA}_\mu = -\frac{\sqrt{3}i}{2}\left( \bar{\zeta}^I\Psi_\mu^J\varepsilon_{IJ} - \bar{\zeta}_I\Psi_{\mu J}\varepsilon^{IJ} \right)\,,\qquad \delta{\cA}_4 =0\,.
\end{equation} 
 $\delta {\cA}_\mu$ should be understood here in terms of the supercovariant field strength. The associated transformation of the supercovariant field strength, \ie the $\mu\nu$-component of (\ref{eq:5d-susy-supercov-field-strength}), is 
\begin{equation}
    \begin{aligned}
        \delta \hat{\cF}_{\mu\nu} &= -\frac{\sqrt{3}}{4}\epsilon_{\mu\nu\rho\sigma}\delta\hat{F}^{\rho\sigma}\\
        &= -\sqrt{3}  i \left(\bar{\epsilon}^I\hat{D}_{[\mu}(\hat{\nu}) \psi^J_{\sigma]}\varepsilon_{IJ}  - \bar{\epsilon}_I\hat{D}_{[\mu}(\hat{\nu}) \psi_{\sigma]J}\varepsilon^{IJ} \right)\,,
    \end{aligned}
\end{equation}
which gives
\begin{equation}
    \begin{aligned}
        \delta\hat{F}_{\mu\nu} &= - i\epsilon_{\mu\nu}{}^{\rho\sigma} \left(\bar{\epsilon}^I\hat{D}_{[\rho}(\hat{\omega}) \psi^J_{\sigma]}\varepsilon_{IJ}  - \bar{\epsilon}_I\hat{D}_{[\rho}(\hat{\omega}) \psi_{\sigma]J}\varepsilon^{IJ} \right)\\
        &= -2\left(\bar{\epsilon}^I\hat{D}_{[\mu}(\hat{\omega}) \psi^J_{\nu]}\varepsilon_{IJ}  + \bar{\epsilon}_I\hat{D}_{[\mu}(\hat{\omega}) \psi_{\nu]J}\varepsilon^{IJ} \right).
    \end{aligned}
    \label{eq:4d-susy-supercov-field}
\end{equation}
To get to the second line, one needs to use the 4d gravitino equation of motion, which implies that $\bar{\epsilon}^I\hat{D}_{[\mu}(\hat{\omega}) \psi^J_{\nu]}$ is self-dual and $\bar{\epsilon}_I\hat{D}_{[\mu}(\hat{\omega}) \psi_{\nu]J}\varepsilon^{IJ}$ is anti-self-dual. We find precisely the transformation of the 4d supercovariant field strength, (\ref{eq:4d-susy-supercov}). Thus, the non-trivial supersymmetry transformation of $\cA_{\mu}$ keeps the reduced supercovariant field strength transforming as (the Hodge-dual of) the 4d supercovariant field strength. We have now shown full consistency of the supersymmetry transformations in the truncation to 4d as well.

\subsection{From IIA to pure 4d supergravity}\label{ConsistencyIIA-4}

In a purely bosonic background, the embedding Ansatz (\ref{eq:IIA-4d-ansztz-corrected}) reduces the equations of motion of $\bC_{(1)}$ and $\bC_{(3)}$, given in  (\ref{eq:IIA-eom-bosonic}), to 
\begin{equation}\label{4dfieldeom}
    \dd*F_{(2)} = 0\,,\qquad \dd F_{(2)}=0\,,
\end{equation}
which are the equation of motion and Bianchi identity for the 4d field strength. Similarly, the Bianchi identity in 10d reduces to the same equations \eqref{4dfieldeom}. The $\bH_{(3)}$ and $\bPhi$ equations vanish identically using the geometric relations in \eqref{relation}. The $\mu\nu$-component of Einstein's equation becomes 
\begin{equation}
    \begin{aligned}
        R_{\mu\nu} =& \frac18\left(F_{\mu\lambda}F_{\nu}{}^\lambda - \frac{1}{8\cdot 2!}g_{\mu\nu}F_{\rho\sigma}F^{\rho\sigma}\right) + \frac{1}{16}\left(6\tilde{F}_{\mu\lambda}\tilde{F}_{\nu}{}^\lambda - \frac{9}{8}g_{\mu\nu}\tilde{F}_{\rho\sigma}\tilde{F}^{\rho\sigma}\right)\,, \\
        =& \frac12F_{\mu\lambda}F_{\nu}{}^\lambda - \frac18 g_{\mu\nu}F_{\rho\sigma}F^{\rho\sigma}\,,
    \end{aligned}
\end{equation}
where $\tilde{F}_{\mu\nu} = \frac12 \epsilon_{\mu\nu\rho\sigma}F^{\rho\sigma}$ and we used the relation $F_{\mu\nu}F^{\mu\nu} = - \tilde{F}_{\mu\nu}\tilde{F}^{\mu\nu}$, and $F_{\mu}{}^\lambda F_{\nu\lambda} = \tilde{F}_{\mu}{}^\lambda\tilde{F}_{\nu\lambda} - \frac12 g_{\mu\nu}\tilde{F}_{\rho\sigma}\tilde{F}^{\rho\sigma}$\,. This is Einstein's equation for pure 4d $\mathcal{N}=2$ supergravity, as expected. Due to its direct product structure, our Ansatz trivially satisfies the $\mu m$-component of Einstein's equation. Finally, the $mn$-component reads
\begin{equation}
    R_{mn} = -\frac12\cdot \frac{1}{64}g_{mn}F_{\mu\nu}F^{\mu\nu} + \frac{1}{12}\left(\frac34 \omega_{m}{}^p\omega_{np} \tilde{F}_{\mu\nu}\tilde{F}^{\mu\nu} -\frac{9}{64}g_{mn} \tilde{F}_{\mu\nu}\tilde{F}^{\mu\nu}\omega_{pq}\omega^{pq}\right).
\end{equation}
The left-hand side vanishes because of the Ricci flatness of CY3, while the right-hand side vanishes using \eqref{relation}. This concludes the proof of consistency at the level of the bosonic equations of motion.

\

Next, we will demonstrate the consistency of the truncation including fermions by matching the supersymmetry transformations (\ref{eq:IIA-susy}).
 Let us begin with the transformation of the fermionic fields. The $\mu$-component of the gravitino is 
\begin{equation}
    \begin{aligned}
        \delta \bPsi_\mu =& \frac{1}{\sqrt{2}}\left( \delta\Psi^1_\mu\otimes\eta_+ + \delta\Psi_{1\mu}\otimes\eta_-  + \delta\Psi_{2\mu}\otimes\eta_+ + \delta\Psi^2_{\mu}\otimes\eta_-\right) \\ 
        =& D_\mu(\hat{\omega})\bep - \frac{1}{64\cdot2!}\hat{F}_{\rho\sigma}\left[\left( \gamma_\mu{}^{\rho\sigma} -14 \delta^\rho_\mu\gamma^\sigma \right)\otimes\mathbf{1}\right]\Gamma^{(10)}\bep\\
        &+\frac{3}{32\cdot 4}\frac{1}{4}\epsilon_{\rho\sigma\lambda\eta}\hat{F}^{\lambda\eta}\left[\left(\gamma_\mu{}^{\rho\sigma} - \frac{10}{3}\delta_\mu^\rho \gamma^\sigma \right) \otimes \omega_{mn}\Sigma^{mn}\right] \bep \,.\\
        =& \frac{1}{\sqrt{2}}\left[D_\mu(\hat{\omega})(\epsilon^1+\epsilon_2) - \frac{1}{8}\hat{F}_{\rho\sigma}\gamma^{\rho\sigma}\gamma_\mu\left(\epsilon^1-\epsilon_2\right) \right]\otimes\eta_+\\
        &+ \frac{1}{\sqrt{2}}\left[D_\mu(\hat{\omega})(\epsilon_1+\epsilon^2) - \frac{1}{8}\hat{F}_{\rho\sigma}\gamma^{\rho\sigma}\gamma_\mu\left(\epsilon_1-\epsilon^2\right) \right]\otimes\eta_-\\
    \end{aligned}
\end{equation}
We have used the fact that the (anti-)self-dual field strength $F_{\mu\nu}^\pm = \frac12 \left(F_{\mu\nu}\pm\frac{i}{2}F_{\rho\sigma}\epsilon^{\rho\sigma}{}_{\mu\nu}\right)$ contracted with gamma matrices is chiral according to
\begin{equation}
    F_{\mu\nu}^\pm \gamma^{\mu\nu} = F_{\mu\nu}\gamma^{\mu\nu} P_\pm, \label{eq:self-dual-chiral}\,.
\end{equation} 
We can then take projections in both external and internal spaces to find the supersymmetry transformation  (\ref{eq:4d-susy}) of the gravitino in 4d. 
As $\eta_\pm$ is a covariantly constant spinor in the internal space, the $m$-component of the gravitino equation is 
\begin{equation}
        \delta\bPsi_m = -\frac{1}{64\cdot2!}\hat{F}_{\rho\sigma}\left(\gamma^{\rho\sigma}i\gamma^{(4)}\otimes\Sigma_m\right)\Gamma^{(10)}\bep + \frac{3}{64\cdot4}\tilde{\hat{F}}_{\rho\sigma}\omega_{pq}\left(\gamma^{\rho\sigma}i\gamma^{(4)}\otimes\left(\Sigma_m{}^{pq}-\frac{10}{3}\delta^p_m\Sigma^q\right)\right) \bep\,,
\end{equation}
where we introduced $\tilde{\hat{F}}_{\rho\sigma} = \frac{1}{2}\varepsilon_{\rho\sigma\lambda\eta}\hat{F}^{\lambda\eta}$ to simplify the expression. We can show that this vanishes using $\left(\omega_{pq}\Sigma_m{}^{pq} - \frac{10}{3}\omega_{mq}\Sigma^q\right)\eta_\pm = \pm i\frac23 \Sigma_m \eta_\pm$ and (\ref{eq:self-dual-chiral}). It is straightforward to use \eqref{eq:omega-gamma-projection} and (\ref{eq:self-dual-chiral}) to prove vanishing of the supersymmetry transformation of the dilatino 
\begin{equation}
        \delta\bla = -\frac{1}{8\cdot2!}\hat{F}_{\mu\nu}\Gamma^{\mu\nu}\bep + \frac{1}{8\cdot4!} \varepsilon_{\mu\nu\rho\sigma}\hat{F}^{\rho\sigma}\omega_{mn}\Gamma^{\mu\nu mn}\Gamma^{(10)} \bep  = 0\,.
\end{equation}

The non-trivial transformations (\ref{eq:IIA-susy}) of the bosonic fields reduce to 
\begin{equation}
    \begin{aligned}
        &\delta \be_\mu{}^{\underline{\nu}} = \delta e_\mu{}^{\underline{\nu}} = \bar{\bep}^1\Gamma^{\underline{\nu}}\bPsi_\mu^1 + \bar{\bep}_2\Gamma^{\underline{\nu}}\bPsi_{2\mu} = \frac12 \left(\bar{\epsilon}^I\gamma^{\underline{\nu}}\psi_{I\mu} + \bar{\epsilon}_I\gamma^{\underline{\nu}}\psi^I_{\mu} \right)\,,\\
        &\delta \bC_\mu = \frac12 \delta A_\mu = -\bar{\bep}^1\bPsi_{2\mu} + \bar{\bep}_2\bPsi^1_{\mu} = -\frac12 \left(\bar{\epsilon}^I\psi_\mu^J\varepsilon_{IJ} +  \bar{\epsilon}_I\psi_{J\mu}\varepsilon^{IJ}\right)\,,
    \end{aligned}
\end{equation}
which are just the 4d transformation rules in (\ref{eq:4d-susy}). 
The other non-trivial transformation of potentials are  $\delta \bB_{\mu\nu},\ \delta\bC_{\mu\nu\rho},\ \delta\bC_{\mu mn}$, which should be understood in terms of the supersymmetry transformations of supercovariant field strengths (\ref{eq:IIA-susy-supercov-field}, \ref{eq:IIA-susy-supercov-field2}). The supersymmetry transformation of the $\mu\nu\rho\sigma$-component of $\hat{\bF}_{(4)}$ is
\begin{equation}
    \begin{aligned}
        \delta \hat{\bF}_{\mu\nu\rho\sigma} &=-12\bar{\bep}\Gamma_{[\nu\rho}\left[D_\mu(\hat{\bom}) - \frac18(\gamma^{\lambda}\otimes\mathbf{1})\Gamma^{(10)} + \frac{1}{48}\left[\left(\gamma^{\lambda\eta}{}_\mu -4\delta^\lambda_\mu \gamma^\eta\right)\otimes \Sigma^{mn}\omega_{mn}\right]\frac{1}{2}\tilde{\hat{F}}_{\lambda\eta} \right] \bPsi_{\sigma]} \\
        &=6\left(\bar{\epsilon}^1\gamma_{[\gamma\rho}\hat{D}_\mu(\hat{\omega})\psi^2_{\sigma]} + \bar{\epsilon}_1\gamma_{[\gamma\rho}\hat{D}_\mu(\hat{\omega})\psi_{2\sigma]} +\bar{\epsilon}_2\gamma_{[\gamma\rho}\hat{D}_\mu(\hat{\omega})\psi_{1\sigma]}
        +\bar{\epsilon}^2\gamma_{[\gamma\rho}\hat{D}_\mu(\hat{\omega})\psi^1_{\sigma]}\right) = 0\,,
    \end{aligned}
\end{equation}
where we have used \eqref{relation}, \eqref{relation3}, (\ref{eq:self-dual-chiral}) as well as the 4d gravitino equation of motion. The non-trivial transformation $\delta \bC_{\mu\nu\rho}$ assures the invariance of this supercovariant field strength component. 
The $\mu\nu mn$-component of $\hat{\bF}_{(4)}$ reduces to 
\begin{equation}
    \begin{aligned}
        \delta\hat{\bF}_{\mu\nu mn} =& \delta\left(\frac14 \epsilon_{\mu\nu\rho\sigma}\hat{F}^{\rho\sigma}\omega_{mn}\right) \\
        =&-2\bar{\bep}\left(\Gamma_{\mu[m}\frac{1}{288}\left(\Gamma^{ABCD}{}_{n]} - 8\delta^{A}_{n]}\Gamma^{BCD}\right)\hat{\bF}_{ABCD} \right)\bPsi_{\nu} - (\mu\leftrightarrow\nu) \\
        &-2\bar{\bep}\Gamma_{mn}\left[D_{[\mu}(\hat{\omega})+\frac14 \Gamma^\lambda \Gamma^{(10)} \hat{F}_{\lambda[\mu} + \frac{1}{288}\left( \Gamma^{ABCD}{}_{[\mu} - 8\delta^{A}_{[\mu}\Gamma^{BCD}\right)\hat{\bF}_{ABCD} \right]\bPsi_{\nu]}\,.
    \end{aligned}
\end{equation}
We can prove that the first line vanishes using $\eta^\dagger_\pm\Sigma_{mn}{}^{pq}\eta_\pm  = \frac12 \varepsilon_{mnpqst}\omega^{st}$ and \eqref{relation}. Then we can get the result from the second line as 
\begin{equation}
    \begin{aligned}
        \delta\hat{\bF}_{\mu\nu mn} =& \delta\left(\frac14 \epsilon_{\mu\nu\rho\sigma}\hat{F}^{\rho\sigma}\omega_{mn}\right) \\
        =&i\left(\bar{\epsilon}^I\hat{D}_{[\mu}(\hat{\omega})\psi^J_{\nu]}\varepsilon_{IJ} - \bar{\epsilon}_I\hat{D}_{[\mu}(\hat{\omega})\psi_{J\nu]}\varepsilon^{IJ}\right)\omega_{mn},
    \end{aligned}
\end{equation}
which leads us to the transformation (\ref{eq:4d-susy-supercov}) of the supercovariant field strength in 4d. The non-trivial transformation of $\bC_{\mu mn}$ guarantees that the reduction of $\hat{\bF}_{\mu\nu mn} $ transforms as (the Hodge-dual of) the 4d supercovariant field strength. 
Finally, the non-trivial transformation of $\bB_{\mu\nu}$ assures vanishing of the supersymmetry transformation of $\hat{\bH}_{\mu\nu\rho}$  as 
\begin{equation}
    \begin{aligned}
        \delta\hat{\bH}_{\mu\nu\rho} =& 6\bar{\bep}\Gamma^{(10)}\left[D_\mu(\hat{\omega}) -\frac18 \Gamma^{\lambda}\Gamma^{(10)}\hat{F}_{\mu\lambda} + \frac{1}{96}\left(\Gamma^{\mu_1\mu_2}{}_\mu - 4\delta^{\mu_1}_\mu\gamma^{\mu_2}\right)\Gamma^{mn}\tilde{\hat{F}}_{\mu_1 \mu_2}\omega_{mn}\right]\bPsi_{\rho]} \\ 
        &-3\bar{\bep}\Gamma_{[\mu\nu|}\left[\frac{1}{16}\hat{F}_{\mu\nu}\Gamma^{\mu\nu}+\frac{1}{96}\Gamma^{\mu_1\mu_2}\Gamma^{(10)}\Gamma^{mn}\tilde{\hat{F}}_{\mu_1\mu_2}\omega_{mn} \right]\bPsi_{\rho]} \\
        & = 3\left(\bar{\epsilon}^1\gamma_{[\nu}\hat{D}_\mu(\hat{\omega})\psi_{\rho]1} + \bar{\epsilon}_1\gamma_{[\nu}\hat{D}_\mu(\hat{\omega})\psi_{\rho]}^1 - \bar{\epsilon}_2\gamma_{[\nu}\hat{D}_\mu(\hat{\omega})\psi_{\rho]}^2 - \bar{\epsilon}^2\gamma_{[\nu}\hat{D}_\mu(\hat{\omega})\psi_{\rho]2}\right)\\
        &=0\,.
    \end{aligned}
\end{equation}
To derive this, we need (\ref{eq:self-dual-chiral}) and the gravitino equation of motion in the final step. 

At this point, we have proven consistency of supersymmetry transformations in our IIA reduction to pure 4d $\mathcal{N}=2$ supergravity.

\subsection{From IIB to pure 4d supergravity}\label{ConsistencyIIB-4}

At purely bosonic level, the non-trivial equations of motion under the Ansatz (\ref{eq:IIB-4-ansatz-corrected}) are 
\begin{equation}
    \bF_{(5)} = *\bF_{(5)}\,,\qquad \bR_{MN} = \frac{1}{4\cdot4!}\bF_{MPQLK}\bF_N{}^{PQLK}\,.
\end{equation}
Due to the direct product structure of 4d spacetime and the CY3, the 10d Hodge duality operator will act as a direct product of the Hodge duality operators in the external and internal spaces. With the relation \eqref{relation2}, it is easy to proof
\begin{equation}
    *\bF_{(5)}= \frac12 i*F^+_{(2)}\wedge \Omega - \frac12 i*F^-_{(2)}\wedge \bar{\Omega} = \frac12 F^+_{(2)}\wedge \Omega + \frac12 F^-_{(2)}\wedge \bar{\Omega} = \bF_{(5)}\,.
    \label{eq:IIB-self-dual-eom}
\end{equation}
The $\mu\nu$-component of Einstein's equation is 
\begin{equation}
        R_{\mu\nu}=\frac{1}{2}\left(F_{\mu\rho}^{+}F_{\nu}^{-\rho}+F_{\mu\rho}^{-}F_{\nu}^{+\rho}\right) = \frac12 F_{\mu\lambda}F_{\nu}{}^\lambda - \frac18 g_{\mu\nu} F_{\rho\sigma}F^{\rho\sigma}\,,
\end{equation}
using the normalisation condition (\ref{eq:CY-geo}). We recognise the correct 4d Einstein's equation. The $\mu m$-component of Einstein's equation is trivial due to the direct product structure of the metric. Finally, the $mn$-component is trivially satisfied due to the Ricci flatness of the CY3. 

The 10d Bianchi identity gives $\dd F^\pm_{(2)} = 0$, which can be recombined to arrive at the appropriate 4d equations
\begin{equation}
    \dd *F_{(2)}=0\,,\qquad \dd F_{(2)}=0\,,
\end{equation}
thus proving consistency at the level of bosonic equations of motion.

\

Next, we will prove the consistency of the embedding Ansatz including fermions by matching the supersymmetry transformations (\ref{eq:IIB-susy}). Starting with the $\mu$-component of the gravitino 
\begin{equation}
    \begin{aligned}
        \delta\bPsi_{\mu}&=\frac{1}{2}\left(\delta\Psi_{\mu}^{1}\otimes\eta_{+}+\delta\Psi_{1\mu}\otimes\eta_{-}\right) + \frac{i}{2}\left(\delta\Psi_{\mu}^{2}\otimes\eta_{+}+\delta\Psi_{2\mu}\otimes\eta_{-}\right),\\
        &=\frac{1}{2}\bigg[ \left(D_{\mu}\left(\hat{\omega}\right)\epsilon^{1}\otimes\eta_{+}+D_{\mu}\left(\hat{\omega}\right)\epsilon_{1}\otimes\eta_{-}\right)\\
        &\qquad -\frac{i}{8}\left(\hat{F}_{\rho\sigma}\gamma^{\rho\sigma}\gamma_{\mu}\epsilon_{1}\otimes\eta_{+}+\hat{F}_{\rho\sigma}\gamma^{\rho\sigma}\gamma_{\mu}\epsilon^{1}\otimes\eta_{-}\right)\bigg]\,,\\
        &+\frac{i}{2}\bigg[ \left(D_{\mu}\left(\hat{\omega}\right)\epsilon^{2}\otimes\eta_{+}+D_{\mu}\left(\hat{\omega}\right)\epsilon_{2}\otimes\eta_{-}\right)\\
        &\qquad -\frac{i}{8}\left(\hat{F}_{\rho\sigma}\gamma^{\rho\sigma}\gamma_{\mu}\epsilon_{2}\otimes\eta_{+}+\hat{F}_{\rho\sigma}\gamma^{\rho\sigma}\gamma_{\mu}\epsilon^{2}\otimes\eta_{-}\right)\bigg]\,,
    \end{aligned}
\end{equation}
we need to use the chirality of the (anti-)self-dual field strength \eqref{eq:self-dual-chiral} and the relations
\begin{equation}
    \begin{aligned}
        &\frac{1}{3!}\Omega_{mnp}\Sigma^{mnp} = -\frac{1}{3!}\eta_-^\dagger \Sigma_{ijk} \eta_+ \Sigma^{ijk} = 8\eta_+\otimes\eta_-^\dagger\,, \\
        &\frac{1}{3!}\bar{\Omega}_{mnp}\Sigma^{mnp} = \frac{1}{3!}\eta_+^\dagger \Sigma_{ijk} \eta_- \Sigma^{ijk}=  -8\eta_-\otimes\eta_+^\dagger\,.\\
    \end{aligned}
\end{equation}
Projecting in the internal space and recombining the results, we get the supersymmetry transformation (\ref{eq:4d-susy}) of the gravitino in 4d. The $m$-component of the transformation vanishes using $\Omega_{mnp}\Sigma^{np}=\frac{1}{2}\Omega_{qnp}\left(\Sigma_{m}\Sigma^{qnp}+\Sigma^{qnp}\Sigma_{m}\right)$ and \eqref{relation3}.

The non-trivial supersymmetry transformation of the vielbein is 
\begin{equation}
    \begin{aligned}
        \delta \be_{\mu}{}^{\underline{\nu}} = \delta e_{\mu}{}^{\underline{\nu}} = 2\left(\bar{\bep}^{1}\Gamma^{\underline{\nu}}\bPsi_{\mu}^{1}-\bar{\bep}^{2}\Gamma^{\underline{\nu}}\bPsi_{\mu}^{2}\right) = \frac{1}{2}\left(\bar{\epsilon}^{I}\gamma^{\underline{\nu}}\psi_{I\mu}+\bar{\epsilon}_{I}\gamma^{\underline{\nu}}\psi_{\mu}^{I}\right),
    \end{aligned}
\end{equation}
which is precisely the 4d supersymmetry transformation (\ref{eq:4d-susy}). The supersymmetry transformation of the scalars, parametrised by $\bu$ and $\bv$, is trivial since the dilatino vanishes. The $\mu\nu$-component of the $\bA_{(2)}$-transformation is non-trivial, which should be understood in terms of the transformation behaviour (\ref{eq:IIB-susy-supercov}) of the $\mu\nu\rho$-component of the supercovariant field strength. Under our Ansatz, 
\begin{equation}
    \begin{aligned}
        \delta\hat{\bF}_{\mu\nu\rho}&=6\bar{\bep}\Gamma_{[\rho}\hat{D}_{\mu}\left(\hat{\bom}\right)\bPsi_{\nu]}        \\
        &=\frac{3}{2}\left(\bar{\epsilon}^{1}\gamma_{[\rho}\hat{D}_{\mu}\left(\hat{\omega}\right)\psi_{1\nu]}+\bar{\epsilon}_{1}\gamma_{[\rho}\hat{D}_{\mu}\left(\hat{\omega}\right)\psi_{\nu]}^{1}\right)+\frac{3}{2}i\left(\bar{\epsilon}^{1}\gamma_{[\rho}\hat{D}_{\mu}\left(\hat{\omega}\right)\psi_{2\nu]}+\bar{\epsilon}_{1}\gamma_{[\rho}\hat{D}_{\mu}\left(\hat{\omega}\right)\psi_{\nu]}^{2}\right)\\
        &\quad+\frac{3}{2}i\left(\bar{\epsilon}^{2}\gamma_{[\rho}\hat{D}_{\mu}\left(\hat{\omega}\right)\psi_{1\nu]}+\bar{\epsilon}_{2}\gamma_{[\rho}\hat{D}_{\mu}\left(\hat{\omega}\right)\psi_{\nu]}^{1}\right)+\frac{3}{2}\left(\bar{\epsilon}^{2}\gamma_{[\rho}\hat{D}_{\mu}\left(\hat{\omega}\right)\psi_{2\nu]}+\bar{\epsilon}_{2}\gamma_{[\rho}\hat{D}_{\mu}\left(\hat{\omega}\right)\psi_{\nu]}^{2}\right)\,,
    \end{aligned}
\end{equation}
which vanishes after imposing the 4d gravitino equation of motion. This shows that the non-trivial transformation of $\bA_{\mu\nu}$ cancels the transformation of the spinor bilinears in the embedding Ansatz. It is easy to prove that $\delta\bA_{\mu m} = 0$ using \eqref{relation3}. The $mn$-component of the $\bA_{(2)}$-transformation is trivial.

The supersymmetry transformation \eqref{eq:IIB-susy} generates a non-trivial $\bC_{\mu\nu\rho\sigma}$-term which is closed in 10d. We may cancel this term by a simultaneous gauge transformation. 
The transformation of $\bC_{\mu\nu\rho m}$ vanishes due to relation \eqref{relation3}.  As for $\bA_{\mu\nu}$, the transformation of $\bC_{\mu\nu mn}$ ensures a vanishing supersymmetry transformation of $\hat{\tilde{\bF}}_{\mu\nu\rho mn}$ in (\ref{eq:IIB-susy-supercov}). After a straightforward calculation, we get  
\begin{equation}
    \begin{aligned}
        \delta\hat{\tilde{\bF}}_{\mu\nu\rho mn}=3i\left[\bar{\epsilon}^{1}\gamma_{[\rho} \left(\hat{D}_{\mu}(\hat{\omega})\psi_{2\nu]}-\hat{D}_{\mu}(\hat{\omega})\psi_{\nu]}^{2}\right)-\bar{\epsilon}^{2}\gamma_{[\rho}\left(\hat{D}_{\mu}(\hat{\omega})\psi_{1\nu]}-\hat{D}_{\mu}(\hat{\omega})\psi_{\nu]}^{1}\right)\right]\omega_{mn}\,,
    \end{aligned}
\end{equation}
which vanishes after imposing the 4d gravitino equation of motion.

Na{\"\i}vely, we would expect the supersymmetry transformation of $\bC_{\mu mnp}$ to reduce to the transformation of $C_\mu$ in 4d. However, as we are correcting the field strength $\bF_{\mu\nu ijk} $ and $\bF_{\mu\nu \bar{\imath}\bar{\jmath}\bar{k}}$ by spinor bilinears, it should be $\delta \left(\bC_{\mu mnp} + \text{spinor bilinears}\right)$ that gives the correct 4d transformation. 
We furthermore know that an appropriate correction to the potential exists at least locally because the field strength in our Ansatz satisfies the Bianchi identity \eqref{eq:IIB-5-form Bianchi}.
Instead of analysing the transformation of the potential, we therefore consider the supersymmetry transformation directly at field-strength level. After a long calculation, we find the transformation of the $\mu\nu mnp$-component of $\hat{\tilde{\bF}}_{(5)}$ 
\begin{equation}
        \delta\hat{\tilde{\bF}}_{\mu\nu ijk} = \frac12 \delta \hat{F}^+_{\mu\nu} \Omega_{ijk} = -\bar{\epsilon}^{I}\left[D_{[\mu}(\hat{\omega})\psi_{\nu]}^{J}+\frac{1}{8}\hat{F}_{\rho\sigma}\gamma^{\rho\sigma}\gamma_{[\mu}\varepsilon^{JK}\psi_{K\nu]}\right]\varepsilon_{IJ}\Omega_{ijk}
\end{equation}
and its complex conjugate. Combining both we end up with
\begin{equation}
    \delta \hat{F}_{\mu\nu} = -2 \left(\bar{\zeta}^{I}\hat{D}_{[\mu}(\hat{\omega})\psi_{\nu]}^{J}\varepsilon_{IJ}+\bar{\zeta}_{I}\hat{D}_{[\mu}(\hat{\omega})\psi_{\nu]J}\varepsilon^{IJ}\right)\,.
\end{equation}
This is the expected supersymmetry transformation (\ref{eq:4d-susy-supercov}) of the 4d supercovariant field strength. With this, we have proven the consistency of all supersymmetry transformations in our IIB reduction to pure 4d $\mathcal{N}=2$ supergravity.

\section{Extremal black holes}\label{BH}

Having demonstrated that a consistent truncation to pure 4d $\mathcal{N}=2$ supergravity is possible, we are now able to uplift any solution thereof to a solution of 11d supergravity or to type IIA or IIB supergravity in 10 dimensions. A trivial example is flat Minkowski space, but we may also consider more interesting backgrounds, such as black holes. Given the presence of the 4d gauge field $A_\mu$, we may moreover study charged black holes. The amount of (electric) charge a black hole can accrue is bounded by the BPS-bound \cite{Gibbons:1982fy}
 \begin{equation}
     G_N Q^2\leq M^2\,,
\end{equation}
where we have reinstated the gravitational constant $G_N$. A black hole saturating this bound is called extremal and preserves half the supersymmetry of the theory. The associated field configuration can be found in \eg \cite{Gibbons:1982fy} and in spherical coordinates $(t,r,\theta,\phi)$ it takes the form
 \begin{align}\label{blackhole}
     \dd s^2_4= - h^{-2}(r)\,\dd t^2+h^2(r)\left(\dd r^2+r^2 \,\dd\Omega_2^2\right)\,,\qquad A_{(1)}=2 h^{-1}(r)\,\dd t\,,\qquad h(r)=1+\frac{r_H}{r}\,.
 \end{align}
One can check explicitly that this solves the equations of motion \eqref{eq:4d-eom}. 

We now want to embed this 4d black hole in the various higher-dimensional scenarios we have discussed. We may then attempt to interpret these solutions as backgrounds generated by extended objects like D-branes wrapping cycles in the CY3 \cite{Strominger:1996sh}.  

To motivate this interpretation, we may start with a D3-brane solution in type IIB supergravity
\begin{equation}
    \dd s_{10}^2=H(r)^{-\frac{1}{2}}\dd x_\parallel^2+H(r)^{\frac 12}\dd x_\perp ^2\,,\qquad r=\abs{x_\perp}\,,\qquad \Delta_{x_\perp}H(r)=\delta(r)\,,
\end{equation}
which furthermore sources self-dual 5-form flux. If we compactify a dimension $x_\parallel^A$ parallel to the brane, we talk about the brane wrapping this dimension, but all other dimensions remain unaffected. If, however, we decide to compactify a dimension $x_\perp^A$ orthogonal to the brane, we change the geometry on which to solve the differential equation for $H(r)$. The simplest solution is to ``smear"\footnote{Such ``smearing'' in dimensional reductions has been discussed in Ref.\ \cite{Lu:1996mg}, where it was termed ``vertical'' dimensional reduction, as opposed to ``wrapping'' a brane in worldvolume directions, which was termed ``diagonal'' dimensional reduction.} the brane in this direction, \ie making $H(r)$ independent of $x_\perp^A$.  Compactification on $T^6$ along 3 parallel and 3 orthogonal directions then results in a 4d extremal black hole solution as discussed above, now interpreted as a D3-brane wrapping a 3-cycle and completely smeared in the other torus-dimensions. To make contact to our earlier discussion, one may simply interpret the $T^6$ as  CY3 (although it preserves even more supersymmetry) and extend this logic to other CY3 manifolds by analogy.

Let us now discuss the different embeddings of our extremal black hole \eqref{blackhole}

\begin{itemize}
    \item \textbf{minimal 5d supergravity}
    \begin{equation}
    \begin{aligned}
        &\dd s_5^2 =- h^{-2}\,\dd t^2+h^2\left(\dd r^2+r^2 \,\dd\Omega_2^2\right) + \left(\dd z+ h^{-1}\,\dd t \right)^2\,,\\
        & \cF_{\theta\phi} = -\sqrt{3}r_H\sin{\theta}\,.
    \end{aligned}
    \end{equation}
    The potential of $\mathcal{F}_{(2)}$ takes the Dirac-monopole form $\mathcal{A}_{(1)}=\sqrt{3}r_H(\cos\theta\pm 1)\dd \phi$, where the sign depends on the coordinate patch of choice. One may interpret this background as generated by a magnetic source, which in 5d is a 2-dimensional extended object, extended in $t$ and $z$ direction. Note furthermore the nontrivial fibration (rotation) of the $z$-circle.
    \item \textbf{11d supergravity}
    \begin{equation}
    \begin{aligned}
        &\dd s_5^2 =- h^{-2}\,\dd t^2+h^2\left(\dd r^2+r^2 \,\dd\Omega_2^2\right) + \left(\dd z+ h^{-1}\,\dd t \right)^2+ \partial_{i}\partial_{\bar{\jmath}}\mathcal{K}\dd y^i \dd\bar{y}^{\bar{\jmath}}\,,\\
        &\bF_{\theta\phi mn}= r_H \sin\theta\cdot\omega_{mn}\,.
    \end{aligned}
    \end{equation}
    This solution may be interpreted as an M5-brane wrapping the $z$-circle as well as $\omega\wedge\omega$.
    \item \textbf{10d IIA supergravity}
    \begin{equation}
    \begin{aligned}
        &\dd\bs^2_{10} = - h^{-2}\,\dd t^2+h^2\left(\dd r^2+r^2 \,\dd\Omega_2^2\right) + \partial_i\partial_{\overline{\jmath}}\mathcal{K}\dd y^i\dd \overline{y}^{\overline{\jmath}}\,,\qquad \bC_{t}=h^{-1}\,, \\
        &\bF_{\theta\phi mn} = r_H\sin{\theta}\cdot \omega_{mn}\,.
    \end{aligned}
    \end{equation}
    This solution can be understood as a circle reduction of the 11d supergravity solution. A possible interpretation of this background is then in terms of D4-branes wrapping $\omega\wedge\omega$. The presence of the $\bC_{(1)}$-potential is the IIA dual of the non-trivially fibred circle in the 5d/11d case. 
    \item \textbf{10d IIB  supergravity}
    \begin{equation}
    \begin{aligned}
        &\dd\bs_{10}^{2}=- h^{-2}\,\dd t^2+h^2\left(\dd r^2+r^2 \,\dd\Omega_2^2\right)+\partial_{i}\partial_{\bar{\jmath}}\mathcal{K}\dd y^{i}\dd y^{\bar{\jmath}}\,,\\
        &\bF_{trijk}=-\frac{r_H}{2h^2r^2}\cdot\Omega_{ijk}\,,\qquad \bF_{\theta\phi ijk}=i\frac{r_H}{2}\sin\theta\cdot\Omega_{ijk}\,,\\
        &\bF_{tr\bar{\imath}\bar{\jmath}\bar{k}}=-\frac{r_H}{2h^2r^2}\cdot\bar{\Omega}_{\bar{\imath}\bar{\jmath}\bar{k}}\,,\qquad \bF_{\theta\phi\bar{\imath}\bar{\jmath}\bar{k}}=-i\frac{r_H}{2}\sin\theta\cdot\bar{\Omega}_{\bar{\imath}\bar{\jmath}\bar{k}}\,.
    \end{aligned}
\end{equation}
This case is the CY3 analogue of the toroidal case mentioned earlier. We may interpret this configuration in terms of D3-branes wrapping $\Omega_3$ and $\bar{\Omega}_3$ in such a way as to generate a real-valued field configuration. 
\end{itemize}

As opposed to black holes arising from branes wrapping individual cycles of the compactification, the consistent truncation forces us to wrap ``overall" cycles of the CY3. This is expected, since a brane wrapping a specific cycle would transform nontrivially under the $SU(3)$ structure group. It would furthermore source gauge fields that have been truncated out of the 4d theory. 

It would be interesting to repeat this exercise with other solutions of the 4d $\mathcal{N}=2$ supergravity, such as \eg non-extremal black holes. Some of these may turn out to be part of the class of solutions studied in \cite{Hulsey:2019xdb}.

\section{Comments on including universal matter multiplets}\label{Matter}
In Ref.\ \cite{Cassani:2019vcl} it is claimed that one may keep a hypermultiplet in the consistent truncation when compactifying 11d supergravity on a CY3, or one hyper- and one vector multiplet when compactifying 11d supergravity on $S^1\times\text{CY3}$. In this section, we propose embedding Ans\"atze using the universal structures of CY3, $\{\eta,\omega,\Omega\}$, up to some constant factor, for the consistent truncation in purely bosonic backgrounds.  We believe that the spinor bilinear corrections can again be identified by comparing supercovariant quantities as we did in Section \ref{Ansatz}. 

\begin{itemize}
    \item \textbf{11d supergravity to 5d  supergravity with 1 hypermultiplet } 
    \begin{equation}
        \begin{aligned}
            &\dd \bs_{11}^2 = e^{\alpha \varrho(x)}\cg_{\mu\nu}(x) \dd x^\mu \dd x^\nu + e^{\beta \varrho(x)}g_{mn}(y)dy^mdy^n ,,\\
            &\bA_{(3)} = -\frac{1}{\sqrt{3}} \cA_{(1)}(x) \wedge \omega + \mathcal{A}(x)\cdot\Omega + \bar{\mathcal{A}}(x)\cdot\bar\Omega + \mathcal{A}_{(3)}\,,\\
            &\bPsi_{\mu} = \Psi_\mu \otimes \eta + \cPsi_\mu \otimes \tilde{\eta}\,,\quad \bPsi_{\bar{\imath}} = \xi \otimes \omega_{\bar{\imath}j}\Sigma^j\eta + c.c.\,.
        \end{aligned}
    \end{equation}
    Here, the K\"ahler structure of the CY3 metric \cite{Candelas:1990pi} is rescaled by a spacetime dependent parameter $\varrho$, sometimes called a volume modulus, and dressed constants $\alpha$ and $\beta$, which are fixed by choosing Einstein frame. The supergravity multiplet is composed of $\{\cg_{\mu\nu},\, \cA_{\mu},\, \Psi_\mu \}$ and the hypermultiplet is composed of $\{\mathcal{A},\, \bar{\mathcal{A}},\, \varrho,\, \mathcal{A}_{(3)},\, \xi\}$. $\mathcal{A}$ is a complex scalar with complex conjugate $\bar{\mathcal{A}}$. The 3-form potential $\mathcal{A}_{(3)}$ in 5d is dual to a scalar. $\xi$ is a 5d Dirac spinor in the hypermultiplet.
    
    \item \textbf{11d supergravity to  4d $\mathcal{N}=2$ supergravity with 1 hyper- and 1 vector multiplet} 
    
    The Ansatz \eqref{eq:5d-4d-1}, \eqref{eq:5d-4d-2}, and \eqref{eq:5d-4d-3} shows how the 5d supergravity multiplet reduces on a circle giving the 4d supergravity multiplet plus a vector multiplet. Furthermore, circle compactification leaves the hypermultiplet invariant, only restricting its coordinate dependence. Hence, the Ansatz for truncating 11d supergravity to 4d $\mathcal{N}=2$ supergravity with a hypermultiplet and a vector multiplet is given by 
    \begin{equation}
        \begin{aligned}
            &\dd \bs_{11}^2 = e^{\alpha\varrho(x)}\left(e^{-\frac{1}{\sqrt{3}}\varphi}\dd s_4^2(x) + e^{\frac{2}{\sqrt{3}}\varphi}\left(\dd z + B_\mu^1 \right)^2\right) + e^{\beta\varrho(x)}g_{mn}(y)dy^mdy^n\,,\\
            &\bA_{(3)} = \frac{1}{\sqrt{3}} \left(B_{(1)}^2 + \chi \dd z\right) \wedge \omega + A(x)\cdot\Omega + \bar{A}(x)\cdot\bar\Omega + A_{(2)}(x) \wedge dz\,,\\
            &\bPsi_\mu = \Psi_\mu \otimes \eta + \cPsi_\mu \otimes \tilde{\eta}\,,\  \bPsi_4 = \Psi_4 \otimes \eta + \cPsi_4 \otimes \tilde{\eta}\,,\ \bPsi_{\bar{\imath}} = \xi \otimes \omega_{\bar{\imath}j}\Sigma^j\eta + c.c.\,,
        \end{aligned}
    \end{equation}
    where
    \begin{equation}
        \Psi_\mu = e^{\frac{1}{4\sqrt{3}}\varphi}\psi_\mu + \frac{1}{2\sqrt{2}}e^{-\frac{\sqrt{3}}{4}\varphi}\gamma_\mu \gamma^{(4)}\lambda\,,\quad \Psi_4 = \frac{i}{\sqrt{2}}e^{-\frac{\sqrt{3}}{4}\varphi}\lambda\,.
    \end{equation}
    Identically to the 5d case, the CY3 metric is deformed from $g_{mn}(y)$ by a K\"ahler structure rescaling $\varrho$. The constants  $\alpha$ and $\beta$ are chosen to put the lower dimensional theory into Einstein frame. The two vectors $\{B^1_\mu,\, B^2_\mu\}$ form a fundamental representation of the electric-magnetic duality group $Sp(4,\mathbb{R})$. Hence, we cannot generically distinguish which vector belongs to the supergravity multiplet. We expect the scalars in the vector multiplet  $\{\chi,\,\varphi\}$ to parametrise a special K\"ahler manifold and the scalars in the hypermultiplet $\{A,\, \bar{A},\, \varrho,\, A_{(2)}\}$ to parametrise a quaternionic-K\"ahler manifold \cite{Andrianopoli:1996vr, Andrianopoli:1996cm}. $A_{(2)}$, reduced from $\mathcal{A}_{(3)}$, is dual to a scalar in 4d.   $\psi_\mu,\, \lambda,\, \xi$ are the Dirac gravitino of the supergravity multiplet, the Dirac spinor of the vector multiplet, and the Dirac spinor of the  hypermultiplet, respectively. 
        
    \item \textbf{IIA supergravity to 4d $\mathcal{N}=2$ supergravity with 1 hyper- and 1 vector multiplet}
    \begin{equation}
        \begin{aligned}
            &\dd \bs_{10}^2 = e^{\alpha\varrho(x)}\cg_{\mu\nu}(x) \dd x^\mu \dd x^\nu + e^{\beta\varrho(x)}g_{mn}(y)dy^mdy^n \,, \\
            &\bPhi = \phi(x),\quad \bA_{(1)} = A_{(1)}(x),\quad \bB_{(2)}= B_{(2)}(x) + B(x)\cdot\omega\,,\\ &\bC_{(3)} = C_{(1)}(x)\wedge\omega + C(x)\cdot\Omega + \bar{C}(x)\cdot\bar{\Omega}\,,\\
            &\bPsi_\mu = \frac{1}{\sqrt{2}}\left(\psi^1_\mu\otimes\eta_+ +\psi_{1\mu}\otimes\eta_- + \psi_{2\mu}\otimes\eta_+ +\psi^2_{\mu}\otimes\eta_-\right)\,,\\ 
            &\bPsi_{\bar{\imath}} = \lambda^1\otimes \omega_{\bar{\imath}j}\Sigma^j\eta_+ + \lambda_2\otimes \omega_{\bar{\imath}j}\Sigma^j\eta_+ + c.c. \\
            &\bla = \xi^1\otimes\eta_+ +\xi_1\otimes\eta_- + \xi^2\otimes\eta_+ +\xi_2\otimes\eta_-\,.
        \end{aligned}
    \end{equation}
    The CY3 metric is deformed from the original one by rescaling the K\"ahler form by a factor $\varrho$. The two constants $\alpha$ and $\beta$ are chosen to put the lower dimensional theory into Einstein frame. The two vectors $\{A_\mu,\, C_\mu\}$ form a fundamental representation of the electric-magnetic duality group $Sp(4,\mathbb{R})$. We expect the scalars in the vector multiplet $\{B,\, \varrho\}$  parametrise a special K\"ahler manifold. The scalars in the hypermultiplet are $\{\phi,\, B_{(2)},\, C,\, \bar{C}\}$, parametrising a quaternionic-K\"ahler manifold.\footnote{We note that the scalars from the metric reduction, \ie $\phi$ and $\varrho$, appear in different multiplets depending on the order in which $S^1$ and CY3 compactification are performed. This should be due to mixing upon changing frames.} The 2-form potential $B_{(2)}$ is dual to a scalar and $C$ is a complex scalar with complex conjugate $\bar{C}$. The hypermultiplet here is the universal hypermultiplet, and we note that the embedding of these fields in 10d only depends on the holomorphic 3-form.
    $\psi^I_\mu,\, \lambda_I,\, \xi^I$ are 2 Weyl gravitini of the supergravity multiplet, 2 Weyl spinors of the vector multiplet, and 2 Weyl spinors of the universal hypermultiplet, respectively, all with positive chirality. A more general Ansatz for this truncation has been constructed in \cite{Terrisse_2019, tsimpis2020}, where they also considered background fluxes and fermion condensation.

    \item \textbf{IIB supergravity to 4d $\mathcal{N}=2$ supergravity with 2 hypermultiplets}
    \begin{equation}
        \begin{aligned}
            &\dd \bs_{10}^2 = e^{\alpha\varrho(x)}g_{\mu\nu}(x) \dd x^\mu \dd x^\nu + e^{\beta\varrho(x)}g_{mn}(y)dy^mdy^n\,, \\
            &\bPhi = \phi(x),\quad \bC = C(x),\quad \bB_{(2)}= B_{(2)}(x) + B(x)\cdot\omega\,,\\
            &\bC_{(2)} = C_{(2)}(x) + \tilde{C}(x)\wedge\omega\,,\quad \bC_{(4)} = C^+_{(1)}(x)\wedge\Omega + C^-_{(1)}(x)\cdot\bar{\Omega} + \tilde{C}_{(2)}\wedge\omega \,,\\
            &\bPsi_{\mu}^{I}=\frac{1}{2}\left(\psi_{\mu}^{I}\otimes\eta_{+}+\psi_{I\mu}\otimes\eta_{-}\right),\quad \bPsi_{\bar{\imath}}^{I}=\lambda^{I}\otimes\omega_{\bar{\imath}j}\Sigma_j \eta_{-} + c.c.\,,\\
            &\bla^{I}=\xi^{I} \otimes \eta_{+}+\xi_{I} \otimes\eta_{-}\,.
        \end{aligned}
    \end{equation}
    The CY3 metric is deformed from the original one by rescaling the K\"ahler form by a warp factor $\varrho$. The two constants $\alpha$ and $\beta$ chosen to put the lower dimensional theory into Einstein frame. We have $F^\pm_{(2)} = \dd C^\pm_{(1)}$ being the (anti-)self-dual field strength. This guarantees the self-duality of the 5-form field strength in 10d \eqref{eq:IIB-self-dual-eom}. The supergravity multiplet comprises $\{g_{\mu\nu},\, C_{(1)},\, \psi_\mu^I\}$. The scalars in the two hypermultiplets are $\{\Phi,\, C,\, B_{(2)},\, C_{(2)}\}$ and $\{\varrho,\, B,\, \tilde{C},\, \tilde{C}_{(2)}\}$ respectively, together parametrising a quaternionic-K\"ahler manifold. The 2-form potentials in 4d are dual to scalars. The first multiplet here is the universal hypermultiplet which is independent of any CY3 structure. The second multiplet results from the volume deformation. $\psi^I_\mu,\, \lambda^I,\, \xi^I$ are 2 Weyl gravitini of the supergravity multiplet, 2 Weyl spinors of the deformation hypermultiplet, and 2 Weyl spinors of the universal hypermultiplet, respectively, all with positive chirality. 
\end{itemize}

The task of checking the consistency of these embedding Ans\"atze will be left to future work. However, we would like to point out some interesting features of our proposed Ans\"atze. The only metric deformation we introduce is the overall rescaling $\varrho(x)$, \ie turning on the volume modulus. We keep the complex structure and the relative K\"ahler structure unchanged.  Besides the universal hypermultiplet, we will keep one more vector multiplet in the IIA reduction and one more hypermultiplet in the IIB case. This is expected for IIA due to the relation to 11d supergravity and for IIB due to the mirror-symmetry to IIA, resulting in an overall consistent duality picture.

\section{Conclusion}\label{conclusion}
In this paper, we have proven that the following truncations are consistent:
\begin{itemize}
    \item 11d supergravity compactified on CY3 $\rightarrow$ pure 5d supergravity,
    \item Pure 5d supergravity compactified on $S^1$ $\rightarrow$ pure 4d $\mathcal{N}=2$ supergravity,
    \item 11d supergravity compactified on CY3$\times S^1$ $\rightarrow$ pure 4d $\mathcal{N}=2$ supergravity,
    \item IIA supergravity compactified on CY3 $\rightarrow$ pure 4d $\mathcal{N}=2$ supergravity,
    \item IIB supergravity compactified on CY3 $\rightarrow$ pure 4d $\mathcal{N}=2$ supergravity.
\end{itemize}
We provided explicit embedding Ans\"atze (Section \ref{Ansatz}) and checked explicitly that the equations of motion of the higher-dimensional theory reduce to the equations of motion of the lower-dimensional theory. We furthermore checked that the supersymmetry transformations reduce consistently (Section \ref{proof}). This explicit proof supports the more implicit reasoning in \cite{Cassani:2019vcl} and furthermore shows that a truncation to pure supergravity is possible, constituting a strict subsector of the $SU(3)$-singlet sector discussed there. We commented on the remaining universal matter multiplets which would complete the $SU(3)$-singlet sector (Section \ref{Matter}). It would be interesting to explicitly demonstrate the consistency of 4d $\mathcal{N}=2$ supergravity including these universal matter multiplets. We expect that, again, fermion bilinears have to be added to our na{\"\i}ve Ansatz in order to restore supercovariance.

It is interesting to note at this point that the embedding Ans\"atze we proposed for 11d supergravity on CY3$\times S^1$ and IIA supergravity on CY3 suggest that circle reduction commutes with our CY3 truncation. In fact one may see this as a consistency check for our Ans\"atze. Furthermore our IIA and IIB Ans\"atze show signs of mirror symmetry, with gauge fields in IIA only wrapping the K\"ahler form while the IIB fields wrap the holomorphic 3-form. It would be interesting to establish the exact mapping, which we assume requires the addition of the matter multiplets discussed in Section \ref{Matter}, which may mix with the supergravity multiplet under mirror symmetry.

We presented a simple application of these results by embedding a non-trivial 4d background, in our case an extremal black hole, in the higher-dimensional theories (Section \ref{BH}). Consistency guarantees that this solves the higher-dimensional equations of motion and therefore suggests that we should interpret 4d extremal black holes as extended BPS objects in the higher-dimensional theories. We gave an interpretation in terms of wrapped branes on overall cycles of the CY3 in analogy to past discussions of extremal black holes in compactification scenarios \cite{Strominger:1996sh}. It would be interesting to establish this interpretation more precisely also in the context of similar embeddings in braneworld scenarios \cite{Leung:2022nhy}. Various other non-trivial backgrounds could be explored such as for example non-extremal or rotating black holes.

A side product of our discussion is the extensive list of equations of motion in Appendix \ref{conventions} which we gathered from various sources and presented in full detail. We expressed them in terms of the basic fields present in the respective supergravity theory, but it stands to reason that more elegant expressions are possible, for example by expressing the equations of motion in terms of supercovariant quantities. However, there may well be more systematic approaches such as the superspace formalism \cite{Wess:1977fn,Howe:1983sra} or generalised geometry \cite{Coimbra:2011nw} to simplify these equations. Recent progress has been made for 10d $\mathcal{N}=1$ theories \cite{Kupka:2024vrd} and it stands to reason that a similar discussion for $\mathcal{N}=2$ is possible.

A more conceptual issue is whether these truncations can be interpreted as some kind of EFT, where massive states have been integrated out and only the supergravity multiplet remains \cite{Duff:1989cr}. In generic compactification scenarios light moduli fields are ubiquitous (reviewed \eg in \cite{Conlon:2006gv,McAllister:2023vgy}), which would not be integrated out, so there is a priori no expectation to find pure supergravity. It would be interesting to investigate whether certain scenarios exist where pure supergravity (or the universally extended $SU(3)$-singlet sector) is in fact the low-energy EFT.

\section*{Acknowledgments}
We are grateful to Costas Bachas, Calvin Chen,  Steven Hsia, Yusheng Jiao, Rahim Leung and Dan Waldram for helpful discussions. The work of KSS was supported in part by the STFC under consolidated grants ST/T000791/1 and ST/X000575/1. The work of TS was supported in part by the President's PhD Scholarships of Imperial College London and the research grant  ``Deforming the integrable structure of string models" of the Department of Physics and Astronomy ``Galileo Galilei" at the University of Padova. TS would like to thank the Cluster of Excellence EXC 2121 Quantum Universe 390833306 and the Collaborative Research
Center SFB1624 for creating a productive research environment at DESY.
\pagebreak

\begin{appendices}
\section{Conventions}\label{conventions}
In this appendix we gather our conventions for fermions as well as the equations of motion and supersymmetry transformations of supergravity theories in various dimensions. Since we chose conventions that make our discussion as homogeneous as possible we will comment on the translation to standard literature wherever necessary. It may also be helpful to state some initial choices:
\begin{itemize}
    \item We work in ``mostly plus" signature $(-,+,+,+,\dots)$. 
    \item We discuss 10d supergravity in Einstein frame.
    \item We denote 11d and 10d quantities by boldface letters with capital latin indices ($\bF_{AB}$). 5d quantities are denoted by calligraphic letters with greek indices ($\mathcal{F}_{\mu\nu}$) and 4d quantities by regular letters with greek indices ($F_{\mu\nu}$). 
    \item The local Lorentz frame is indicated by the underlined indices ($\underline{A}, \underline{B}, \dots$ for 11d and 10d, and $\underline{\mu}, \underline{\nu},\dots$ for 5d and 4d). 
    \item The CY3 can be parametrised by real coordinates indicated by ($m,n,\dots$) or by complex coordinates indicated by ($i,j,\bar{\imath},\bar{\jmath},\dots$).
    \item The 11d and 10d Clifford algebra is denoted by $\Gamma$, while for 5d and 4d we use $\gamma$. To clearly distinguish the Clifford algebra on the CY3 space, we use $\Sigma$ for the relevant 6d Euclidean Clifford algebra.
    \item We define the top rank gamma matrix as $\Gamma^{(d)}\equiv\Gamma_{\underline{0}}\Gamma_{\underline{1}}\cdots\Gamma_{\underline{d-1}}$.
    \item (Anti-)Symmetrisation of indices is denoted by (square) brackets and is normalised ``weight one'', \eg
    \begin{equation}
    \bF_{[ABCD]}=\frac{1}{4!}\sum_{\sigma\in S_4}\sgn[\sigma]\,\bF_{\sigma(A)\sigma(B)\sigma(C)\sigma(D)}\,.
    \end{equation}
    Indices can be excluded from (anti-)symmetrisation by vertical lines as for example $\bF_{[A\vert B\vert C]}=\tfrac{1}{2}(\bF_{ABC}-\bF_{CBA})$.
    \item The covariant Levi-Civita tensor is totally antisymmetric and normalised as $\epsilon_{01\dots d} = \sqrt{\abs{g}}$.
    \item Spinor components appearing in spinor bilinears are to be understood as Grassmann numbers, changing sign when one changes their order. 
\end{itemize}

\subsection{Fermions across dimensions}\label{conventionsFermions}

The Clifford algebra $\text{Cliff}(p,q; \mathbb{R})$ is defined by the relation $\{\Gamma_{\underline{A}}, \Gamma_{\underline{B}}\} = 2\eta_{\underline{AB}}$, where $\eta_{\underline{AB}}$ is a $d$-dimensional metric with signature $(p,q)$. Starting from a representation $\Gamma_{\underline{A}}$ we can build eight alternative representations
\begin{equation}
    \{\Gamma_{\underline{A}}\,,\ -\Gamma_{\underline{A}}\,,\ \Gamma_{\underline{A}}^\dagger\,,\ -\Gamma_{\underline{A}}^\dagger\,,\ \Gamma_{\underline{A}}^T\,,\ -\Gamma_{\underline{A}}^T\,,\ \Gamma_{\underline{A}}^*\,,\ -\Gamma_{\underline{A}}^*\}\,.
\end{equation}
Depending on the dimension $d$, they fall into different classes of irreducible representations. In even dimensions, all the irreducible representations are equivalent, while they compose two equivalence classes in odd dimensions. We can introduce intertwiners to relate the representations in the same equivalence class.

In even dimension, we can use the top rank gamma matrix $\Gamma^{(d)}\equiv\Gamma_{\underline{0}}\Gamma_{\underline{1}}\cdots\Gamma_{\underline{d-1}}$ to relate negative representation as 
\begin{equation}
    \Gamma^{(d)}\Gamma_{\underline{A}}(\Gamma^{(d)})^{-1}=-\Gamma_{\underline{A}}\,.
\end{equation}
However, in odd dimensions $\Gamma^{(d)} \propto \mathbf{1}$ cannot serve as an intertwiner. In even dimensions, we can also introduce (linearly dependent) intertwiners as 
\begin{equation}
    \begin{aligned}
        &A\Gamma_{\underline{A}}A^{-1} = \Gamma_{\underline{A}}^\dagger\,,\qquad C\Gamma_{\underline{A}}C^{-1}=\Gamma^T_{\underline{A}}\,,\qquad D\Gamma_{\underline{A}}D^{-1}=\Gamma_{\underline{A}}^*\,,\\
        &\tilde{A}\Gamma_{\underline{A}}\tilde{A}^{-1} = -\Gamma_{\underline{A}}^\dagger\,,\qquad \tilde{C}\Gamma_{\underline{A}}\tilde{C}^{-1}=-\Gamma^T_{\underline{A}}\,,\qquad \tilde{D}\Gamma_{\underline{A}}\tilde{D}^{-1}=-\Gamma_{\underline{A}}^*\,.
    \end{aligned}
\end{equation}
The intertwiners for negative representations  can be related to the normal ones by successive action with $\Gamma^{(d)}$ and a potential factor, \eg $\tilde{D} = \alpha D\Gamma^{(d)}$. In odd dimensions, only half of the intertwiners exist for each representation. The intertwiners have to satisfy the consistency conditions
\begin{equation}
    A=\alpha A^\dagger\,,\qquad \Gamma^{(d)}=\beta(\Gamma^{(d)})^{-1}\,,\qquad C=\eta C^T\,,\qquad DD^*=\delta\,,
\end{equation}
in order to generate involution maps. The constants introduced here are arbitrary non-zero numbers satisfying 
\begin{equation}
    \alpha\alpha^* = \eta^2 = 1\,,\quad \delta = \delta^*\,.
\end{equation}
In most cases, we choose to normalise $\delta = \pm 1$.  More details can be found in the appendix of ref.\cite{SOHNIUS198539, Coimbra:2012af}.

We have two different ways to define the charge conjugation of a spinor using different intertwiners,  $\Psi^c=(D\Psi)^*$ and $\Psi^c=(\tilde{D}\Psi)^*$. If the $D$ intertwiner satisfies $DD^*=1$, a Majorana spinor satisfying $\Psi=\Psi^c$ exists for the corresponding charge conjugation, while if $DD^*=-1$, we can only define the symplectic Majorana spinor satisfying $(\Psi^c)^c= - \Psi$. A pair of symplectic Majorana spinors is equivalent to a Dirac spinor. The same argument applies for charge conjugation mediated by $\tilde{D}$. The possibility of defining Majorana (M) or symplectic Majorana (SM) fermions in different dimensions is summarised in table \ref{table:charge-conjugation}. Similarly, we have two ways to define Majorana conjugation $\bar{\Psi} = \Psi^TC$ and $\bar{\Psi} = \Psi^T\tilde{C}$, and two ways to define Dirac conjugation $\Psi^{\text{D.C.}} = \Psi^\dagger A$ and $\Psi^{\text{D.C.}} = \Psi^\dagger \tilde{A}$.
\begin{table}[h]
\begin{center}
\begin{tabular}{|c|c|c|c|c|c|c|c|c|}
\hline
$p-q\ \mathrm{mod}\ 8$&    0  &   1   &   2   &   3   &   4   &   5   &   6   &   7    \\
\hline
\phantom{\Big(}$D$                     &    M  &   M   &   M   &       &   SM  &   SM  &   SM  &         \\  
\hline
\phantom{\Big(}$\tilde{D}$           &    M  &       &   SM  &   SM  &   SM  &       &   M   &   M    \\
\hline
\end{tabular}
\caption{Existence of Majorana (M) or symplectic Majorana (SM) representations in different dimensions with signature $(p,q)$.\label{table:charge-conjugation}}
\end{center}
\end{table}

The Fierz identity in general dimension is 
\begin{equation}
    M = 2^{-m}\sum_{k=0}^{[d]} \frac{1}{k!} \Gamma_{\underline{A_1\cdots A_k}}\text{Tr}\left(\Gamma^{\underline{A_k\cdots A_1} }M\right),
    \label{eq:Fierz-identity}
\end{equation}
where $[d] = d = 2m$ for even dimensions $d$, and $[d] = (d-1)/2 = m$ for odd $d$.
 
 \

We now summarise explicitly our conventions in the dimensions relevant to this paper. We will not use intertwiners $A$ and $\tilde{A}$ in this paper and will therefore not list them here. For more details, see \cite{SOHNIUS198539, Coimbra:2012af}.
\begin{itemize}
    \item \textbf{11d:} The equivalence class including $\Gamma_{\underline{A}}$ is
        \begin{equation}
            \left\{\Gamma_{\underline{A}}\,,\ -\Gamma_{\underline{A}}^{\dagger}\,,\ -\Gamma_{\underline{A}}^{T}\,,\ \Gamma_{\underline{A}}^{*}\right\}\,,
        \end{equation}
        and we can only consistently define the intertwiners 
        \begin{equation}
            \tilde{C}^{(11)}=-\tilde{C}^{(11)T}\,,\qquad D^{(11)}D^{(11)*}=1\,.
            \label{eq:intertwiner-consitence-11}
        \end{equation}
        We define the charge conjugation as $\bla^c= \left(D^{(11)}\bla\right)^* $.  Majorana conjugation is defined as $\bar{\bla}=\bla^T\tilde{C}^{(11)}\,.$ When we compactify 11d supergravity on CY3, we need the factorisation of gamma matrices and intertwiners 
        \begin{equation}
            \begin{aligned}
                &\Gamma_{\mu}=\gamma_\mu\otimes -i\Sigma^{(6)}\,,\qquad \Gamma_m = \mathbf{1}\otimes\Sigma_m\,,\\
                &\tilde{C}^{(11)} = C^{(5)}\otimes\tilde{C}^{(6)}\,,\qquad  D^{(11)}=\tilde{D}^{(5)}\otimes D^{(6)}\,.
            \end{aligned}
            \label{eq:11-5intertwiner}
        \end{equation}
    \item \textbf{10d:} In even dimensions, all Clifford algebra representations are equivalent. In 10d spacetime, we can define Majorana-Weyl spinors. The intertwiners we use in 10d are\footnote{The intertwiner we choose for the Majorana conjugation is aligned with the choice in \cite{freedman_van_proeyen_2012}.}
        \begin{equation}
            \tilde{C}^{(10)} = -\tilde{C}^{(10)T}\,,\quad D^{(10)}D^{(10)*} = 1\,.
        \end{equation}
        We define the charge conjugation as $\bla^c=(D^{(10)}\bla)^*$ and Majorana conjugation as $\bar{\bla}=\bla^T\tilde{C}^{(10)}\,.$  The projection operator to a left- and right-handed Weyl-spinor basis is given by
        \begin{equation}
            P_\pm = \frac12 \left(1 \pm \Gamma^{(10)}\right)\,.
        \end{equation} 
        In this paper, fermionic type IIA supergravity fields are taken to be Majorana spinors while fermionic type IIB supergravity fields are taken to be Weyl spinors. We factorise the gamma matrices and intertwiners as 
        \begin{equation}
            \begin{aligned}
                &\Gamma^{\mu}=\gamma^\mu\otimes\mathbf{1}\,,\qquad \Gamma^m = i\gamma^{(4)}\otimes\Sigma^m\,.\\
                &D^{(10)}=D^{(4)}\otimes \tilde{D}^{(6)},\qquad  \tilde{C}^{(10)}=\tilde{C}^{(4)}\otimes\tilde{C}^{(6)}.
            \end{aligned}
            \label{eq:10-4intertwiner}
        \end{equation}
        when we compactify 10d supergravity on CY3.
    \item \textbf{5d:} The equivalence class including $\gamma_\mu$ is 
        \begin{equation}
            \left\{\gamma_{\underline{\mu}}\,,\  -\gamma_{\underline{\mu}}^{\dagger}\,,\ \gamma_{\underline{\mu}}^{T}\,,\  -\gamma_{\underline{\mu}}^{*}\right\}\,,
        \end{equation}
        The intertwiners we use in 5d are
        \begin{equation}
            C^{(5)}=-C^{(5)T}\,,\qquad \tilde{D}^{(5)}\tilde{D}^{(5)*}=-1\,.
            \label{eq:intertwiner-consitence-5}
        \end{equation}
         The charge conjugation is defined as $\cep^c = (\tilde{D}^{(5)}\cep)^*$. The Majorana conjugate is defined as $\bar{\cep}=\cep^TC^{(5)}.$
         Note that the intertwiner $\tilde{D}^{(5)}$ can only define symplectic Majorana spinors, as acting with the charge conjugation twice results in a sign change. 
        When we compactify 5d supergravity on a circle, we choose the following convention for the gamma matrices:
        \begin{equation}
            \gamma^{(d=5)}_\mu = - \gamma^{(d=4)}_\mu\,,\qquad \gamma^{(d=5)}_4 = -i \gamma^{(4)}\,.
            \label{eq:5-4-gamma matrices}
        \end{equation}
        Notice that the appropriate intertwiners in 5d are $\tilde{D}$ and $C$.  The 4d theory resulting from the dimension reduction will inherit them. We need to relate them to the usual 4d intertwiners by 
        \begin{equation}
            D^{(4)}=\tilde{D}^{(4)}\gamma^{(4)}\,,\qquad \tilde{C}^{(4)} = iC^{(4)}\gamma^{(4)}\,.
            \label{eq:5-4-intertwiners}
        \end{equation}
    \item \textbf{4d:} We can define Majorana spinors using the intertwiner $D^{(4)}$ and symplectic Majorana spinors using $\tilde{D}^{(4)}$. Charge conjugation will change the chirality of a Weyl spinor in 4d. Hence we cannot define Majorana-Weyl spinors as in 10d. In this paper, we use the intertwiners 
        \begin{equation}
            \tilde{C}^{(4)}=-\tilde{C}^{(4)T}\,,\quad D^{(4)}D^{(4)*} = 1\,.
        \end{equation}
    We define the charge conjugation for 4d spinors as $\lambda^c = (\tilde{D}^{(4)}\lambda)^*\,$ and Majorana conjugation as $\bar{\lambda}=\bar{\lambda}^TC$. In this paper, we use Weyl spinors in 4d and indicate their chirality with the placement of the $SU(2)_R$ indices. We define the projection operator onto the Weyl basis as
    \begin{equation}
        P_\pm = \frac12\left(1 \pm i\gamma^{(4)}\right)\,.
    \end{equation} 
    We define the chiral gravitino and supersymmetry transformation parameter as 
    \begin{equation}
        P_+ \psi^I_\mu = \psi^I_\mu\,, \quad\ P_- \psi_{I\mu} = \psi_{I\mu}\,,\quad\text{and}\quad P_+ \epsilon^I = \epsilon^I\,, \quad P_- \epsilon_I = \epsilon_I\,,
    \end{equation}
    and the chiral dilatino as
    \begin{equation}
        P_+ \lambda_I = \lambda_I\,, \qquad P_- \lambda^I = \lambda^I\,.
    \end{equation}
    Note that this indicates an opposite chirality for the dilatino which simplifies the expressions for supersymmetry transformations \cite{freedman_van_proeyen_2012}.
    \item \textbf{6d Euclidean:} We can define Majorana spinors using the intertwiner $\tilde{D}^{(6)}$ and symplectic Majorana  spinors using $D^{(6)}$. The intertwiners we use in this paper are 
    \begin{equation}
        \tilde{C}^{(6)}=\tilde{C}^{(6)T}\,,\quad \tilde{D}^{(6)}\tilde{D}^{(6)*} = 1\,,\quad D^{(6)}D^{(6)*} = -1\,.
    \end{equation}
    We reserve the label $\eta^c$ for the charge conjugation associated to Majorana spinors as $\eta^c=(\tilde{D}^{(6)}\eta)^*$, while define $\tilde{\eta} = (D^{(6)}\eta)^*$. Charge conjugation in 6d Euclidean space changes the chirality. We define Majorana conjugation as $\bar{\eta}=\eta^T\tilde{C}^{(6)}\,$. We may choose a purely imaginary representation for the Clifford algebra \cite{Tomasiello:2022dwe}, in which $A^{(6)}=\tilde{C}^{(6)}=\tilde{D}^{(6)}=\mathbf{1}$ and hence $\eta^c=\eta^*$. We will use this imaginary representation unless otherwise stated. The only spinor we use in 6d Euclidean space is the covariantly constant spinor of the CY3, $\eta_+\,,$ and its charge conjugation $\eta_-\,.$ In the purely imaginary representation, the Majorana conjugation acts as  $\overline{\eta_\pm}=\eta_{\mp}^\dagger.$ The projection operator onto the Weyl basis is given by
    \begin{equation}
        P_\pm = \frac12 \left(1 \mp i\Sigma^{(6)} \right)\,.
    \end{equation}
    and $P_\pm \eta_\pm = \eta_\pm\,.$
\end{itemize}

\subsection{11d supergravity}\label{conventions11d}
The field content of 11d supergravity comprises the graviton $\bg_{AB}$, a Majorana gravitino $\bPsi_A$ and a 3-form potential $\bA_{ABC}$ with curvature $\bF_{ABCD}=4\partial_{[A}\bA_{BCD]}$. The gravitino is the connection for the gauged supersymmetry and it is useful to introduce a supercovariant field strength tensor and spin connection
\begin{equation}
    \begin{aligned}
        \hat{\bF}_{ABCD}&=\bF_{ABCD}-3\bbPsi_{[A}\Gamma_{BC}\bPsi_{D]}\,,\\
        \hat{\bom}_{A\underline{BC}}&=\bom_{A\underline{BC}}(\be)-\frac14\left(\bar{\bPsi}_A\Gamma_{\underline{C}}\bPsi_{\underline{B}} - \bar{\bPsi}_{\underline{B}}\Gamma_A\bPsi_{\underline{C}} + \bar{\bPsi}_{\underline{C}}\Gamma_{\underline{B}}\bPsi_A\right)\,\\
    \end{aligned}
    \label{eq:11d-supercov}
\end{equation}
whose supersymmetry transformations do not involve derivatives of the transformation parameter $\bep$. $\bom_{M{\underline{AB}}}(\be)$ is the spin connection associated to the Levi-Civita connection.
The action is \cite{Cremmer:1978km, Cremmer:1980ru, freedman_van_proeyen_2012} 
\begin{equation}
    \begin{aligned}
        S_{11}=\frac{1}{2\kappa_{11}^2} \int d^{11}x \be&\left[\bR(\bom)-\bar{\bPsi}_A\Gamma^{ABC}D_B\left(\frac12(\bom+\hat{\bom})\right)\bPsi_C - \frac{1}{2\cdot4!}\bF^{ABCD}\bF_{ABCD}\right.\\
        &\ \ +\frac{1}{192}\bar{\bPsi}_E\left(\Gamma^{A\dots F}+12\Gamma^{AB}\bg^{CE}\bg^{DF} \right)\bPsi_F\left(\bF_{ABCD}+\hat{\bF}_{ABCD}\right)\\
        &\ \ + \left.\frac{\be^{-1}}{144^2}\epsilon^{A\dots K}\bF_{ABCD}\bF_{EFGH}\bA_{IJK}\right]
    \end{aligned}
\end{equation}
with equations of motion in the form 
\begin{equation}
    \begin{aligned}
        0=&\Gamma^{ABC}D_B(\hat{\bom}) \bPsi_C-\frac{1}{288}\hat{\bF}_{DEFG}\Gamma^{ABC}\left(\Gamma^{DEFG}{}_B-8\delta^D_B\Gamma^{EFG}\right)\bPsi_C\,,\\
        0=&\nabla_A \bF^{ABCD}-\frac{1}{4}\nabla_A\left(\bbPsi_E \left(\Gamma^{A\dots F}+12\Gamma^{[AB}\eta^{C|E}\eta^{|D]F}\right)\bPsi_F\right)+\frac{1}{2\cdot 4!^2}\epsilon^{B\dots L}\bF_{EFGH}\bF_{IJKL}\,,\\
        \bR_{AB}(\bom)=&\frac{1}{2\cdot3!}\left(\bF_{ACDE}\bF_{B}{}^{CDE}-\frac{1}{12}\bg_{AB}\bF^{CDEF}\bF_{CDEF}\right)\\
        &+\frac{3}{2}\bar{\bPsi}_{[A|}\Gamma_{B}{}^{CD}D_{|C}\left(\frac{1}{2}(\bom+\hat{\bom})\right)\bPsi_{D]}-\frac{1}{18}\bg_{AB}\bar{\bPsi}_{C}\Gamma^{CDE}D_{D}\left(\frac{1}{2}(\bom+\hat{\bom})\right)\bPsi_{E}\\
        &-\frac{1}{8}\bar{\bPsi}_{(A}\Gamma^{CD}\bPsi^{E}\left(\bF_{B)ECD}+\hat{\bF}_{B)ECD}\right)
        -\frac{1}{16}\bar{\bPsi}^{D}\Gamma_{B}{}^{C}\bPsi^{E}\left(\bF_{ACDE}+\hat{\bF}_{ASCD}\right)\\
        &+\frac{1}{72}\bg_{AB}\bar{\bPsi}^{C}\Gamma^{EF}\bPsi^{D}\left(\bF_{CDEF}+\hat{\bF}_{CDEF}\right)\\
        &-\frac{1}{64}\left[\bar{\bPsi}_{[F|}\Gamma_{B}{}^{C\dots G}\bPsi_{|G}\left(\bF_{ACDE]}+\hat{\bF}_{ACDE]}\right)-\frac{2}{27}\bg_{AB}\bar{\bPsi}_{G}\Gamma^{C\dots H}\bPsi_{H}\left(\bF_{CDEF}+\hat{\bF}_{CDEF}\right)\right]\\
        &-\frac{1}{64}\left(\bar{\bPsi}_{C}\Gamma^{CDE}\Gamma^{GH}\bPsi_{E}\right)\left(\bar{\bPsi}_{[A|}\Gamma_{B}{}^{F}{}_{DGH}\bPsi_{|F]}\right)\\
        &  +\frac{1}{9\cdot 64}\bg_{AB}\left(\bar{\bPsi}_{C}\Gamma^{CDE}\Gamma^{HI}\bPsi_{E}\right)\left(\bar{\bPsi}_{[F|}\Gamma^{FG}{}_{DHI}\bPsi_{|G]}\right)\,.
    \end{aligned}
    \label{eq:11d-eom}
\end{equation}
The spin connection in the equation of motion is defined as
\begin{equation}
        \bom_{A\underline{BC}}=\hat{\bom}_{A\underline{BC}} +\frac18\bar{\bPsi}_D\Gamma^{DE}{}_{A\underline{BC}}\bPsi_E
    \label{eq:11d-spin-connection}
\end{equation}
which is the one derived by the 1\textsuperscript{st} order formalism.\footnote{In other words, the variation of the spin connection $\bom$ in the action provides an algebraic equation for the torsion, which generates fermion bilinears in the connection.} 
The coordinate covariant derivative $\nabla$ in the equations of motion is defined as acting on both Lorentz and spacetime indices, \ie $\nabla = \partial + \bom(\be) + \bGa(\bg) = D(\be) + \bGa(\bg)$, where $\bGa(\bg)$ is the Levi-Civita connection. In contrast, the local Lorentz derivative $D(\hat{\bom})$ only acts on Lorentz indices. It acts on spinors as
\begin{equation}
    D_A(\hat{\bom}) \bPsi_B = \partial_A \bPsi_B + \frac14 \hat{\bom}_{A\underline{CD}}\Gamma^{\underline{CD}}\bPsi_B\,.
    \label{eq:derivative on spinor}
\end{equation}

The equations of motion have to be supplemented by the Bianchi identity
\begin{align}\label{11dbianchi}
\partial_{[A} \bF_{BCDE]}=0\,.
\end{align}
The supersymmetry transformations with parameter $\bep$ are given by \cite{Cremmer:1978km, freedman_van_proeyen_2012}
\begin{equation}
    \begin{aligned}
        &\delta \be_{A}{}^{\underline{B}}=\frac12 \bbep\Gamma^{\underline{B}}\bPsi_{A}\,,\qquad \delta \bA_{ABC}=\frac{3}{2}\bbep\Gamma_{[AB}\bPsi_{C]}\,,\\
        &\delta \bPsi_A=D_A(\hat{\bom})\bep-\frac{1}{288}\hat{\bF}_{BCDE}\left(\Gamma^{BCDE}{}_A-8\delta^B_A\Gamma^{CDE}\right)\bep\,.
    \end{aligned}
    \label{eq:11d-susy}
\end{equation}
The gravitino equation of motion and supersymmetry transformation have a similar structure. We can define a supercovariant derivative on spinors as 
\begin{equation}
    \hat{D}_A(\hat{\bom}) = D_A(\hat{\bom}) - \frac{1}{288}\hat{\bF}_{BCDE}\left(\Gamma^{BCDE}{}_A-8\delta^B_A\Gamma^{CDE}\right)\,.
\end{equation}
which simplifies the supersymmetry transformation of the supercovariant field strength as 
\begin{equation}
    \delta\hat{\bF}_{ABCD}=6\bar{\bep}\Gamma_{[AB}\hat{D}_C(\hat{\bom})\bPsi_{D]}\,.
    \label{eq:11d-susy-supercov}
\end{equation}
\subsection{IIA supergravity}\label{conventionsIIA}
The field content of 10d type IIA supergravity comprises the graviton $\bg_{AB}$, a Majorana gravitino $\bPsi_A$, form fields $\bC_{A},~\bB_{AB},~ \bC_{ABC}$ with their respective curvatures $\bF_{(2)}=\dd \bC_{(1)},~\bH_{(3)}=\dd\bB_{(2)},~ \bF_{(4)}=\dd\bC_{(3)}$, a Majorana dilatino $\bla$ and a dilaton $\bPhi$. As usual we define the gauge covariant curvature
\begin{equation}
\tilde{\bF}_{ABCD}=\bF_{ABCD}+4\bC_{[A}\bH_{BCD]}\,.
\end{equation}
We also define the supercovariant quantities
\begin{equation}
    \begin{aligned}
        \hat{D}_A\bPhi&=\partial_A\bPhi-\frac{3}{2}\bbPsi_A\Gamma^{(10)}\bla\,,\\
        \hat{\bF}_{AB}&=\bF_{AB} - e^{-\frac{3}{4}\bPhi}\bbPsi_A\Gamma^{(10)}\bPsi_B-\frac{9}{4}e^{-\frac{3}{4}\bPhi}\bbla\Gamma_{[A}\bPsi_{B]}\,,\\
        \hat{\bH}_{ABC}&=\bH_{ABC}-3e^{\frac{1}{2}\bPhi}\bbPsi_{[A}\Gamma_B\Gamma^{(10)}\bPsi_{C]}-\frac{9}{4}e^{\frac{1}{2}\bPhi}\bbla\Gamma_{[AB}\bPsi_{C]}\,,\\
        \hat{\tilde{\bF}}_{ABCD}&=\tilde{\bF}_{ABCD}+6 e^{-\frac{1}{4}\bPhi}\bbPsi_{[A}\Gamma_{BC}\bPsi_{D]}-\frac{3}{2}e^{-\frac{1}{4}\bPhi}\bbla\Gamma^{(10)}\Gamma_{[ABC}\bPsi_{D]}\,,\\
        \hat{\bom}_{A\underline{BC}}&={\bom}_{A\underline{BC}}(\be)-\frac{1}{2}\left(\bbPsi_A\Gamma_{\underline{C}}\bPsi_{\underline{B}}-\bbPsi_{\underline{B}}\Gamma_A\bPsi_{\underline{C}}+\bbPsi_{\underline{C}}\Gamma_{\underline{B}}\bPsi_A\right)\,.
    \end{aligned}
\end{equation}
Similar to 11d supergravity, we define the spin connection solving the 1\textsuperscript{st} order equations associated to the kinetic terms\footnote{Note here that the kinetic term of the dilatino is not canonically normalised, which could however be done by rescaling $\lambda \rightarrow \frac{4}{3\sqrt{2}}\lambda$.}
\begin{equation}
    \mathcal{L} \supset \bR\left(\bom\right)-2\bar{\bPsi}_{M}\Gamma^{MNP}D_{N}\left(\bom\right)\bPsi_{P}-\frac{9}{4}\bar{\bla}\Gamma^{M}D_{M}\left(\bom\right)\bla
\end{equation}
as
\begin{equation}
    \bom_{A\underline{BC}}
    =\hat{\bom}_{A\underline{BC}}
    +\frac{1}{4}\bar{\bPsi}_{D}\Gamma^{DE}{}_{A\underline{BC}}\bPsi_{E}-\frac{9}{16}\bar{\bla}\Gamma_{A\underline{BC}}\bla\,.
    \label{eq:IIA-spin connection}
\end{equation}
The full action of the type IIA supergravity can be found in \cite{Campbell:1984zc}.
The bosonic equations of motion are 
\begin{align}
    &\nabla_A\nabla^A\bPhi=\frac{3}{4\cdot2!}e^{\frac{3}{2}\bPhi}\bF_{AB}\bF^{AB}-\frac{1}{2\cdot 3!}e^{-\bPhi}\bH_{ABC}\bH^{ABC}+\frac{1}{4\cdot4!}e^{\frac{1}{2}\bPhi}\tilde{\bF}_{ABCD}\tilde{\bF}^{ABCD} \notag \\
    &\qquad\quad-\frac{3}{16}e^{\frac{3}{4}\bPhi}\left[\bbPsi_A\Gamma^{(10)}\Gamma^{ABCD}\bPsi_B+2\bbPsi^C\Gamma^{(10)}\bPsi^D-\frac{9}{4}\bbla\Gamma^A\Gamma^{CD}\bPsi_A-\frac{45}{32}\bbla\Gamma^{(10)}\Gamma^{CD}\bla\right]\bF_{CD} \notag\\
    &\qquad\quad+\frac{1}{24}e^{-\frac{1}{2}\bPhi}\left[\bbPsi_A\Gamma^{(10)}\Gamma^{A\dots E}\bPsi_B-6\bbPsi^C\Gamma^{(10)}\Gamma^{D}\bPsi^E+\frac{3}{2}\bbla\Gamma^A\Gamma^{CDE}\bPsi_A\right]\bH_{CDE} \notag\\
    &\qquad\quad+\frac{1}{192}e^{\frac{1}{4}\bPhi}\left[\bbPsi_A\Gamma^{A\dots F}\bPsi_B+12\bbPsi^C\Gamma^{DE}\bPsi^F+\frac{3}{4}\bbla\Gamma^{(10)}\Gamma^A\Gamma^{C\dots F}\bPsi_A-\frac{27}{32}\bbla\Gamma^{C\dots F}\bla\right]\tilde{\bF}_{CDEF}\,, \notag \allowdisplaybreaks\\ 
    &\nabla_A\left(e^{\frac{3}{2}\bPhi}\bF^{AB}\right)=\frac{1}{3!}e^{\frac{1}{2}\bPhi}\bH_{CDE}\tilde{\bF}^{BCDE}+\frac{1}{2}\nabla_A\bigg(e^{\frac{3}{4}\bPhi}\bigg[\bbPsi_C\Gamma^{(10)}\Gamma^{ABCD}\bPsi_D \notag\\
    &\qquad\quad+2\bbPsi^{[A}\Gamma^{\vert (10)\vert}\bPsi^{B]}-\frac{9}{4}\bbla\Gamma^{C}\Gamma^{AB}\bPsi_C-\frac{45}{32}\bbla\Gamma^{(10)}\Gamma^{AB}\bla\bigg]\bigg)\,, \notag \allowdisplaybreaks\\ 
    &\nabla_{A}\left(e^{-\bPhi}\bH^{ABC}\right)=\frac{1}{2}e^{\frac{1}{2}\bPhi}\tilde{\bF}^{ABCD}\bF_{AD}-\frac{1}{2\cdot\left(4!\right)^{2}}\epsilon^{B\cdots K}\tilde{\bF}_{DEFG}\tilde{\bF}_{HIJK} \notag\\
    &\qquad\quad+\frac{1}{2}\nabla_{A}\left(e^{\frac{1}{2}\bPhi}\left[\bar{\bPsi}_{D}\Gamma^{(10)}\Gamma^{A\dots E}\bPsi_{E}-6\bar{\bPsi}^{[A}\Gamma^{(10)}\Gamma^{B}\bPsi^{C]}+\frac{3}{2}\bar{\bla}\Gamma^{D}\Gamma^{ABC}\bPsi_{D}\right]\right) \notag\\
    &\qquad\quad+\frac{1}{4}\bF_{AD}\left(e^{\Phi/4}\left[\bar{\bPsi}_{E}\Gamma^{A\dots F}\bPsi_{F}+12\bar{\bPsi}^{[A}\Gamma^{BC}\bPsi^{D]}+\frac{3}{4}\bar{\bla}\Gamma^{(10)}\Gamma^{E}\Gamma^{ABCD}\bPsi_{E}-\frac{27}{32}\bar{\bla}\Gamma^{ABCD}\bla\right]\right)\,, \notag \allowdisplaybreaks\\
    &\nabla_A\left(e^{\frac{1}{2}\bPhi}\tilde{\bF}^{ABCD}\right)=-\frac{1}{6\cdot 4!}\epsilon^{B\dots K}\bF_{EFGH}\bH_{IJK}-\frac{1}{2}\nabla_A\bigg(e^{\frac{1}{4}\bPhi}\bigg[\bbPsi_E\Gamma^{A\dots F}\bPsi_F \notag\\
    &\qquad\quad+12\bbPsi^{[A}\Gamma^{BC}\bPsi^{D]}+\frac{3}{4}\bbla\Gamma^{(10)}\Gamma^{E}\Gamma^{ABCD}\bPsi_E-\frac{27}{32}\bbla\Gamma^{ABCD}\bla\bigg]\bigg)\,, \notag \allowdisplaybreaks\\
    &\bR_{AB}\left(\bom\right)=\frac{1}{2}\partial_{A}\bPhi\partial_{B}\bPhi+\frac{1}{2}e^{\frac32\bPhi}\left(\bF_{A}{}^{D}\bF_{BD}-\frac{1}{8\cdot2!}\bg_{AB}\bF_{CD}\bF^{CD}\right) \notag\\
    &\qquad+\frac{1}{4}e^{-\frac56\bPhi}\left(\bH_{ACD}\bH_{B}{}^{CD}-\frac{1}{2\cdot3!}\bg_{AB}\bH^{CDE}\bH_{CDE}\right) \notag\\
    &\qquad+\frac{1}{2\cdot3!}e^{\frac23\bPhi}\left(\tilde{\bF}_{ACDE}\tilde{\bF}_{B}{}^{CDE}-\frac{9}{4\cdot4!}\bg_{AB}\tilde{\bF}^{CDEF}\tilde{\bF}_{CDEF}\right) \notag \allowdisplaybreaks\\
    &\qquad+3\bar{\bTh}_{[A|}\Gamma_{B}{}^{CD}\left[D_{|C}\left(\frac{1}{2}\left(\hat{\bom}+\bom\right)\right)+\mathcal{P}_{|C}\right]\bTh_{D]} +2\bar{\bTh}_{[A|}\Gamma_{B}{}^{C}\Gamma^{(10)}\left[D_{|C]}\left(\frac{1}{2}\left(\hat{\bom}+\bom\right)\right)+\mathcal{P}_{C]}\right]\bla \notag\\
    &\qquad +2\bar{\bla}\Gamma_{B}{}^{C}\Gamma^{(10)}\left[D_{[A}\left(\frac{1}{2}\left(\hat{\bom}+\bom\right)\right)+\mathcal{P}_{[A}\right]\bTh_{C]}        
    +2\bar{\bTh}_{[A|}\Gamma_{B}{}^{C}\Gamma^{(10)}\mathcal{Q}\bTh_{|C]} \notag\\   
    &\qquad-\frac{1}{9}\bg_{AB}\left\{ \bar{\bTh}_{C}\Gamma^{CDE} \left[ D_{D}\left(\frac{1}{2}\left(\hat{\bom}+\bom\right)\right) +\mathcal{P}_{D}\right]\bTh_{E} \right. +\bar{\bTh}_{C}\Gamma^{CD}\Gamma^{(10)}\left[ D_{D}\left(\frac{1}{2}\left(\hat{\bom}+\bom\right)\right)+\mathcal{P}_{D} \right] \bla\notag\\ 
    &\qquad\left.+\bar{\bla}\Gamma^{CD}\Gamma^{(10)}\left[ D_{C}\left(\frac{1}{2}\left(\hat{\bom}+\bom\right)\right) + \mathcal{P}_{C} \right]\bTh_{D}+\bar{\bTh}_{C}\Gamma^{CD}\Gamma^{(10)}\mathcal{Q}\bTh_{D}\right\} \notag\\       
    &\qquad-\frac{1}{8}D_{A}\left(\bom\right)\left(\bar{\bPsi}_{D}\Gamma^D{}_{B}\Gamma^{(10)}\bla\right)+\frac{1}{8}D^{D}\left(\bom\right)\left(\bar{\bPsi}_{A}\Gamma_{BD}\Gamma^{(10)}\bla\right)\notag\\ 
    &\qquad+\frac{1}{8}D^{D}\left(\bom\right)\left(\bar{\bPsi}_{G}\Gamma^{G}{}_{ABD}\Gamma^{(10)}\bla\right)+\frac{1}{8}\bg_{AB}D^{D}\left(\bom\right)\left(\bar{\bPsi}_{D}\Gamma^{(10)}\bla\right) \notag \allowdisplaybreaks \\    
    &\qquad+\frac{1}{12}\left(\bg_{BE}\hat{D}_{D}\bPhi-\bg_{DE}\hat{D}_{B}\bPhi\right)\left(\bar{\bPsi}_{A}\Gamma^{E}\bPsi^{D}+\frac{1}{2}\bar{\bPsi}_{G}\gamma^{GEH}{}_A{}^D\bPsi_{H}+\frac{9}{8}\bar{\bla}\Gamma^E{}_{A}{}^{D}\bla\right)\notag\\ 
    &\qquad -\frac34 \partial_A \bPhi \left(\bar{\bPsi}\Gamma^{(10)}\bla\right) + \frac{1}{12}\bg_{AB}\partial_C\bPhi \left(\bar{\Psi}^C \Gamma_D \bPsi^D\right)\notag \allowdisplaybreaks\\ 
    &\qquad+e^{\frac34\bPhi}\bF_{A}{}^{D}\left[-\frac{1}{2}\bar{\bPsi}_{B}\Gamma^{(10)}\bPsi_{D}-\frac{1}{4}\left(\bar{\bPsi}_{G}\Gamma^{GH}{}_{BD}\Gamma^{(10)}\bPsi_{H}\right)-\frac{7}{16}\left(\bar{\bPsi}_{G}\Gamma^{G}{}_{BD}\bla\right)\right.\notag\\
    &\qquad\left.-\frac{9}{8}\bar{\bla}\Gamma_{[B}\bPsi_{D]}+\frac{45}{128}\bar{\bla}\Gamma^{(10)}\Gamma_{BD}\bla\right] + \frac{1}{64}\bg_{AB}e^{\frac34\bPhi}\left[\bar{\bPsi}_{EF}\Gamma^{(10)}\Gamma^{CDEF}\bPsi_{F}\right.\notag\\
    &\qquad+\left.2\bar{\bPsi}^{C}\Gamma^{(10)}\bPsi^{D}-\frac{9}{4}\bar{\bla}\Gamma^{E}\Gamma^{CD}\bPsi_{E}-\frac{45}{32}\bar{\bla}\Gamma^{(10)}\Gamma^{CD}\bla\right]\bF_{CD}\notag \allowdisplaybreaks\\
    &\qquad+\frac{1}{24}\left(\bar{\bTh}_{F}\Gamma_{B}{}^{CDEF}\Gamma^{(10)}\bTh_{A}\right)\mathcal{T}_{CDE}-\frac{1}{16}\left(\bar{\bTh}_{F}\Gamma_{B}{}^{CDEF}\Gamma^{(10)}\bTh_{C}\right)\mathcal{T}_{ADE}\notag\\
    &\qquad+\frac{1}{8}\left[\left(\bar{\bTh}^{C}\Gamma_{B}\Gamma^{(10)}\bTh^{D}\right)+\left(\bar{\bTh}^{D}\Gamma_{B}{}^{C}\bla\right)-\left(\bar{\bla}\Gamma_{B}{}^{C}\bTh^{D}\right)\right]\mathcal{T}_{ACD}\notag\\
    &\qquad-\frac{1}{2}\left[\left(\bar{\bTh}_{(A}\Gamma^{C}\Gamma^{(10)}\bTh^{D}\right)-\frac{1}{2}\left(\bar{\bTh}_{(A}\Gamma^{CD}\bla\right)\right]\mathcal{T}_{B)CD}\notag\\
    &\qquad-\frac{1}{108}\bg_{AB}\left(\bar{\bTh}_{G}\Gamma^{CDEGH}\Gamma^{(10)}\bTh_{H}\right)\mathcal{T}_{CDE}+\frac{1}{18}\bg_{AB}\left[\left(\bar{\bTh}^{E}\Gamma^{C}\Gamma^{(10)}\bTh^{D}\right)-\left(\bar{\bTh}^{E}\Gamma^{CD}\bla\right)\right]\mathcal{T}_{CDE}\notag\\
    &\qquad-\frac{1}{288}\bg_{AB}e^{-\frac12\bPhi}\left[\bar{\bPsi}_{F}\Gamma^{(10)}\Gamma^{CDEFG}\bPsi_{G}-6\bar{\bPsi}^{C}\Gamma^{(10)}\Gamma^{D}\bPsi^{E}+\frac{3}{2}\bar{\bla}\Gamma^{G}\Gamma^{CDE}\bPsi_{G}\right]\bH_{CDE}\notag \allowdisplaybreaks\\
    &\qquad+\frac{1}{32}\left(\bar{\bTh}_{[F}\Gamma_{B}{}^{CDEFG}\bTh_{|G}\right)\mathcal{S}_{ACDE]}+\frac{1}{24}\left(\bar{\bTh}_{F}\Gamma_{B}{}^{CDEF}\Gamma^{(10)}\bla\right)\mathcal{S}_{ACDE}\notag\\
    &\qquad+\frac{1}{96}\left(\bar{\bTh}_{A}\Gamma_{B}{}^{CDEF}\Gamma^{(10)}\bla\right)\mathcal{S}_{CDEF} +\frac{1}{8}\left(\bar{\bTh}^{D}\Gamma_{B}{}^{C}\bTh^{E}\right)\mathcal{S}_{ACDE}+\frac{1}{4}\left(\bar{\bTh}_{(A}\Gamma^{CD}\bTh^{E}\right)\mathcal{S}_{B)CDE}\notag\\
    &\qquad-\frac{1}{432}\bg_{AB}\left(\bar{\bTh}_{G}\Gamma^{CDEFGH}\bTh_{H}\right)\mathcal{S}_{CDEF}-\frac{1}{216}\bg_{AB}\left(\bar{\bTh}_{G}\Gamma^{CDEFG}\Gamma^{(10)}\bla\right)\mathcal{S}_{CDEF}\notag\\
    &\qquad-\frac{1}{36}\bg_{AB}\left(\bar{\bTh}^{E}\Gamma^{CD}\bTh^{F}\right)\mathcal{S}_{CDEF}-\frac{1}{2304}\bg_{AB}e^{\Phi/4}\left[\bar{\bPsi}_{G}\Gamma^{CDEFGH}\bPsi_{H}+12\bar{\bPsi}^{C}\Gamma^{DE}\bPsi^{F}\right.\notag\\
    &\qquad+\left.\frac{3}{4}\bar{\bla}\Gamma^{(10)}\Gamma^{G}\Gamma^{CDEF}\bPsi_{G}-\frac{27}{32}\bar{\bla}\Gamma^{CDEF}\bla\right]\tilde{\bF}_{CDEF}\notag \allowdisplaybreaks\\
    &\qquad-\frac{1}{16}\left[\left(\bar{\bTh}_{C}\Gamma^{CDE}\Gamma^{GH}\bTh_{E}\right)+\left(\bar{\bla}\Gamma^{DE}\Gamma^{(10)}\Gamma^{GH}\bTh_{E}\right)+\left(\bar{\bTh}_{C}\Gamma^{CD}\Gamma^{(10)}\Gamma^{GH}\bla\right)\right] \notag\\
    &\qquad \qquad \cdot \left[\left(\bar{\bTh}_{A}\Gamma_{B}{}^{F}{}_{DGH}\bTh_{F}\right)-\left(\bar{\bTh}_{A}\Gamma_{BDGH}\Gamma^{(10)}\bla\right)\right]-\frac{1}{8}\left(\bar{\bTh}_{A}\Gamma_{B}{}^{F}{}_{DG}\Gamma^{(10)}\bTh_{F}\right)\notag\\
    &\qquad\quad \cdot \left[\left(\bar{\bTh}_{C}\Gamma^{CDE}\Gamma^{G}\Gamma^{(10)}\bTh_{E}\right)+\left(\bar{\bla}\Gamma^{CD}\Gamma^{G}\bTh_{C}\right)-\left(\bar{\bTh}_{C}\Gamma^{CD}\Gamma^{G}\bla\right)-\frac{1}{2}\left(\bar{\bTh}_{C}\Gamma^{CE}\Gamma^{(10)}\Gamma^{DG}\bTh_{E}\right)\right]\notag\\
    &\qquad+\frac{1}{144}\bg_{AB}\left[\left(\bar{\bTh}_{C}\Gamma^{CDE}\Gamma^{HI}\bTh_{E}\right)+\left(\bar{\bla}\Gamma^{(10)}\Gamma^{CD}\Gamma^{HI}\bTh_{D}\right)+\left(\bar{\bTh}_{C}\Gamma^{CD}\Gamma^{(10)}\Gamma^{HI}\bla\right)\right] \notag\\
    &\qquad\qquad \cdot \left[\left(\bar{\bTh}_{F}\Gamma^{FG}{}_{DHI}\bTh_{G}\right)-2\left(\bar{\bTh}_{F}\Gamma^{F}{}_{DHI}\Gamma^{(10)}\bla\right)\right]  +\frac{1}{72}\bg_{AB}\left(\bar{\bTh}_{F}\Gamma^{FG}{}_{DH}\Gamma^{(10)}\bTh_{G}\right) \notag\\
    &\qquad\quad\cdot \left[\left(\bar{\bTh}_{C}\Gamma^{CDE}\Gamma^{H}\Gamma^{(10)}\bTh_{E}\right)+\left(\bar{\bla}\Gamma^{CD}\Gamma^{H}\bTh_{C}\right)-\left(\bar{\bTh}_{C}\Gamma^{CD}\Gamma^{H}\bla\right)-\frac{1}{2}\left(\bar{\bTh}_{C}\Gamma^{CE}\Gamma^{(10)}\Gamma^{DH}\bTh_{E}\right)\right]\notag\\
    &\qquad+\frac{1}{8}\left[\left(\bar{\bPsi}_{E}\Gamma_{BD}\Gamma^{(10)}\bla\right)+\left(\bar{\bPsi}_{G}\Gamma^{G}{}_{EBD}\Gamma^{(10)}\bla\right)\right]\left(\bar{\bPsi}_{A}\Gamma^{E}\bPsi^{D}+\frac{1}{2}\bar{\bPsi}_{G}\Gamma^{GEH}{}_{A}{}^{D}\bPsi_{H}+\frac{9}{8}\bar{\bla}\Gamma^{E}{}_{A}{}^{D}\bla\right)
    \label{eq:IIA-eom-bosonic}
\end{align}
where we have defined \begin{equation}\hspace{-0.7cm}
    \begin{aligned}
        &\mathcal{P}_{A}\equiv\frac{1}{4}\left[-\frac{1}{6}\bg_{AB}\hat{D}_{C}\bPhi-\frac{1}{8}\left(\bar{\bPsi}_{A}\Gamma_{BC}\Gamma^{(10)}\bla\right)-\frac{1}{16}\left(\bar{\bPsi}_{D}\Gamma^D{}_{ABC}\Gamma^{(10)}\bla\right)-\frac{1}{256}\bar{\bla}\Gamma_{ABC}\bla\right]\Gamma^{BC}\\
        &\ -\frac{1}{4}\left[e^{\frac34\bPhi}\hat{\bF}_{BA}+2\bar{\bla}\Gamma_{B}\bPsi_{A}-\frac{1}{4}\bar{\bPsi}_{C}\Gamma^{CD}{}_{AB}\Gamma^{(10)}\bPsi_{D}-\frac{7}{16}\bar{\bPsi}_{C}\Gamma^{C}{}_{AB}\bla+\frac{7}{32}\bar{\bla}\Gamma^{(10)}\Gamma_{AB}\bla\right]\Gamma^{B}\Gamma^{(10)}+\frac{1}{24}\partial_{A}\bPhi\,,\\
        &\mathcal{Q}\equiv\frac{1}{8}\left[e^{\frac34\bPhi}\hat{\bF}_{AB}-\frac{1}{4}\bar{\bPsi}_{C}\Gamma^{CD}{}_{AB}\Gamma^{(10)}\bPsi_{D}-\frac{7}{16}\bar{\bPsi}_{C}\Gamma^{C}{}_{AB}\bla+\frac{31}{64}\bar{\bla}\Gamma^{(10)}\Gamma_{AB}\bla\right]\Gamma^{AB}+\frac{1}{3}\hat{D}_{A}\bPhi\Gamma^{A}\Gamma^{(10)}\,,\\
        &\bTh_{A}\equiv\bPsi_{A}-\frac{1}{8}\Gamma_{A}\Gamma^{(10)}\bla,\quad\bar{\bTh}_{A}\equiv\bar{\bPsi}_{A}-\frac{1}{8}\bar{\bla}\Gamma^{(10)}\Gamma_{A}\,,\\
        &\mathcal{S}_{ABCD}\equiv e^{\frac14\bPhi}\tilde{\bF}_{ABCD}+e^{\frac14 \bPhi}\hat{\tilde{\bF}}_{ABCD}+\frac{3}{32}\bar{\bla}\Gamma_{ABCD}\bla\,,\\
        &\mathcal{T}_{ABC}\equiv e^{-\bPhi/2}\bH_{ABC}+e^{-\bPhi/2}\hat{\bH}_{ABC}-\frac{21}{64}\bar{\bla}\Gamma^{(10)}\Gamma_{ABC}\bla\,.
    \end{aligned}
\end{equation}
Instead of varying the second-order action in \cite{Campbell:1984zc}, we obtained the Einstein equation by dimension reduction from the 11d one, \eqref{eq:11d-eom}, following \cite{Campbell:1984zc}.  $\mathcal{P}_{A}$ and $\mathcal{Q}$ descend from the action of the 11d covariant derivative with spin connection \eqref{eq:11d-spin-connection} on spinors. $\bTh_A$ descends from the 11d gravitino $\bPsi_A$. $\mathcal{S}_{ABCD}$ and $\mathcal{T}_{ABC}$ are the reduction of $\tilde{\bF}_{ABCD} + \hat{\tilde{\bF}}_{ABCD} $ in 11d. 

The fermionic equations of motion are similarly determined by dimensional reduction, which is easier than varying the action to get the equation of motion, which would generally require the use of  Fierz identities \eqref{eq:Fierz-identity}. The dilatino equation of motion is
\begin{equation}
    \begin{aligned}
        0=&\Gamma^{A}D_{A}\left(\hat{\bom}\right)\bla+\frac{5}{32}e^{\frac34\bPhi}\hat{\bF}_{AB}\Gamma^{AB}\Gamma^{(10)}\bla + \frac{1}{256}\left[e^{\frac14 \bPhi}\hat{\tilde{\bF}}_{ABCD}+\frac{3}{32}\bar{\bla}\Gamma_{ABCD}\bla\right]\Gamma^{ABCD}\bla\\
        &+\frac{1}{3}\Gamma^{A}\hat{D}_{B}\bPhi\Gamma^{B}\Gamma^{(10)}\bPsi_{A}+\frac{1}{8}\Gamma^{A}e^{\frac34 \bPhi}\hat{\bF}_{BC}\Gamma^{BC}\bPsi_{A}\\
        &-\frac{1}{36}\Gamma^{A}\left[e^{-\frac12 \bPhi}\hat{\bH}_{BCD}-\frac{21}{64}\bar{\bla}\Gamma^{(10)}\Gamma_{BCD}\bla\right]\Gamma^{BCD}\bPsi_{A}\\
        &-\frac{1}{288}\Gamma^{A}\left[e^{\frac14 \bPhi}\hat{\tilde{\bF}}_{BCDE}+\frac{3}{32}\bar{\bla}\Gamma_{BCDE}\bla\right]\Gamma^{BCDE}\Gamma^{(10)}\bPsi_{A}\\
        &+\left[\left(\bar{\bla}\Gamma^{B}\bPsi_{A}\right)\Gamma^{A}+\frac{1}{8}\left(\bar{\bla}\Gamma_{CD}\Gamma^{(10)}\bla\right)\left[\frac{17}{64}\Gamma^{BCD}-\frac{15}{32}g^{BC}\Gamma^{D}\right]\right]\bPsi_{B}\\
        &+\left[ \frac{1}{16}\left(\bar{\bPsi}_{A}\Gamma^{(10)}\bla\right)\Gamma^{A}-\frac{5}{8}\left(\bar{\bla}\Gamma_{B}\bPsi_{A}\right)\Gamma^{A}\Gamma^{B}\Gamma^{(10)}-\frac{1}{32}\left(\bar{\bPsi}_{A}\Gamma_{BC}\Gamma^{(10)}\bla\right)\Gamma^{A}\Gamma^{BC}\right] \bla\\
        &-\frac{1}{512}\left[\frac{75}{4}\left(\bar{\bla}\Gamma_{AB}\Gamma^{(10)}\bla\right)\Gamma^{AB}\Gamma^{(10)}+\left(\bar{\bla}\Gamma_{ABC}\bla\right)\Gamma^{ABC}\right]\bla\,,
    \end{aligned}
    \label{eq:IIA-eom-dilatino}
\end{equation}
and finally, the gravitino equation of motion is given by
\begin{equation}\hspace{-0.7cm}
    \begin{aligned}
        0=&\Gamma^{CAB}D_{A}\left(\hat{\bom}\right)\bPsi_{B}-\frac{1}{64}\Gamma^{CAB}e^{\frac{3}{4}\bPhi}\hat{\bF}_{GH}\left(\Gamma_{A}{}^{GH}-14\delta_{A}^{G}\Gamma^{H}\right)\Gamma^{(10)}\bPsi_{B}\\
        &+\frac{1}{96}\Gamma^{CAB}\left( \Gamma_{A}{}^{EFG}-9\delta_{A}^{E}\Gamma^{FG}\right) \Gamma^{(10)}\bPsi_{B}\left[e^{-\frac12 \bPhi}\hat{\bH}_{EFG}-\frac{21}{64}\bar{\bla}\Gamma^{(10)}\Gamma_{EFG}\bla\right]\\
        &+\frac{1}{256}\Gamma^{CAB}\left( \Gamma_{A}{}^{EFGH}-\frac{20}{3}\delta_{A}^{E}\Gamma^{FGH}\right) \bPsi_{B}\left[e^{\frac14 \bPhi}\hat{\tilde{\bF}}_{EFGH}+\frac{3}{32}\bar{\bla}\Gamma_{EFGH}\bla\right]\\
        &-\Gamma^{CAB}\frac{1}{192}\hat{D}_{g}\bPhi\Gamma_B\left(\Gamma_{A}{}^{G}-17\delta_{A}^{G}\right)\Gamma^{(10)}\bla-\frac{1}{512}\Gamma^{CAB}e^{\frac34 \bPhi}\hat{\bF}_{GH}\left(\Gamma_{AB}{}^{GH}+32\delta_{A}^{G}\Gamma_{B}{}^{H}-254\delta_{A}^{G}\delta_{B}^{H}\right)\bla\\
        &+\frac{1}{2304}\Gamma^{CAB}\left(\Gamma_{BA}{}^{EFG}+24\delta_{A}^{E}\Gamma_{B}{}^{FG}-102\delta_{A}^{E}\delta_{B}^{F}\Gamma^{G}\right)\bla\left[e^{-\frac12 \bPhi}\hat{\bH}_{EFG}-\frac{21}{64}\bar{\bla}\Gamma^{(10)}\Gamma_{EFG}\bla\right]\\
        &+\frac{1}{18432}\Gamma^{CAB}\left(\Gamma_{BA}{}^{EFGH}+64\delta_{B}^{E}\Gamma_{A}{}^{FGH}+372\delta_{A}^{E}\delta_{B}^{F}\Gamma^{Gh}\right)\Gamma^{(10)}\bla\left[e^{\frac14 \bPhi}\hat{\tilde{\bF}}_{EFGH}+\frac{3}{32}\bar{\bla}\Gamma_{EFGH}\bla\right]\\
        &+\Gamma^{CAB}\left\{ \frac{1}{16}\left(\bar{\bPsi}_{A}\Gamma^{(10)}\bla\right)-\frac{1}{4}\left[\frac{1}{8}\left(\bar{\bPsi}_{A}\Gamma_{GH}\Gamma^{(10)}\bla\right)+\frac{1}{128}\bar{\bla}\Gamma_{AGH}\bla\right]\Gamma^{GH}\right.\\
        &\left.-\frac{1}{2}\left[\bar{\bla}\Gamma_{G}\Psi_{A}+\frac{1}{128}\bar{\bla}\Gamma_{GA}\Gamma^{(10)}\bla\right]\Gamma^{G}\Gamma^{(10)}-\frac{17}{4096}\Gamma_{a}\Gamma^{(10)}\left(\bar{\bla}\Gamma_{GH}\Gamma^{(10)}\bla\right)\Gamma^{GH}\right\} \bPsi_{B}\\
        &+\Gamma^{CAB}\left[\frac{1}{8}\left(\bar{\bPsi}_{A}\Gamma^{G}{}_{B}\Gamma^{(10)}\bla\right)+\frac{1}{128}\bar{\bla}\Gamma_{A}{}^G{}_{B}\bla+\frac{1}{8}\left(\bar{\bla}\Gamma^{G}\bPsi_{A}-\frac{1}{8}\bar{\bla}\Gamma^G{}_{A}\Gamma^{(10)}\bla\right)\Gamma_{B}\Gamma^{(10)}\right]\bPsi_{G}\\
        &+\frac{1}{64}\Gamma^{CAB}\left[\left(\bar{\bla}\Gamma_{G}\bPsi_{A}\right)\left(5\Gamma_{B}{}^{G}-63\delta_{B}^{G}\right)-\left(\bar{\bla}\Gamma_{GH}\Gamma^{(10)}\bla\right)\left(\frac{17}{512}\Gamma_{AB}{}^{GH}+\frac{3}{32}\Gamma_{B}{}^{G}\delta_{A}^{H}+\frac{177}{256}\delta_{A}^{G}\delta_{B}^{H}\right)\right]\bla\,.
    \end{aligned}
    \label{eq:IIA-eom-gravitino}
\end{equation}

 These equations of motion have to be supplemented by the Bianchi identities
\begin{align}
\partial_{[A} \bF_{BC]}=0\,,\qquad \partial_{[A} \bH_{BCD]}=0\,,\qquad \partial_{[A} \bF_{BCDE]}=0\,.
\end{align}
The supersymmetry transformations with parameter $\bep$ are given by \cite{Campbell:1984zc}
\begin{align}
    &\delta\be_A{}^{\underline{B}}=\bar{\bep}\Gamma^{\underline{B}}\bPsi_A \quad \delta\bPhi=-\frac{3}{2}\bbla\Gamma^{(10)}\bep\,, \notag \allowdisplaybreaks\\
    &\delta\bC_A=-e^{-\frac{3}{4}\bPhi}\bbPsi_A\Gamma^{(10)}\bep-\frac{9}{8}e^{-\frac{3}{4}\bPhi}\bbla\Gamma_A\bep\,,  \notag \allowdisplaybreaks\\
    &\delta\bB_{AB}=2e^{\frac{1}{2}\bPhi}\bbPsi_{[A}\Gamma_{B]}\Gamma^{(10)}\bep+\frac{3}{4}e^{\frac{1}{2}\bPhi}\bbla\Gamma_{AB}\bep\,,\notag \allowdisplaybreaks\\
    &\delta \bC_{ABC}=3 e^{-\frac{1}{4}\bPhi}\bbPsi_{[A}\Gamma_{BC]}\bep-\frac{3}{8}e^{-\frac{1}{4}\bPhi}\bbla\Gamma^{(10)}\Gamma_{ABC}\bep +\frac{3}{2}\bC_{[A}\delta \bB_{BC]} \,,  \notag \allowdisplaybreaks\\
    &\delta\bla=-\frac{1}{3}\hat{D}_A\bPhi\Gamma^A\Gamma^{(10)}\bep-\frac{1}{4\cdot 2!}e^{\frac{3}{4}\bPhi}\hat{\bF}_{AB}\Gamma^{AB}\bep+\frac{1}{6\cdot 3!}e^{-\frac{1}{2}\bPhi}\hat{\bH}_{ABC}\Gamma^{ABC}\bep\notag \\
    &\qquad+\frac{1}{12\cdot 4!}e^{\frac{1}{4}\bPhi}\hat{\tilde{\bF}}_{ABCD}\Gamma^{ABCD}\Gamma^{(10)}\bep-\frac{9}{256}\bigg[\left(\bbla\bla\right)\Gamma^{(10)}-\frac{1}{2!}\left(\bbla\Gamma^{(10)}\Gamma_{AB}\bla\right)\Gamma^{AB}\notag\\
    &\qquad-\frac{1}{3!}\left(\bbla\Gamma^{(10)}\Gamma_{ABC}\bla\right)\Gamma^{ABC}+\frac{1}{3!}\left(\bbla\Gamma_{ABC}\bla\right)\Gamma^{ABC}\Gamma^{(10)}\bigg]\bep\,, \notag \allowdisplaybreaks\\ 
    &\delta\bPsi_A=D_A(\hat{\bom})\bep-\frac{1}{32\cdot 2!}e^{\frac{3}{4}\bPhi}\hat{\bF}_{BC}\left(\Gamma_A{}^{BC}-14\delta_A^B\Gamma^C\right)\Gamma^{(10)}\bep\notag\\
    &\qquad+\frac{1}{16\cdot 3!}e^{-\frac{1}{2}\bPhi}\hat{\bH}_{BCD}\left(\Gamma_A{}^{BCD}-9\delta_A^B\Gamma^{CD}\right)\Gamma^{(10)}\bep\notag\\
    &\qquad+\frac{3}{32\cdot 4!}e^{\frac{1}{4}\bPhi}\hat{\tilde{\bF}}_{BCDE}\left(\Gamma_A{}^{BCDE}-\frac{20}{3}\delta_A^B\Gamma^{CDE}\right)\bep\notag \allowdisplaybreaks\\
    & \qquad+\frac{1}{2048}\bigg[72\left(\bbla\bla\right)\Gamma_A+\left(\bbla\Gamma^{(10)}\Gamma_{BC}\bla\right)\left(18\Gamma_A{}^{BC}-74\delta_A^B\Gamma^{C}\right)\Gamma^{(10)}\notag\\
    &\qquad+\left(\bbla\Gamma_{BCD}\bla\right)\left(11\Gamma_A{}^{BCD}-47\delta_A^B\Gamma^{CD}\right)-\left(\bbla\Gamma^{(10)}\Gamma_{BCD}\bla\right)\left(20\Gamma_A{}^{BCD}-92\delta_A^B\Gamma^{CD}\right)\Gamma^{(10)}\phantom{\bigg[}\notag\\
    &\qquad+\frac{1}{12}\left(\bbla\Gamma_{BCDE}\bla\right)\left(41\Gamma_A{}^{BCDE}-192\delta_A^B\Gamma^{CDE}\right)\bigg]\bep\notag\\
    &\qquad-\frac{1}{32}\bigg[-2\left(\bbla\Gamma^{(10)}\bPsi_A\right)+16\left(\bla\Gamma_B\bPsi_A\right)\Gamma^B\Gamma^{(10)}+\left(\bbla\Gamma^{(10)}\Gamma_{BC}\bPsi_A\right)\Gamma^{BC}\bigg]\bep\notag\\
    &\qquad-\frac{1}{8}\bigg[8 \left(\bbPsi_A\Gamma^{(10)}\bep\right)+\left(\bbPsi_A\Gamma_B\bep\right)\Gamma^B\Gamma^{(10)}\bigg]\bla\notag\\
    &\qquad-\frac{1}{32}\bigg[2\left(\bbla\Gamma^{(10)}\bep\right)-16\left(\bbla\Gamma_{B}\bep\right)\Gamma^B\Gamma^{(10)}-\left(\bbla\Gamma^{(10)}\Gamma_{BC}\bep\right)\Gamma^{BC}\bigg]\bPsi_A\,.
    \label{eq:IIA-susy}
\end{align}
We furthermore require the supersymmetry transformation of the supercovariant field strength in the $\bPhi = \bla = 0$ background 
\begin{equation}
    \begin{aligned}
        \delta\hat{\bF}_{ABCD}=&-12\bar{\bep}\Gamma_{[AB}\left[D_C(\hat{\bom})-\frac14 \Gamma^{E}\Gamma^{(10)}\hat{\bF}_{C|E|}\right.\\
        &\qquad-\frac{1}{72}\left(\Gamma_C{}^{EFG}-6\delta^{E}_C \Gamma^{FG}\right)\Gamma^{(10)}\hat{\bH}_{|EFG|}\\
        &\qquad\left.+\frac{1}{288}\left(\Gamma^{EFGH}{}_C - 8\delta^{E}_C\Gamma^{FGH}\right)\hat{\bF}_{|EFGH|}\right]\bPsi_{D]}\,,\\
    \end{aligned}
    \label{eq:IIA-susy-supercov-field}
\end{equation}
\begin{equation}
    \begin{aligned}
        \delta\hat{\bH}_{ABC}=&-6\bar{\bep}\Gamma^{(10)}\Gamma_{[A}\left[D_B(\hat{\bom})-\frac14 \Gamma^{D}\Gamma^{(10)}\hat{\bF}_{B|D|}\right.\\
        &\qquad-\frac{1}{72}\left(\Gamma^{DEF}{}_B-6\delta^{D}_B\Gamma^{EF}\right)\Gamma^{(10)}\hat{\bH}_{|DEF|}\\
        &\qquad\left.+\frac{1}{288}\left(\Gamma^{DEFG}{}_B-8\delta^{D}_B\Gamma^{EFG}\right)\hat{\bF}_{|DEFG|}\right]\bPsi_{C]}\\
        &-3\bar{\bep}\Gamma_{[AB|}\left[\frac18\hat{\bF}_{DE}\Gamma^{DE}-\frac{1}{36}\Gamma^{DEF}\hat{\bH}_{DEF}\right.\\
        &\qquad\left.+\frac{1}{288}\Gamma^{DEFG}\Gamma^{(10)}\hat{\bF}_{DEFG}\right]\bPsi_{|C]}\,.
    \end{aligned}
     \label{eq:IIA-susy-supercov-field2}
\end{equation}
 When comparing to \cite{Campbell:1984zc}, some fields have to be rescaled and renamed as
\begin{equation}\begin{split}
\left(\sigma\right)_{\text{\cite{Campbell:1984zc}}}:=\frac{2}{3}\bPhi\,,\qquad\left(\lambda\right)_{\text{\cite{Campbell:1984zc}}}:=-\frac{3\sqrt{2}}{4}\bla\,,\qquad\left(\eta\right)_{\text{\cite{Campbell:1984zc}}}:=\bep\,,\qquad\left(G_{AB}\right)_{\text{\cite{Campbell:1984zc}}}:=-\frac{1}{2}\tilde{\bF}_{AB}\,,\\\left(F_{ABC}\right)_{\text{\cite{Campbell:1984zc}}}:=-\frac{1}{2}\tilde{\bH}_{ABC}\,,\qquad
\left(F'_{ABCD}\right)_{\text{\cite{Campbell:1984zc}}}:=\frac{1}{2}\tilde{\bF}_{ABCD}\,,\qquad \left(F_{ABCD}\right)_{\text{\cite{Campbell:1984zc}}}:=\frac{1}{2}\bF_{ABCD}\,.\\
\end{split}\end{equation} 
Additionally, one should use the opposite signature by replacing, \cite{West_2012}
\begin{equation}
    \eta_{\underline{AB}}\rightarrow -\eta_{\underline{AB}}\,,\qquad \Gamma^{\underline{A}} \rightarrow i\Gamma^{\underline{A}}\,,\qquad \be_A{}^{\underline{B}}\rightarrow\be_A{}^{\underline{B}}\,,\qquad \Gamma^{11}\rightarrow -i\Gamma^{(10)}\,.
    \label{eq:signature-change}
\end{equation}

\subsection{IIB supergravity}\label{conventionsIIB}

The field content of 10d type IIB supergravity comprises the graviton $\bg_{AB}$, a left-handed Weyl gravitino $\bPsi_A$, form fields $\bC$, $\bB_{AB}$, $\bC_{AB}$, $\bC_{ABCD}$ with their respective curvatures $\bF_{(1)}=\dd\bC$, $\bH_{(3)}=\dd\bB_{(2)}$, $\bF_{{(3)}}=\dd\bB_{(2)}$, $\bF_{(4)}=\dd\bC_{(4)}$, a right-handed Weyl dilatino $\bla$ and a dilaton $\bPhi$\cite{Schwarz:1983wa, Schwarz:1983qr, Howe:1983sra, DallAgata:1998ahf}. The 0-form gauge field $\bC$ and the dilaton $\bPhi$ are often combined to form a complex scalar
\begin{equation}
    \btau=\bC+ie^{-\bPhi}\,,
\end{equation}
which transforms as a modular parameter under the $SL(2,\mathbb{R})$ global symmetry of type IIB supergravity. Alternatively, we can introduce a symmetric representation of the form 
\begin{equation}
    \bM=e^\bPhi\left(\begin{array}{cc}
    1 & -\bC\\
    -\bC & \abs{\btau}^2
    \end{array}\right)\,,
\end{equation}
The 2-form field $\bB_{AB}$ and $\bC_{AB}$ transform as a doublet under the global $SL(2,\mathbb{R})$. A generic $SL(2,\mathbb{R})$ transformation $\Lambda$ acts on these quantities as
\begin{equation}
    \Lambda = \left(\begin{array}{cc}
    a & b\\
    c & d
    \end{array}\right):\quad \btau\to\frac{a \btau +b}{c\btau + d}\,,\qquad \bM\to (\Lambda^{-1})^{\dagger}\bM\Lambda^{-1}\,,\quad \left(\begin{array}{cc}
    \bC_{AB} \\
    \bB_{AB} 
    \end{array}\right)\to \Lambda \left(\begin{array}{cc}
    \bC_{AB} \\
    \bB_{AB} 
    \end{array}\right)\,.
\end{equation}

The most complete discussion of the equations of motion and supersymmetry transformations we could find,  \cite{Howe:1983sra}, parametrises the scalar manifold instead as an $SU(1,1)/U(1)$ coset space spanned by coordinates 
\begin{equation}
    \bV=\left(\begin{array}{cc}
    \bu & \bv\\
    \bar{\bv} & \bar{\bu}
    \end{array}\right)
\end{equation}
with $\bu\bar{\bu} - \bv\bar{\bv} = 1$ and Maurer-Cartan differential form 
\begin{equation}
    \bV^{-1}\dd\bV = 
    \left(\begin{array}{cc}
    2i\bQ & \bP\\
    \bar{\bP} & -2i\bQ
    \end{array}\right).
\end{equation}
We can establish the isomorphism between $SU(1,1)$ and $SL(2,\mathbb{R})$ explicitly by identifying
\begin{equation}\begin{split}
    U = \frac{1}{\sqrt{2}}\left(\begin{array}{cc}
    1 & i\\
    1 & -i
    \end{array}\right):\qquad
    \Lambda_{SU(1,1)}&=U\Lambda_{SL(2,\mathbb{R})}U^{-1}\,,\quad \bV= U\bM U^{-1}\,,\\
    \left(\begin{array}{cc}
    \bA_{AB} \\
    \bar{\bA}_{AB} 
    \end{array}\right)&= U \left(\begin{array}{cc}
    \bC_{AB} \\
    \bB_{AB} 
    \end{array}\right)\,,\quad \bA_{AB}=\frac{1}{\sqrt{2}}(\bC_{AB}+i \bB_{AB})\,.
\end{split}
\end{equation}
Explicitly, we can express $\bu,\bv,\bP$ and $\bQ$ through standard fields $\bC$ and $\bPhi$ as \cite{Becker:2006dvp}
\begin{equation}\begin{split}
    \bu&=\frac{1}{2}e^{\bPhi}(1+\abs{\btau}^2)\,,\qquad \bv=\frac{1}{2}e^{\bPhi}\left(1-\abs{\btau}^2-2i \bC\right)\\
    \bP_A&=\frac{i}{2}\left((i+\bC)^2e^{2\bPhi}-1\right)\partial_A\bC+(1-i\bC)\partial_A\bPhi\,,\\
    \bQ_A&=\frac{1}{4}\left(\left(1-(1+\bC^2)e^{2\bPhi}\right)\partial_A \bC+2 \bC \partial_A\bPhi\right)\,,
\end{split}\end{equation}
and show that the curvature associated to this $SU(1,1)$-connection vanishes. We note that this parametrisation constitutes a convenient $U(1)$-gauge choice with real-valued $\textbf{u}$. It allows us to identify the canonical fields $\bC$ and $\bPhi$. The discussion in \cite{Howe:1983sra} allows for more general parametrisations, but we will not require them here. 

We define the field strength $\bG_{ABC}=3\partial_{[A}\bA_{BC]}$ satisfying the corresponding Bianchi identity. An important detail to note is that this 3-form and its conjugate transforms as an $SU(1,1)$ doublet and has a dual representation that transforms with $\Lambda^{-1}$ given by 
\begin{equation}
    \left(\begin{array}{cc}
    \bF_{ABC} \\
    \bar{\bF}_{ABC} 
    \end{array}\right)=\bV \left(\begin{array}{cc}
    \bG_{ABC} \\
    \bar{\bG}_{ABC} 
    \end{array}\right)\,.
\end{equation}
It is this representation that will appear in the equations of motion.

The real self-dual 5-form field strength is a $SU(1,1)$-singlet and defined as 
\begin{equation}
    \tilde{\bF}_{ABCDE} = \bF_{ABCDE} + 20i \left(\bA_{[AB}  \bar{\bG}_{CDE]} - \bar{\bA}_{[AB}\bG_{CDE]} \right),
\end{equation}
where $\bF_{ABCDE} = 5\partial_{[A}\bC_{BCDE]}$. 

The formulation of an action for type IIB supergravity is complicated by the self-duality of the 5-form field strength, which is complicated to encode. One may instead write down actions that are invariant under the  $SL(2,\mathbb{R})$ global symmetry \cite{Bergshoeff:1995as, Bergshoeff:1995sq, Polchinski:1998rr, DallAgata:1998ahf} and impose the self-duality by hand. Proposals to encode the self-duality in the action itself can be found in \cite{Sen_2020, Mkrtchyan:2022xrm, Hull:2023dgp}.

The full equations of motion for the fermions are given by\footnote{There are different results given in the main section and summary of \cite{Howe:1983sra}. Here, we use the results from Sections 7 and 9 there. We rewrite the 16 component gamma matrices into 32-component ones, change the intertwiner as $C \rightarrow \tilde{C} = -C\Gamma^{(10)}$, and change the signature according to (\ref{eq:signature-change}).} \cite{Howe:1983sra} 
\begin{equation}
    \begin{aligned}
        &\Gamma^{A}\hat{D}_{A}\bla=-\frac{1}{32}\left(\bar{\bla}\Gamma_{ABC}\bla\right)\Gamma^{ABC}\bla^{c}-\frac{i}{320}\Gamma^{ABCDE}\left(\frac{1}{3}\hat{\tilde{\bF}}_{ABCDE}+\frac{i}{4}\bar{\bla}\Gamma_{ABCDE}\bla^{c}\right)\bla\,,\\
        & \Gamma^{AB}\hat{D}_{A}\bPsi_{B} = -\frac{1}{24}\hat{\bar{\bF}}_{BCD}\Gamma^{BCD}\bla
        +i\hat{\bP}_{B}\Gamma^{B}\bla^c\,,
        \end{aligned}
\end{equation}
and the bosonic equations of motion take the form
\begin{equation}
    \begin{aligned}
        &\hat{\nabla}^{A}\hat{\bar{\bP}}_{A}=\frac{1}{6}\hat{\bar{\bF}}_{ABC}\hat{\bar{\bF}}^{ABC} +\frac{9}{4}\overline{\bla^c}\Gamma_{A}\bla\hat{\bar{\bP}}^{A} + \frac{1}{16}\left(\overline{\bla^c}\Gamma_{ABC}\bla^c\right)\hat{\bF}^{ABC}\,,\\
        &\hat{\nabla}^{C}\hat{\bar{\bF}}_{ABC}=-\hat{\bF}_{ABC}\hat{\bar{\bP}}^{C} - \frac{i}{6}\hat{\tilde{\bF}}_{ABCDE}\hat{\bar{\bF}}^{CDE} -\left(\overline{\bla^c}\Gamma_{ABC}\bla^c\right)\hat{\bP}^{C} -\overline{\bla^c}\hat{D}_{[B}\bPsi_{C]} \\
        &\qquad\qquad +\frac{5}{4}\hat{\bar{\bF}}_{ABC}\left(\overline{\bla^c}\Gamma^{C}\bla\right)-\frac{1}{4}\hat{\bar{\bF}}_{[A}{}^{CD}\left(\bar{\bla}\Gamma_{B]CD}\bla^c\right) - \frac{1}{24}\left(\overline{\bla^c}\Gamma_{ABCDE}\bla\right)\hat{\bar{\bF}}^{CDE}\,, \\
        &\hat{\bR}_{AB}=2\hat{\bar{\bP}}_{(A}\hat{\bP}_{B)} + \hat{\bar{\bF}}_{(A}{}^ {CD}\hat{\bF}_{B)CD}-\frac{1}{12}g_{AB}\hat{\bar{\bF}}_{CDE}\hat{\bF}^{CDE}+\frac{1}{96}\hat{\tilde{\bF}}_{A}{}^{CDEF}\hat{\tilde{\bF}}_{BCDEF}\\
        &\qquad\quad -\frac{5}{32}\left(\bar{\bla}\Gamma_{(A}{}^{CD}\bla\right)\left(\overline{\bla^c}\Gamma_{B)CD}\bla^c\right)-\frac{1}{64}g_{AB}\left(\bar{\bla}\Gamma_{CDE}\bla\right)\left(\overline{\bla^c}\Gamma^{CDE}\bla^c\right)\,,\\
        &\hat{\tilde{\bF}}_{ABCDE} = *\hat{\tilde{\bF}}_{ABCDE} - i \overline{\bla^c}\Gamma_{ABCDE}\bla\,.
    \end{aligned}
\end{equation}
Here, we have defined the supercovariant quantities as 
\begin{align}
        &\hat{\bom}_{A\underline{BC}} = \bom(\be)_{A\underline{BC}} - \left(\overline{\bPsi^c}_{[A|}\Gamma_{\underline{C}}\bPsi_{|\underline{B}]} - \overline{\bPsi^c}_{[\underline{B}|}\Gamma_A\bPsi_{|\underline{C}]} + \overline{\bPsi^c}_{[\underline{C}|}\Gamma_{\underline{B}}\bPsi_{|A]} \right), \notag \allowdisplaybreaks\\
        &\hat{\bP}_{A}= \bP_{A}-2\bar{\bPsi}_{A}\bla,\notag\allowdisplaybreaks\\
        &\hat{\bF}_{ABC}= \bF_{ABC}+3\overline{\bPsi^c}_{[A}\Gamma_{BC]}\bla+3\bar{\bPsi}_{[A}\Gamma_{B}\bPsi_{C]}, \notag\allowdisplaybreaks\\
        &\hat{\tilde{\bF}}_{ABCDE} = \tilde{\bF}_{ABCDE}-20i\overline{\bPsi^c}_{[A}\Gamma_{BCD}\bPsi_{E]},\notag\allowdisplaybreaks\\
        &\hat{D}_{A}\bla = \left(D_{A}(\hat{\bom})+3i\bQ_{A}\right)\bla-\frac{1}{2}\hat{\bP}_{B}\Gamma^{B}\bPsi^c_{A}+\frac{1}{24}\hat{\bF}^{BCD}\Gamma_{BCD}\bPsi_{A}, \notag\allowdisplaybreaks\\ 
        &\hat{D}_{[A}\bPsi_{B]} = 
        \left(D_{A}(\hat{\bom})+i\bQ_{A}\right)\bPsi_{B}+\frac{1}{48} \left(\hat{\bF}^{CDE} \Gamma_{[A|CDE}  -9\hat{\bF}_{[A|CD} \Gamma^{CD}\right) \bPsi^c_{|B]} \notag\\
        &\qquad\qquad +\frac{i}{192}\hat{\tilde{\bF}}_{[A|CDEF}\Gamma^{CDEF}\bPsi_{|B]} + \frac{1}{2}\left(\overline{\bPsi^c}_{A}\Gamma^{C}\bPsi^c_{B}\right)\Gamma_{C}\bla - \left(\overline{\bPsi^c}_{[B}\bla\right) \bPsi^c_{A]} \notag\\
        &\qquad\qquad + \frac{21}{32} \left(\overline{\bla^c}\Gamma_{[A}\bla\right)\bPsi_{|B]}
        -\frac{3}{32}\left(\overline{\bla^c}\Gamma^{C}\bla\right)\left(\Gamma_{[A|C}\bPsi_{|B]}\right) \notag\\
        &\qquad\qquad -\frac{5}{64}\left(\overline{\bla^c}\Gamma_{[A|CD}\bla\right)\left(\Gamma^{CD}\bPsi_{|B]}\right)
         +\frac{1}{64}\left(\overline{\bla^c}\Gamma^{CDE}\bla\right)\left(\Gamma_{[A|CDE}\bPsi_{|B]}\right) \notag\\
        &\qquad\qquad +\frac{1}{256}\left(\overline{\bla^c}\Gamma_{[A|CDEF}\bla\right)\left(\Gamma^{CDEF}\bPsi_{|B]}\right)\,,\notag\allowdisplaybreaks\\
        &\hat{\nabla}^{A}\hat{\bar{\bP}}_{A} = \left(\nabla^{A}-4i\bQ^{A}\right)\left(\bar{\bP}_{A}-2\overline{\bPsi^c}_A\bla^c\right),\notag\allowdisplaybreaks\\
        &\hat{\nabla}^{C}\hat{\bar{\bF}}_{ABC} = \left(\nabla^{C}-2i\bQ^{C}\right)\left(\hat{\bar{\bF}}_{ABC}-3\overline{\bPsi}_{[A}\Gamma_{BC]}\bla^c + 3\overline{\bPsi^c}_{[A}\Gamma_{B}\bPsi^c_{C]}\right)\,,\notag\allowdisplaybreaks\\
        &\hat{\bR}_{AB} =\bR_{AB}(\hat{\bom})-\left[\bar{\bPsi}_{[A}\Gamma_{D]}\left(\hat{D}_{B}\bPsi^{D}\right)^{c}-\frac{1}{2}\bar{\bPsi}_{D}\Gamma_{B}\left(\hat{D}_{A}\bPsi^{D}\right)^{c}-\bar{\bPsi}_{[A}\Gamma^{D}\left(\hat{D}_{D]}\bPsi_{B}\right)^{c}\right] \notag\\
        &\qquad\quad-\left[\overline{\bPsi^{c}}_{[A}\Gamma_{D]}\hat{D}_{B}\bPsi^{D}-\frac{1}{2}\overline{\bPsi^{c}}_{D}\Gamma_{B}\hat{D}_{A}\bPsi^{D}-\overline{\bPsi^{c}}_{[A}\Gamma^{D}\hat{D}_{D]}\bPsi_{B}\right]\notag\\
        &\qquad\quad +\frac{3}{4}\bar{\bPsi}_{[A|}\left[\Gamma^{E}\hat{\bar{\bF}}_{B}{}^{D}{}_{E}+\frac{1}{18}\Gamma_{B}{}^{D}{}_{EFG}\hat{\bar{\bF}}^{EFG}\right]\bPsi_{|D]}+\frac{3}{4}\overline{\bPsi^{c}}_{[A|}\left[\Gamma^{E}\hat{\bF}_{B}{}^{D}{}_{E}+\frac{1}{18}\Gamma_{B}{}^{D}{}_{EFG}\hat{\bF}^{EFG}\right]\bPsi_{|D]}^{c}\notag\\
        &\qquad\quad -\overline{\bPsi^{c}}_{[A|}\left[\frac{3}{8}\Gamma_{B}{}^{D}{}_{E}\left(\bar{\bla}\Gamma^{E}\bla\right)-\frac{5}{8}\Gamma^{E}\left(\bar{\bla}\Gamma_{B}{}^{D}{}_{E}\bla\right)-\frac{1}{16}\Gamma_{B}{}^{D}{}_{EFG}\left(\bar{\bla}\Gamma^{EFG}\bla\right)\right.\notag\\
        &\qquad\quad \left.-\frac{i}{12}\Gamma^{EFG}\left( \hat{\tilde{\bF}}_{B}{}^{D}{}_{EFG}+\frac{3i}{4}\overline{\bla^{c}}\Gamma_{B}{}^{D}{}_{EFG}\bla\right) \right]\Psi_{|D]}\,.
\end{align}

These equations of motion have to be supplemented by the Bianchi identities
\begin{align}
\partial_{[A} \bG_{BCD]}=0\,,\qquad \partial_{[A} \bF_{BCDEF]}=0\,.
\end{align}
The supersymmetry transformations are given by \cite{Howe:1983sra}
\begin{equation}
    \begin{aligned}
        &\delta \be_{A}{}^{\underline{B}} = \overline{\bep^c}\Gamma^{\underline{B}}\bPsi_A + \bar{\bep}\Gamma^{\underline{B}}\bPsi^c_A\,,\qquad\delta \bu = 2\overline{\bep^c}\bla^c \bv\,,\qquad \delta\bv = -2 \bar{\bep}\bla \bu\,,\\
        &\delta  \bA_{AB} =  \left(-\bar{\bep}\Gamma_{AB}\bla^c + 2 \overline{\bep^c}\Gamma_{[A}\bPsi^c_{B]}\right)\bv-\left( \overline{\bep^c}\Gamma_{AB}\bla + 2\bar{\bep}\Gamma_{[A}\bPsi_{B]}\right)\bu\,,\\
        &\delta \bC_{ABCD} = -4i\bar{\bep}\Gamma_{[ABC}\bPsi^c_{D]} + 4i\overline{\bep^c}\Gamma_{[ABC}\bPsi_{D]} + 12i(\bA_{[AB}\delta \bar{\bA}_{CD]} - \bar{\bA}_{[AB}\delta \bA_{CD]})\\
        &\delta \bla = \frac{1}{24} \hat{\bF}_{ABC}\Gamma^{ABC}\bep - \frac{1}{2}\hat{\bP}_A \Gamma^{A}\bep^c\,,\\
        &\delta \bPsi_A = \left(D_A(\hat{\bom})+i\bQ_A\right)\bep + \frac{1}{48}\left(\hat{\bF}^{BCD}\Gamma_{ABCD} - 9\hat{\bF}_{ABC}\Gamma^{BC} \right)\bep^c + \frac{i}{192}\hat{\tilde{\bF}}_{ABCDE}\Gamma^{BCDE}\bep \\
        &\qquad\qquad + \frac{1}{16}\left(\frac{21}{2}\overline{\bla^c}\Gamma_A\bla - \frac32 \overline{\bla^c}\Gamma^B\bla \Gamma_{AB} - \frac54 \overline{\bla^c}\Gamma_{ABC}\bla\Gamma^{BC}\right. \\
        &\qquad\qquad\left. +\frac14 \overline{\bla^c}\Gamma^{BCD}\bla\Gamma_{ABCD} +\frac{1}{16} \overline{\bla^c}\Gamma_{ABCDE}\bla\Gamma^{BCDE}\right)\bep\,.
    \end{aligned}
    \label{eq:IIB-susy}
\end{equation}

It is straightforward to deduce the supersymmetry transformation of the supercovariant field strengths for the case $\bla=\bP=0$ as 
\begin{equation}
    \begin{aligned}
        &\delta\hat{\bF}_{ABC}=6\bar{\bep}\Gamma_{[A}\hat{D}_{B}\left(\hat{\bom}\right)\bPsi_{C]}\,,\\
        &\delta\hat{\tilde{\bF}}_{ABCDE}=-20i\overline{\bep^{c}}\Gamma_{[ABC}\left[D_{D}(\hat{\bom})\bPsi_{E]}+\frac{i}{192}\hat{\tilde{\bF}}_{D|FGHI}\Gamma^{FGHI}\bPsi_{|E]}\right]\\
        &\qquad\qquad\qquad+20i\bar{\bep}\Gamma_{[ABC}\left[D_{D}(\hat{\bom})\bPsi_{E]}^{c}-\frac{i}{192}\hat{\tilde{\bF}}_{D|FGHI}\Gamma^{FGHI}\bPsi_{|E]}^{c}\right]\\
        &\qquad\qquad\qquad-\delta\left(\bB_{(2)}\wedge \bF_{(3)}\right)\,.\\
    \end{aligned}
    \label{eq:IIB-susy-supercov}
\end{equation}
We do not need the transformation of the 2-form potential for our embedding Ansatz.

\subsection{Pure 5d supergravity}\label{conventions5d}
The field content of pure minimal 5d supergravity comprises the graviton $\cg_{\mu\nu}$, a Dirac gravitino $\Psi_\mu$, or equivalently a pair of symplectic Majorana gravitini $\{\Psi_\mu, \Psi_\mu^c\}$, and a 1-form potential $\cA_{\mu}$ with curvature $\cF_{\mu\nu}=2\partial_{[\mu} \cA_{\nu]}$. Let us also introduce the supercovariant field strength tensor and spin connection
\begin{equation}
    \begin{aligned}
        \hat{\cF}_{\mu\nu}&=\cF_{\mu\nu}+\frac{\sqrt{3}}{2}\left(\bar{\Psi}_{ [\mu}{\Psi^{\text{c}}}_{\nu]}-{{\bar{\Psi}}^{\text{c}}}{}_{[\mu}\Psi_{\nu]}\right)\,,\\
        \hat{\com}_{\mu\underline{\nu\rho}}&=\com_{\mu\underline{\nu\rho}}(\ce)-\frac{i}{4}\left(\cbPsi_\mu\gamma_{\underline{\rho}}\Psi_{\underline{\nu}}-\bar{\Psi}_\mu\gamma_{\underline{\rho}}\cPsi_{\underline{\nu}}-\cbPsi_{\underline{\nu}}\gamma_\mu\Psi_{\underline{\rho}}+\bar{\Psi}_{\underline{\nu}}\gamma_\mu\cPsi_{\underline{\rho}}+\cbPsi_{\underline{\rho}}\gamma_{\underline{\nu}}\Psi_{\mu}-\bar{\Psi}_{\underline{\rho}}\gamma_{\underline{\nu}}\cPsi_{\mu}\right)\,.
    \end{aligned}
    \label{eq:5d-supercov}
\end{equation}
Here, $\com_{\mu \underline{\nu\rho}}(\ce)$ is the spin connection associated to the Levi-Civita connection.

The action of pure 5d supergravity is given by\cite{Chamseddine:1980mpx,  Gunaydin:1983bi, Gunaydin:1984ak, Ceresole_2000, lauria2020cal} 
\begin{equation}
    \begin{aligned}
        S=&\frac{1}{2\kappa_5^2}\int d^5x \ce\left( \cR(\com) - i\cbPsi_\mu \gamma^{\mu\nu\rho}D_\nu\left(\frac{\com+\hat{\com}}{2}\right)\Psi_\rho + i\bar{\Psi}_\mu \gamma^{\mu\nu\rho}D_\nu\left(\frac{\com+\hat{\com}}{2}\right)\Psi^\text{c}_\rho\right. \\
        &\left.-\frac{1}{4} \cF^{\mu\nu}\cF_{\mu\nu} + \frac{\sqrt{3}}{8}\cbPsi_{\mu}X^{\mu\nu\rho\sigma}\Psi_\nu(\cF_{\rho\sigma}+\hat{\cF}_{\rho\sigma}) + \frac{1}{12\sqrt{3}} \epsilon^{\mu\nu\rho\sigma\lambda}\cF_{\mu\nu}\cF_{\rho\sigma}\cA_{\lambda}\right)\,,
    \end{aligned}
\end{equation}
with the equations of motion 
\begin{equation}
    \begin{aligned}
        0=&\gamma^{\mu\nu\rho}D_\nu(\hat{\com})\Psi_\rho + \frac{i}{8\sqrt{3}}\hat{\cF}_{\lambda\sigma}\gamma^{\mu\nu\rho}\left(\gamma^{\lambda\sigma}{}_\nu-4\delta^\lambda_\nu\gamma^\sigma\right)\Psi_\rho \,, \\
        0=&\nabla_\mu\cF^{\mu\nu} - \frac{\sqrt{3}}{2} \nabla_\mu\left(\cbPsi_\rho X^{\rho\sigma\mu\nu}\Psi_\sigma\right) + \frac{1}{4\sqrt{3}}\epsilon^{\nu\mu\lambda\rho\sigma}\cF_{\rho\sigma}\cF_{\mu\lambda}\,,\\
        \cR_{\mu\nu}(\com)=&\frac{1}{2}\cF_{\mu\lambda}\cF_{\nu}{}^{\lambda}-\frac{1}{12}\cg_{\mu\nu}\cF^{\rho\sigma}\cF_{\rho\sigma}\\
        &+\frac{3i}{2}\left[\overline{\Psi^{c}}_{[\mu|}\gamma_{\nu}{}^{\lambda\rho}D_{|\lambda}\left(\frac{\com+\hat{\com}}{2}\right)\Psi_{\rho]}-\bar{\Psi}_{[\mu|}\gamma_{\nu}{}^{\lambda\rho}D_{|\lambda}\left(\frac{\com+\hat{\com}}{2}\right)\Psi_{\rho]}^{\text{c}}\right]\\
        &-\frac{i}{6}\cg_{\mu\nu}\left[\overline{\Psi^{c}}_{\sigma}\gamma^{\sigma\lambda\rho}D_{\lambda}\left(\frac{\com+\hat{\com}}{2}\right)\Psi_{\rho}-\bar{\Psi}_{\sigma}\gamma^{\sigma\lambda\rho}D_{\lambda}\left(\frac{\com+\hat{\com}}{2}\right)\Psi_{\rho}^{\text{c}}\right]\\
        &-\frac{\sqrt{3}}{4}\left[\overline{\Psi^{c}}_{[\mu|}\gamma_{\nu}{}^{\lambda\rho\sigma}\Psi_{|\lambda}\left(\cF_{\rho\sigma]}+\hat{\cF}_{\rho\sigma]}\right)+\left(\overline{\Psi^{c}}_{(\mu}\Psi^{\rho}-\overline{\Psi^{c}}^{\rho}\Psi_{(\mu}\right)\left(\cF_{\nu)\rho}+\hat{\cF}_{\nu)\rho}\right)\right]\\
        &+\frac{1}{8\sqrt{3}}\cg_{\mu\nu}\left[\overline{\Psi^{c}}_{\lambda}\gamma^{\lambda\eta\rho\sigma}\Psi_{\eta}\left(\cF_{\rho\sigma}+\hat{\cF}_{\rho\sigma}\right)+2\overline{\Psi^{c}}^{\rho}\Psi^{\sigma}\left(\cF_{\rho\sigma}+\hat{\cF}_{\rho\sigma}\right)\right]\\
        &+\frac{1}{32}\left[\overline{\Psi^{c}}_{\sigma}\gamma^{\sigma\lambda\rho}\gamma^{cd}\Psi_{\rho}-\bar{\Psi}_{\sigma}\gamma^{\sigma\lambda\rho}\gamma^{cd}\Psi_{\rho}^{c}\right]\left(\overline{\Psi^{c}}_{[\mu|}\gamma_{\nu}{}^{\eta}{}_{\lambda cd}\Psi_{|\eta]}\right)\\
        &-\frac{1}{96}\cg_{\mu\nu}\left[\overline{\Psi^{c}}_{\sigma}\gamma^{\sigma\lambda\rho}\gamma^{cd}\Psi_{\rho}-\bar{\Psi}_{\sigma}\gamma^{\sigma\lambda\rho}\gamma^{cd}\Psi_{\rho}^{c}\right]\left(\overline{\Psi^{c}}_{\tau}\gamma^{\tau\eta}{}_{\lambda cd}\Psi_{\eta}\right)\,,
    \end{aligned}
    \label{eq:5d-eom}
\end{equation}
where $X^{\mu\nu\rho\sigma}=\gamma^{\mu\nu\rho\sigma}+\cg^{\mu\rho}\cg^{\nu\sigma}-\cg^{\mu\sigma}\cg^{\nu\rho}$. 
The spin connection 
\begin{equation}
    \com_{\mu \underline{\nu\rho}}=\hat{\com}_{\mu \underline{\nu\rho}}+\frac{i}{4}\cbPsi_\lambda\gamma^{\lambda\eta}{}_{\mu \underline{\nu\rho}}\Psi_\eta
    \label{eq:5d-spin connection}
\end{equation}
is the one associated to the 1\textsuperscript{st} order formalism. The coordinate covariant derivative $\nabla$ is defined, similar to the 11d case, as $\nabla \equiv \partial + \com(\ce) + \Gamma(\cg) = D(\ce) + \Gamma(\cg)$ with $\Gamma(\cg)$ the Levi-Civita connection. The coordinate covariant derivative acts on both spacetime and Lorentz indices. In contrast, the Lorentz covariant derivative $D(\com)$  acts only on Lorentz indices. Its action on spinors is similar to \eqref{eq:derivative on spinor}.

These equations of motion have to be supplemented by the Bianchi identity
\begin{align}
\partial_{[\mu}\cF_{\nu\rho]}=0\,.
\end{align}
The supersymmetry transformations with parameter $\cep$ are given by 
\begin{equation}
    \begin{aligned}
        &\delta \ce_\mu{}^{\underline{\nu}}=\frac{i}{2}\left(-\bar{\cep}\gamma^{\underline{\nu}}\Psi_{\mu}^{c}+\overline{\cep^{c}}\gamma^{\underline{\nu}}\Psi_{\mu}\right)\,,\\
        &\delta\cA_\mu=\frac{\sqrt{3}}{2}\left(\overline{\cep^{c}}\Psi_{\mu}-\bar{\cep}\cPsi_{\mu}\right)\,,\\
        &\delta\Psi_{\mu}=D_\mu(\hat{\com})\cep+\frac{i}{8\sqrt{3}}\hat{\cF}_{\nu\rho}\left(\gamma_\mu{}^{\nu\rho}-4\delta_\mu^\nu\gamma^\rho\right)\cep\,.
    \end{aligned}
    \label{eq:5d-susy-trans}
\end{equation}
We observe that the gravitino equation of motion and supersymmetry transformation have a similar structure and we can define a supercovariant derivative acting on spinors as 
\begin{equation}
    \hat{D}(\hat{\com}) = D(\hat{\com}) + \frac{i}{8\sqrt{3}}\hat{\cF}_{\lambda\sigma}\gamma^{\mu\nu\rho}\left(\gamma^{\lambda\sigma}{}_\nu-4\delta^\lambda_\nu\gamma^\sigma\right)\,,
    \label{eq:5d-supercov-derivative}
\end{equation}
to simplify the supersymmetry transformation of the supercovariant field strength: 
\begin{equation}
    \delta \hat{\cF}_{\mu\nu} = \sqrt{3}\left(\overline{\zeta^c}\hat{D}_{[\mu}(\hat{\com})\Psi_{\nu]} -  \bar{\zeta}\hat{D}_{[\mu}(\hat{\com})\Psi^c_{\nu]}\right).
    \label{eq:5d-susy-supercov-field-strength}
\end{equation}
\subsection{Pure 4d \texorpdfstring{$\mathcal{N}=2$}{N=2} supergravity}\label{conventions4d}
The field content of 4d $\mathcal{N}=2$ supergravity comprises the graviton $g_{\mu\nu}$, two Weyl gravitini $\psi_{I\mu}$ and a 1-form potential $A_{\mu}$ with curvature $F_{\mu\nu}=2\partial_{[\mu}A_{\nu]}$. Here, we use the position of the index to indicate chirality as $P_+\psi^I_\mu \equiv \frac12\left(1+i\gamma^{(4)}\right)\psi^I_\mu = \psi^I_\mu$. We also introduce the supercovariant field strength and spin connection 
\begin{equation}
    \begin{aligned}
        \hat{F}_{\mu\nu}&=F_{\mu\nu}+\left(\bar{\psi}_{I\mu}\psi_{J\nu}\varepsilon^{IJ}+\bar{\psi}^I_{\mu}\psi^J_{\nu}\varepsilon_{IJ}\right)\,,\\
        \hat{\omega}_{\mu\nu\rho}&=\omega_{\mu\nu\rho}(e)-\frac{1}{4}\left(\bar{\psi}_{I\mu}\gamma_{\rho}\psi_{\nu}^I+\bar{\psi}^I_{\mu}\gamma_{\rho}\psi_{I\nu}-\bar{\psi}_{I\nu}\gamma_{\mu}\psi_{\rho}^I-\bar{\psi}^I_{\nu}\gamma_{\mu}\psi_{I\rho}+\bar{\psi}_{I\rho}\gamma_{\nu}\psi_{\mu}^I+\bar{\psi}^I_{\rho}\gamma_{\nu}\psi_{I\mu}\right)\,.
        \label{eq:4d-supercov}
    \end{aligned}
\end{equation}
In 4d, the supercovariant spin connection is exactly the one associated to the 1\textsuperscript{st} order formalism. 
Here we take $\varepsilon^{12} = 1$ and $\varepsilon_{12} = 1$, generating inner products on different $SU(2)$ representations. 

The 4d action is\footnote{Because of the $SU(2)$ duality of $\mathcal{N}=2$ supergravity, the field-strength equation of motion can be transformed into different forms. Here, we choose the one corresponding to standard Einstein-Maxwell theory with $\theta=0$.} \cite{Ferrara:1976fu,  Andrianopoli:1996vr, Andrianopoli:1996cm, freedman_van_proeyen_2012, lauria2020cal}
\begin{equation}
    \begin{aligned}
        S_4 =& \frac{1}{2\kappa^2}\int d^4x e\left(R(\hat{\omega}) - \left(\bar{\psi}^I_\mu \gamma^{\mu\nu\lambda}D_\nu(\hat{\omega})\psi_{I\lambda}+\bar{\psi}_{I\mu} \gamma^{\mu\nu\lambda}D_\nu(\hat{\omega})\psi^I_{\lambda}\right) - \frac14 F_{\mu\nu}F^{\mu\nu}\right.\\
        &\qquad\qquad\qquad-F^-_{\mu\nu}\bar{\psi}^{I\mu}\psi^{J\nu}\varepsilon_{IJ} - F^+_{\mu\nu}\bar{\psi}^{\mu}_I\psi^{\nu}_J\varepsilon^{IJ} - \bar{\psi}_\mu^I \psi_\nu^J \bar{\psi}^\mu_I \psi^\nu_J \\
        &\qquad\qquad\qquad\left.-\frac12\left(\bar{\psi}_\mu^I \psi_\nu^J\right)^-\left(\bar{\psi}^{K\mu} \psi^{J\nu}\right)^- \varepsilon_{IJ}\varepsilon_{KL} 
        -\frac12\left(\bar{\psi}_{I\mu} \psi_{J\nu}\right)^+\left(\bar{\psi}^{\mu}_K \psi^{\nu}_L\right)^+ \varepsilon^{IJ}\varepsilon^{KL} \right)\,,
    \end{aligned}
\end{equation}
where we also define (anti-) selfdual combinations
\begin{equation}
    F^\pm_{(2)} \equiv \frac12 \left(F_{(2)} \pm i*F_{(2)}\right)\,,\quad \left(\bar{\psi}_{\mu}^{I}\psi^{J}_{\nu}\right)^{\pm} = \frac12\left( \bar{\psi}_{\mu}^{I}\psi^{J}_{\nu} \pm \frac{i}{2}\epsilon_{\mu\nu}{}^{\rho\sigma} \bar{\psi}_{\mu}^{I}\psi^{J}_{\nu}\right)\,.
\end{equation}
The equations of motion are given by 
\begin{equation}
    \begin{aligned}
        0=&\nabla_{\mu}\left[F^{\mu\nu}+\left(\bar{\psi}^{I\mu}\psi^{J\nu}\varepsilon_{IJ}+\bar{\psi}^\mu_I\psi^\nu_J\varepsilon^{IJ}\right)-\frac{i}{2}\varepsilon^{\mu\nu\rho\sigma}\left(\bar{\psi}^I_\rho\psi^J_\sigma\varepsilon_{IJ} - \bar{\psi }_{I\rho}\psi_{J\sigma}\varepsilon^{IJ}\right)\right]\,,\\
        0=&\gamma^{\mu\nu\rho}D_{\nu}(\hat{\omega})\psi^I_{\rho}+\frac{1}{8}\hat{F}_{\sigma\tau}\gamma^{\mu\nu\rho}\gamma^{\sigma\tau}\gamma_{\nu}\varepsilon^{IJ}\psi_{J\rho}\,,\\
        R_{\mu\nu}\left(\hat{\omega}\right)=&\frac{1}{2}F_{\mu\rho}F_{\nu}{}^{\rho}-\frac{1}{8}g_{\mu\nu}F_{\rho\sigma}F^{\rho\sigma}\\
        &+\frac{3}{2}\left(\bar{\psi}_{[\mu|}^{I}\gamma_{\nu}{}^{\rho\lambda}D_{|\rho}\left(\hat{\omega}\right)\psi_{I\lambda]}+\bar{\psi}_{I[\mu|}\gamma_{\nu}{}^{\rho\lambda}D_{|\rho}\left(\hat{\omega}\right)\psi_{\lambda]}^{I}\right)\\
        &-\frac{1}{4}g_{\mu\nu}\left(\bar{\psi}_{\rho}^{I}\gamma^{\rho\sigma\lambda}D_{\sigma}\left(\hat{\omega}\right)\psi_{I\lambda}+\bar{\psi}_{I\rho}\gamma^{\rho\sigma\lambda}D_{\sigma}\left(\hat{\omega}\right)\psi_{\lambda}^{I}\right)\\
        &+2\bar{\psi}_{(\nu}^{I}\psi^{J\lambda}\varepsilon_{IJ}F_{\mu)\lambda}^{-}+2\bar{\psi}_{I(\nu}\psi_{J}^{\lambda}\varepsilon^{IJ}F_{\mu)\lambda}^{+}-\frac{1}{2}g_{\mu\nu}\left(\bar{\psi}^{I\rho}\psi^{J\sigma}\varepsilon_{IJ}F_{\rho\sigma}^{-}+\bar{\psi}_{I}^{\rho}\psi_{J}^{\sigma}\varepsilon^{IJ}F_{\rho\sigma}^{+}\right)\\
        &+2\bar{\psi_{(\mu|}^{I}}\psi_{\lambda}^{J}\bar{\psi}_{I|\nu)}\psi_{J}^{\lambda}+\left(\bar{\psi_{(\mu|}^{I}}\psi_{\lambda}^{J}\right)^{-}\left(\bar{\psi}_{|\nu)}^{K}\psi^{L\lambda}\right)^{-}\varepsilon_{IJ}\varepsilon_{KL}+\left(\bar{\psi}_{(\mu|I}\psi_{\lambda J}\right)^{+}\left(\bar{\psi}_{K|\nu)}\psi_{L}^{\lambda}\right)^{+}\varepsilon^{IJ}\varepsilon^{KL}\\
        &-\frac{1}{2}g_{\mu\nu}\left[\bar{\psi}_{\rho}^{I}\psi_{\sigma}^{J}\bar{\psi}_{I}^{\rho}\psi_{J}^{\sigma}+\frac{1}{2}\left(\bar{\psi}_{\rho}^{I}\psi_{\sigma}^{J}\right)^{-}\left(\bar{\psi}^{K\rho}\psi^{L\sigma}\right)^{-}\varepsilon_{IJ}\varepsilon_{KL}+\frac{1}{2}\left(\bar{\psi}_{\rho I}\psi_{\sigma J}\right)^{+}\left(\bar{\psi}_{K}^{\rho}\psi_{L}^{\sigma}\right)^{+}\varepsilon^{IJ}\varepsilon^{KL}\right]\,.
    \end{aligned}
    \label{eq:4d-eom}
\end{equation}
The derivatives appearing in these equations are defined as in the 5d case. 

We have to supplement the equations of motion by the Bianchi identity
\begin{align}
\partial_{[\mu}F_{\nu\rho]}=0\,.
\label{eq:4d-bainchi}
\end{align}
The supersymmetry transformations with parameters $\epsilon^I$ are given by \cite{Ferrara:1976fu,  Andrianopoli:1996vr, Andrianopoli:1996cm, freedman_van_proeyen_2012, lauria2020cal}
\begin{equation}
    \begin{aligned}
        &\delta e_\mu{}^{\underline{\nu}}=\frac12\left(\bar{\epsilon}^I\gamma^{\underline{\nu}}\psi_{I\mu} + \bar{\epsilon}_I\gamma^{\underline{\nu}}\psi_\mu^I\right)\,, \\
        &\delta A_{\mu}=-\left(\bar{\epsilon}^I\psi_{\mu}^J\varepsilon_{IJ}+\bar{\epsilon}_I\psi_{J\mu}\varepsilon^{IJ}\right)\,,\\
        &\delta\psi^I_{\mu}=D_\mu(\hat{\omega})\epsilon^I+\frac{1}{8}\hat{F}_{\nu\rho}\gamma^{\nu\rho}\gamma_{\mu}\varepsilon^{IJ}\epsilon_J\,.
    \end{aligned}
    \label{eq:4d-susy}
\end{equation}
We can again introduce a supercovariant derivative for spinors as 
\begin{equation}
    \hat{D}_\mu(\hat{\omega})\varepsilon^I \equiv D_\mu(\hat{\omega})\varepsilon^I + \frac18 \hat{F}_{\nu\rho}\gamma^{\nu\rho}\gamma_\mu \varepsilon^{IJ}\epsilon_J\,,
\end{equation}
which simplifies the supersymmetry transformation of the supercovariant field strength 
\begin{equation}
    \delta \hat{F}_{\mu\nu} = -2\left(\bar{\epsilon}^I\hat{D}_{[\mu}(\hat{\omega}) \psi^J_{\nu]}\varepsilon_{IJ}  + \bar{\epsilon}_I\hat{D}_{[\mu}(\hat{\omega}) \psi_{\nu]J}\varepsilon^{IJ} \right)\,.
    \label{eq:4d-susy-supercov}
\end{equation}

\section{Reduction of equations of motion at fermionic order}\label{fermionEoM}
In this appendix, we outline the proof that the embedding Ans\"atze introduced in Section \ref{Ansatz} solve the equations of motion including all fermion terms. Given that the various equations of motion are connected by supersymmetry transformations which we have matched consistently in Section \ref{proof}, we choose to omit Einstein's equation from this appendix but instead discuss all other equations of motion. This concludes the proof of consistency.

\subsection{From 11d to pure 5d and 4d supergravity}

We start by considering the Bianchi identity. The $\mu\nu\rho mn$-component of the 11d Bianchi identity gives the 5d Bianchi identity as mentioned in Section \ref{Consistency11-4}. However, the $\mu\nu\rho\sigma\lambda$-component of the Bianchi identity needs further scrutiny. We insert the embedding Ansatz and argue that the remaining fermion terms indeed vanish as 
\begin{equation}\label{template}
    -3\nabla_{[\mu}\left(\bar{\Psi}_\nu \gamma_{\rho\sigma}\Psi_{\lambda]} + \overline{\cPsi}_\nu \gamma_{\rho\sigma}\Psi^c_{\lambda]}\right) = -\frac32 i\left(\bar{\Psi}_{[\nu}\gamma_\rho{}^\beta \Psi_\lambda + \overline{\cPsi}_{[\nu}\gamma_\rho{}^\beta\Psi^c_\lambda\right) \left(\overline{\cPsi}_{\mu|}\gamma_\beta\Psi_{\sigma]} -\bar{\Psi}_{\mu|}\gamma_\beta\Psi^c_{\sigma]}\right) = 0\,.
\end{equation}
We used here the 5d gravitino equation in the first equality and then use the 5d Fierz identity 
\begin{equation}
    \Psi_\mu\bar{\Psi}_\nu = -\frac14\left[\bar{\Psi}_\nu\Psi_\mu + \gamma_\lambda\bar{\Psi}_\nu\gamma^\lambda\Psi_\mu  - \frac12 \gamma_{\lambda\sigma}\bar{\Psi}_\nu\gamma^{\lambda\sigma}\Psi_\mu\right]
\end{equation}
to prove  that the quartic terms vanish.
This indicates that the spinor bilinear correction of the field strength \eqref{eq:11d-ansatz-corrected} is closed on-shell using the 5d gravitino equation of motion. All other components of the Bianchi identity vanish trivially.

We now consider the field strength equation of motion. Using our embedding Ansatz,  the $\mu\nu\eta$-component of the 11d field strength equation (\ref{eq:11d-eom}) becomes
\begin{equation}
    \nabla_\lambda \bF^{\lambda\mu\nu\eta} - 3\nabla_\lambda \left[\bar{\bPsi}_\rho \Gamma^{[\lambda\mu}\bg^{\nu|\rho}\bg^{|\eta]\sigma}\bPsi_\sigma \right] =0\,,
\end{equation}
which is satisfied precisely because of the fermion bilinear correction in \eqref{eq:11d-ansatz-corrected}.
It is straightforward to get the field strength equation (\ref{eq:5d-eom}) from the $\mu mn$-component of (\ref{eq:11d-eom}) without any further spinor bilinear contributions. We regard this as an independent check of our embedding strategy using supercovariant quantities. The $\mu\nu m$- and $mnp$-components vanish trivially.

We can see that the gravitino supersymmetry transformation and equation of motion are related to each other both in 11d (\eqref{eq:11d-eom} and \eqref{eq:11d-susy-supercov}) and in 5d (\eqref{eq:5d-eom} and \eqref{eq:5d-susy-trans}), \ie 
\begin{equation}\label{eq:gravitino-eom}
     \delta \bPsi_A=\hat{D}_A(\hat{\bom})\bep\qquad \leftrightarrow\qquad 0=\Gamma^{ABC}\hat{D}_B(\hat{\bom})\bPsi_C\,.
 \end{equation}
We furthermore note that our embedding Ansatz \eqref{eq:11d-5d-fermionic-ansatz} decomposes the gravitino and the supersymmetry parameter in the same way. This implies that the proof of consistency of the gravitino equation of motion is exactly parallel to the proof of consistency of the supersymmetry transformation shown in Section \ref{proof} and does not require repetition.

This concludes our proof of consistency for the embedding of pure 5d supergravity into 11d supergravity.

\

Consider now the reduction from 5d to 4d $\mathcal{N}=2$ supergravity. The  only non-trivial component of the 5d Bianchi identity is the $\mu\nu\lambda$-component
\begin{equation}
    \begin{aligned}
        &\nabla_{[\lambda}\left\{-\frac12\epsilon_{\rho\sigma]\mu\nu}\left[F^{\mu\nu} - \frac{i}{2}\epsilon^{\mu\nu\alpha\beta}\left(\bar{\psi}^I_\alpha\psi_\beta^J\varepsilon_{IJ} - \bar{\psi}_{I\alpha}\psi_{J\beta}\varepsilon^{IJ}\right) +  \left(\bar{\psi}^{I\mu}\psi^{J\nu}\varepsilon_{IJ} + \bar{\psi}_I^\mu\psi^\nu_{J}\varepsilon^{IJ}\right)\right]\right\} \\
        &\sim\epsilon_{\lambda\rho\sigma\nu}\nabla_\mu \left[F^{\mu\nu} - \frac{i}{2}\epsilon^{\mu\nu\alpha\beta}\left(\bar{\psi}^I_\alpha\psi_\beta^J\varepsilon_{IJ} - \bar{\psi}_{I\alpha}\psi_{J\beta}\varepsilon^{IJ}\right) +  \left(\bar{\psi}^{I\mu}\psi^{J\nu}\varepsilon_{IJ} + \bar{\psi}_I^\mu\psi^\nu_{J}\varepsilon^{IJ}\right)\right] = 0\,,
    \end{aligned}
    \label{eq:5-4 eom including fermions}
\end{equation}
which is the 4d field strength equation \eqref{eq:4d-eom}. 
The $\mu$-component of the field strength equation is
\begin{equation}
    \begin{aligned}
        0=&\nabla_\mu \mathcal{F}^{\mu\nu} -\frac{\sqrt{3}}{2}\nabla_{\mu}\left(\overline{\Psi^c}_\rho\gamma^{\mu\nu\rho\sigma}\Psi_\sigma + \overline{\Psi^c}^\mu\Psi^\nu - \overline{\Psi^c}^\nu\Psi^\mu\right) \\
        &-\frac{\sqrt{3}}{2}\nabla_{4}\left(\overline{\Psi^c}_\rho\gamma^{4\nu\rho\sigma}\Psi_\sigma + \overline{\Psi^c}^4\Psi^\nu - \overline{\Psi^c}^\nu\Psi^4\right) + \frac{2}{\sqrt{6}}\epsilon^{\nu\rho\sigma\lambda4}\mathcal{F}_{\rho\sigma}\mathcal{F}_{\lambda 4}\\
        =&\frac{\sqrt{3}}{2}\nabla_a\left(-\frac12 \varepsilon^{ab}{}_{cd} F^{cd} \right)\,.
    \end{aligned}
\end{equation}
which is the Bianchi identity of the vector field \eqref{eq:4d-bainchi}. The $4$-component of the field strength equation of motion is the most involved. It  can be shown to vanish algebraically using the 4d Fierz identity 
\begin{equation}
    \psi_{eJ}\bar{\psi}_{Ia} = -\frac12 \bar{\psi}_{aI}\psi_{eJ}P_- + \frac18 \gamma^{gh}\left(\bar{\psi}_{aI}\gamma_{gh}\psi_{eJ}\right)\,. 
    \label{eq:4d-Fierz identity}
\end{equation}
Similarly to the 11d to 5d case, the gravitino equation of motion is similar to the supersymmetry transformation of the gravitino both in 5d (\eqref{eq:5d-eom} and \eqref{eq:5d-susy-trans}) and 4d (\eqref{eq:4d-eom} and \eqref{eq:4d-susy}), \ie
\begin{equation}
    \delta \Psi_\mu = \hat{D}_\mu(\hat{\com})\epsilon\qquad  \leftrightarrow \qquad 0=\gamma^{\mu\nu\rho}\hat{D}_\nu(\hat{\com})\Psi_\rho\,.
\end{equation}
We furthermore note that our embedding Ansatz \eqref{eq:5d-4d-ansatz-corrected} decomposes the gravitino and the supersymmetry parameter in the same way. This implies that the proof of consistency of the gravitino equation of motion is exactly parallel to the proof of consistency of the supersymmetry transformation shown in Section \ref{proof} and does not require repetition. This concludes our proof of consistency for the embedding of pure 4d supergravity into 5d supergravity.

\subsection{From IIA to pure 4d supergravity}
We again start with the Bianchi identities. The $\mu\nu\rho\sigma m$-component is satisfied due to \eqref{relation3}. Using the same trick as in \eqref{eq:5-4 eom including fermions} to change a partial antisymmetrisation into a product of antisymmetric tensors, the $\mu\nu\rho mn$-component gives 
\begin{equation}
    \begin{aligned}
        0=&\nabla_{[\mu}\bF_{\nu\rho mn]}\sim\nabla_{[\mu}\bF_{\nu\rho]mn}\\
        =&\nabla_{[\mu}\left\{ \frac{1}{4}\epsilon_{\nu\rho]\lambda\eta}\left[F^{\lambda\eta} -\frac{i}{2}\epsilon^{\lambda\eta\kappa\tau}\left(\bar{\psi}_{\kappa}^{I}\psi_{\tau}^{J}\varepsilon_{IJ}-\bar{\psi}_{I\kappa}\psi_{\tau J}\varepsilon^{IJ}\right)+\left(\bar{\psi}^{\lambda I}\psi^{\eta J}\varepsilon_{IJ}+\bar{\psi}_{I}^{\lambda}\psi_{J}^{\eta}\varepsilon^{IJ}\right)\right]\right\} \omega_{mn}\\
        \sim&\varepsilon_{\mu\nu\rho\sigma}\nabla_{\lambda}\left[F^{\lambda\sigma}-\frac{i}{2}\varepsilon^{\lambda\sigma\kappa\tau}\left(\bar{\psi}_{\kappa}^{I}\psi_{\tau}^{J}\varepsilon_{IJ}-\bar{\psi}_{I\kappa}\psi_{\tau J}\varepsilon^{IJ}\right)+\left(\bar{\psi}^{\lambda I}\psi^{\sigma J}\varepsilon_{IJ}+\bar{\psi}_{I}^{\lambda }\psi_{J}^{\sigma}\varepsilon^{IJ}\right)\right]\omega_{mn}\,,
    \end{aligned}
\end{equation}
which is the 4d field strength equation \eqref{eq:4d-eom}. The components of the Bianchi identity of $\bF_{(4)}$ with more than 3 internal indices are trivially satisfied. The only non-trivial component of the $\bH_{(3)}$ Bianchi identity is the $\mu\nu\rho\sigma$-component, which needs the 4d gravitino equation of motion 
and the 4d Fierz identity \eqref{eq:4d-Fierz identity} to show that it is satisfied, similarly to the steps in \eqref{template}. Again, we see that the spinor bilinear correction of the field strength is closed under the 4d gravitino equation of motion. Finally, the $\mu\nu\rho$-component of the $\bF_{(2)}$ Bianchi identity reduces to the 4d Bianchi identity and all other components vanish.

We now move on to the equations of motion for bosonic fields. It is straightforward to prove that the $\nu\rho\sigma$-component of the $\bF_{(4)}$ equation of motion is satisfied. The $\mu\nu m$-component is satisfied due to \eqref{relation3}. The $\mu mn$-component of the $\bF_{(4)}$ equation of motion reduces to 
\begin{equation}
    \begin{aligned}
        0 =&\nabla_{\nu}\left[\frac{1}{4}\epsilon^{\nu\mu}{}_{\lambda\eta}\left( F^{\lambda\eta}-\frac{i}{2}\epsilon^{\lambda\eta\rho\sigma }\left(\bar{\psi}^{I}_{\rho}\psi_{f]}^{J}\varepsilon_{IJ}-\bar{\psi}_{I\rho}\psi_{\sigma J}\varepsilon^{IJ}\right)+\left(\bar{\psi}^{I\lambda}\psi^{\eta J}\varepsilon_{IJ}+\bar{\psi}_{I}^{\lambda}\psi_{J}^{\eta}\varepsilon^{IJ}\right)\right) \right]\omega^{mn}\\
        &-\frac{1}{4}\nabla_{\nu}\left[\epsilon^{\nu\mu\lambda\eta}\left( \bar{\psi}^{I}_{\lambda}\psi_{\eta}^{J}\varepsilon_{IJ}+\bar{\psi}_{I\lambda}\psi_{J\eta}\varepsilon^{IJ}\right) +2i\left(\bar{\psi}^{I\nu}\psi^{\mu J}\varepsilon_{IJ}-\bar{\psi}_{I}^{\nu}\psi_{J}^{\mu}\varepsilon^{IJ}\right)\right]\omega^{mn}\\
        =&\nabla_{\nu}\left(\frac{1}{4}\epsilon^{\nu\mu}{}_{\rho\sigma}F^{\rho\sigma}\right)\omega^{mn}\,,
    \end{aligned}
\end{equation}
which is the 4d Bianchi identity.  Similarly to  the $\bH_{\mu\nu\rho}$ Bianchi identity, we need the 4d gravitino equation and the 4d Fierz identity to prove that the $mnp$-component of the $\bF_{(4)}$ equation of motion and the $mn$-component of the $\bH_{(3)}$ equation of motion are satisfied. One may check directly that the $\rho\sigma$-component of the $\bH_{(3)}$ equation of motion is satisfied.  Again, relation \eqref{relation3} guarantees the triviality of the $\mu m$-component of the $\bH_{(3)}$ equation of motion. The $\nu$-component of the $\bF_{(2)}$ equation of motion takes the form 
\begin{equation}
    \begin{aligned}
        0=&\frac{1}{2}\nabla_{\mu}\left[F^{\mu\nu}-\frac{1}{2}i\varepsilon^{\rho\sigma\mu\nu}\left(\bar{\psi}_{\rho}^{I}\psi_{\sigma}^{J}\varepsilon_{IJ}-\bar{\psi}_{\rho I}\psi_{\sigma J}\varepsilon^{IJ}\right)+\left(\bar{\psi}^{I\rho}\psi^{\sigma J}\varepsilon_{IJ}+\bar{\psi}_{I}^{\rho}\psi_{J}^{\sigma}\varepsilon^{IJ}\right)\right]\\
        &+3\left(\bar{\psi}_{[\mu}^{1}\gamma_{\rho}\psi_{\sigma]1}-\bar{\psi}_{[\mu}^{2}\gamma_{\rho}\psi_{\sigma]2}\right)\left(\bar{\psi}^{1[\mu}\gamma^{\rho\sigma}\psi^{\nu]2}+\bar{\psi}_{1}^{[\mu}\gamma^{\rho\sigma}\psi_{2}^{\nu]}\right)\,.
    \end{aligned}
\end{equation}
We can use the Fierz identity to prove the vanishing of the second line, while the first line is the 4d field strength equation. Finally, a straightforward calculation shows that the dilaton equation of motion is satisfied.

For the fermionic equations of motion, we start with the dilatino equation. The residual part after imposing our Ansatz is 
\begin{equation}
    \begin{aligned}
        0=&\frac{1}{4\cdot2!}\Gamma^A\hat{\bF}_{BC}\Gamma^{BC}\bPsi_A-\frac{1}{12\cdot4!}\Gamma^A\hat{\tilde{\bF}}_{BCDE}\Gamma^{BCDE}\Gamma^{(10)}\bPsi_A\\
        =&\left[\frac{1}{4\cdot2!}\frac{1}{2}\hat{F}_{\mu\nu}\left(\gamma^\lambda\gamma^{\mu\nu}\otimes\mathbf{1}\right)-\frac{1}{2\cdot4!}\frac{1}{4}\epsilon_{\mu\nu\rho\sigma}\hat{F}^{\rho\sigma}\left(\gamma^\lambda\gamma^{\mu\nu}\gamma^{(4)}\otimes\omega_{mn}\Sigma^{mn}\Sigma^{(6)}\right)\right] \\
        &\cdot \frac{1}{\sqrt{2}}\left(\psi_\lambda^{1}\otimes\eta_{+}+\psi_{\lambda1}\otimes\eta_{-}+\psi_\lambda^{2}\otimes\eta_{-}+\psi_{\lambda2}\otimes\eta_{+}\right)\,.
    \end{aligned}
\end{equation}
We may use the identity \eqref{eq:omega-gamma-projection} derived from the Fierz identity in the internal space and basic Clifford algebra to prove that this expression vanishes. After setting $\bla=0$, the IIA gravitino equation \eqref{eq:IIA-eom-gravitino} reduces to a similar form as the gravitino supersymmetry transformation \eqref{eq:IIA-susy}. On the other hand, the 4d gravitino supersymmetry transformation and equation are similar as we argued in the last section. We may therefore employ a similar argument as given after \eqref{eq:gravitino-eom} to conclude our proof of consistency.

\subsection{From IIB to pure 4d supergravity}
We again start with the Bianchi identities. The $\mu\nu\lambda ijk$-component of the $\bF_{(5)}$ Bianchi identity is 
\begin{equation}
    \begin{aligned}
        \partial_{[\lambda}\bF_{\mu\nu ijk]}&\sim\partial_{[\lambda}\left(F_{\mu\nu]}^{+}+\frac{1}{2}\left(\bar{\psi}_{I\mu}\psi_{J\nu]}\varepsilon^{IJ}-\bar{\psi}_{\mu}^{I}\psi_{\nu]}^{J}\varepsilon_{IJ}\right)+\frac{i}{4}\epsilon_{\mu\nu]\rho\sigma}\left(\bar{\psi}_{I}^{\rho}\psi_{J}^{\sigma}\varepsilon^{IJ}+\bar{\psi}^{I\rho}\psi^{J\sigma}\varepsilon_{IJ}\right)\right)\Omega_{ijk}=0\,.\\
    \end{aligned}
    \label{eq:IIB-5-form Bianchi}
\end{equation}
This is a combination of the 4d field strength equation (\ref{eq:4d-eom}) and Bianchi identity (\ref{eq:4d-bainchi}) and imposing the complex conjugate equation, we can decouple the two and show overall consistency. 
The other non-trivial part of the Bianchi identities are the $\mu\nu\rho\sigma mn$-component for $\bF_{(5)}$ and the $\mu\nu\rho\sigma$-component for $\bF_{(3)}$, which need the 4d gravitino equation of motion and Fierz identity to close, similarly to the treatment of \eqref{template}.

Moving on to the equations of motion, the only equations that are not satisfied immediately are 
\begin{equation}
    \begin{aligned}
        \hat{\bF}_{(5)}=*\hat{\bF}_{(5)}\,,\qquad 0=\Gamma^{AB}\left(D_{A}\left(\hat{\bom}\right)\bPsi_{B}+\frac{i}{192}\hat{\tilde{\bF}}_{ACDEF}\Gamma^{CDEF}\bPsi_{B}\right)\,.
    \end{aligned}
\end{equation}
The supercovariant self-dual 5-form field strength under the embedding Ansatz gives 
\begin{equation}
    \hat{\tilde{\bF}}_{(5)} = \frac12 \hat{F}^+_{(2)}\wedge\Omega_{(3)} + \frac12 \hat{F}^-_{(2)}\wedge\bar{\Omega}_{(3)}\,.
\end{equation}
The ${\bF}_{(5)}$ equation of motion is therefore satisfied like \eqref{eq:IIB-self-dual-eom}, once we impose (anti)-self-duality of the supercovariant field strength $\hat{F}^\pm_{(2)}$. The gravitino equation of motion has the same structure as the gravitino supersymmetry transformation \eqref{eq:IIB-susy} under our Ansatz. We may therefore employ a similar argument as after \eqref{eq:gravitino-eom} to conclude our proof of consistency.
    
\end{appendices}

\bibliographystyle{JHEP}

\baselineskip 11pt

\bibliography{CY3.bib}

\end{document}